\documentclass[
preprint,
aps,
prc,
showpacs,
a4paper,
nofootinbib,
superscriptaddress,
floatfix
]{revtex4-1}

\usepackage[%
	pdftitle={Heavy quark production at RHIC and LHC within a partonic transport model},%
	pdfauthor={Jan Uphoff, Oliver Fochler, Zhe Xu, and Carsten Greiner},%
	pdfsubject={Heavy quark production at RHIC and LHC within a partonic transport model},
	pdfstartview=FitH,
  pdfpagemode=UseNone,
	bookmarksopen=true
	]{hyperref}

\usepackage[
   centertags, % (default) center tags vertically
   sumlimits,  %(default) Place the subscripts and superscripts of summation
   intlimits,  % Like sumlimits, but for integral symbols.
   namelimits, % (default) Like sumlimits, but for certain 'operator names' such as
]{amsmath} %
\usepackage{amssymb}

\makeatletter
\def\tagform@#1{\maketag@@@{\ignorespaces#1\unskip\@@italiccorr}}
\let\orgtheequation\theequation
\def\theequation{(\orgtheequation)}
\makeatother

\usepackage[all]{hypcap} % Links to figures point actually on figures itself, not on caption
\usepackage{dcolumn}
\usepackage{multirow}

\usepackage[english]{babel} 

\usepackage[%
]{graphicx}

\newcommand{\beq}{\begin{equation}}
\newcommand{\eeq}{\end{equation}}
\newcommand{\umbruch}{\nonumber \\}

\newcommand{\gnuplotwidth}{0.76\textwidth}% \newcommand{\gnuplotwidth}{1.0\columnwidth}

\begin{document}

\title{Heavy quark production at RHIC and LHC within a partonic transport model}

\author{Jan Uphoff}
\email[E-mail: ]{uphoff@th.physik.uni-frankfurt.de}
\affiliation{Institut f\"ur Theoretische Physik, Johann Wolfgang 
Goethe-Universit\"at Frankfurt, Max-von-Laue-Str. 1, 
D-60438 Frankfurt am Main, Germany}

\author{Oliver Fochler}
\affiliation{Institut f\"ur Theoretische Physik, Johann Wolfgang 
Goethe-Universit\"at Frankfurt, Max-von-Laue-Str. 1, 
D-60438 Frankfurt am Main, Germany}

\author{Zhe Xu}
\affiliation{Frankfurt Institute for Advanced Studies, Ruth-Moufang-Str. 1, D-60438 Frankfurt am Main, Germany}
\affiliation{Institut f\"ur Theoretische Physik, Johann Wolfgang 
Goethe-Universit\"at Frankfurt, Max-von-Laue-Str. 1, 
D-60438 Frankfurt am Main, Germany}

\author{Carsten Greiner}
\affiliation{Institut f\"ur Theoretische Physik, Johann Wolfgang 
Goethe-Universit\"at Frankfurt, Max-von-Laue-Str. 1, 
D-60438 Frankfurt am Main, Germany}

\date{\today}

\begin{abstract}
The production and space-time evolution of charm and bottom quarks in nucleus-nucleus collisions at RHIC and LHC are investigated with the partonic transport model  BAMPS (\emph{Boltzmann Approach of MultiParton Scatterings}). Heavy quarks, produced in primary hard parton scatterings during nucleon-nucleon collisions, are sampled using the Monte Carlo event generator PYTHIA or the leading order mini-jet model in conjunction with the Glauber model, revealing a strong sensitivity on the parton distribution functions, scales, and heavy quark mass. In a comprehensive study exploring different charm masses, K factors, and possible initial gluon conditions, secondary production and the evolution of heavy quarks are examined within a fully dynamic BAMPS simulation for central heavy ion collisions at RHIC and LHC.
Although charm production in the quark-gluon plasma can be neglected at RHIC, it is significant at LHC but very sensitive to the initial conditions and the charm mass. Bottom production in the quark-gluon plasma, however, is negligible both at RHIC and LHC.
\end{abstract}


\maketitle

\section{Introduction}
Experimental observations from relativistic heavy ion collisions at the BNL Relativistic Heavy Ion Collider (RHIC) indicate the production of a hot and dense partonic medium \cite{Adams:2005dq,Adcox:2004mh,Arsene:2004fa,Back:2004je}, commonly referred to as the quark-gluon plasma (QGP). A good agreement between ideal hydrodynamic simulations \cite{Kolb:2000sd,Heinz:2001xi,Huovinen:2001cy} and measurements \cite{Adams:2003am,Adler:2003kt} of the elliptic flow hints at fast thermalization of this system, which behaves like a nearly perfect fluid, that is, possesses a very small shear viscosity to entropy density ratio \cite{Romatschke:2007mq,Xu:2007jv}.

Heavy quarks, that is, charm and bottom, are a unique probe for this medium. 
Due to their large mass ($M_Q \gg \Lambda_{\rm QCD}$, $Q=c,b$), a large amount of energy is needed to produce heavy quarks. Such high energy densities are primarily found at the early stage of heavy ion collisions: in hard  scatterings of partons in the nucleons of the heavy ions or during the early phase of the QGP. In that energy domain, the running coupling of the strong interaction $\alpha_s$ is small and nearly constant \cite{Bethke:2006ac}. Therefore, heavy quark production should be describable within the framework of perturbative QCD (pQCD) \cite{Nason:1987xz,Nason:1989zy}, even for small transverse momenta.

Another theoretically proposed implication of the large mass is the ``dead cone effect'' \cite{Dokshitzer:2001zm,Zhang:2003wk}, which implies a smaller energy loss via gluon radiation compared to light quarks and delays the thermalization of heavy quarks by a factor of $\sim M_Q/T$ \cite{Rapp:2008qc}, resulting in a thermalization time of about the lifetime of the QGP.  In contrast, the  experimentally observed energy loss \cite{Abelev:2006db,Adare:2006nq} and elliptic flow \cite{Adare:2006nq} of heavy flavor is comparable to that of light quarks, the reason for this puzzle being under investigation \cite{Armesto:2005mz,vanHees:2005wb,Moore:2004tg,Wicks:2005gt,Peigne:2008nd,Gossiaux:2008jv}. 
Due to the -- in principle -- unique identification of heavy quarks because of their flavor and roughly known production time during the early stage of the collision, their distributions can reveal information about the interaction history, rendering them as an ideal probe of the medium.

In principle, heavy quarks can be produced at three stages of the collision: during initial hard parton scatterings, in the QGP, or during the hadronic phase. As we will show in this article, most of the heavy quarks are created in hard parton scatterings during initial nucleon-nucleon collisions in the heavy ion collision at RHIC. However, at the CERN Large Hadron Collider (LHC) secondary production during the QGP phase becomes important. Here, one often distinguishes between prethermal and thermal production depending on whether the partonic medium is already thermalized or not.
In the present article, we focus on the first two possibilities of heavy quark production and neglect the hadronic phase, which hardly contributes to the heavy quark yield.

An interesting signature in heavy ion collisions is the suppression \cite{Matsui:1986dk} or enhancement of heavy quarkonia like $J/\psi$ (hidden charm) \cite{Andronic:2006ky}.
In this article, however, we examine only open heavy quark production and postpone the investigation of hidden charm or bottom to future investigations. For the simulation of the heavy quark production in the QGP, we use  the partonic transport model called the \emph{Boltzmann Approach of MultiParton Scatterings} (BAMPS) and study the impact of various initial gluon distributions such as from PYTHIA \cite{Sjostrand:2006za}, the mini-jet model \cite{Kajantie:1987pd,Eskola:1988yh}, and the color glass condensate \cite{Iancu:2003xm,Drescher:2006ca}. The initial heavy flavor yield in hard parton scattering is obtained with PYTHIA and compared with that from leading order (LO) pQCD.

Most of the time we will only refer to charm quarks instead of heavy quarks in general. However, the majority of the concepts and findings apply qualitatively also to bottom quarks if one takes the larger mass into account. For  LHC calculations, we will explicitly mention bottom production and make predictions on the bottom yield.

This article is organized as follows. First we introduce our model BAMPS and the partonic cross sections for heavy quark production. In Section \ref{sec:box} we use this model to estimate the chemical equilibration time of a static system filled with gluons and charm quarks and compare it to the analytic solution. The next two sections address charm and bottom production in heavy ion collisions. After investigating  the initial heavy quark distributions from primary nucleon-nucleon scatterings in Section \ref{sec:ini_dist}, we study the heavy quark production during the QGP phase in Section \ref{sec:prod_qgp}. 
Here, the charm quark fugacity is also calculated and related to the estimated chemical equilibration time from Section \ref{sec:box}.
Finally, we conclude with a short summary in
Section \ref{sec:conclusion}.

\section{Parton cascade BAMPS}
\label{sec:bamps}
For the simulation of the QGP phase we use the partonic transport model BAMPS \cite{Xu:2004mz,Xu:2007aa}, which stands for \emph{Boltzmann Approach of MultiParton Scatterings}.
BAMPS simulates the fully  $3+1$ space-time evolution of the QGP at heavy ion collisions by solving the Boltzmann equation,
\begin{equation}
\label{boltzmann}
\left ( \frac{\partial}{\partial t} + \frac{{\mathbf p}_i}{E_i}
\frac{\partial}{\partial {\mathbf r}} \right )\, 
f_i({\mathbf r}, {\mathbf p}_i, t) = {\cal C}_i^{2\rightarrow 2} + {\cal C}_i^{2\leftrightarrow 3}+ \ldots  \ ,
\end{equation}
dynamically for on-shell partons with a stochastic transport algorithm and pQCD interactions. ${\cal C}_i$ are the relevant collision integrals, and $f_i({\mathbf r}, {\mathbf p}_i, t)$ the one-particle distribution function of species $i=g,\, c,\, \bar{c}$, since light quarks are not included yet. Bottom quarks can also be investigated if one changes $c \rightarrow b$ and $M_c \rightarrow M_b$. In addition to the binary collisions $2\rightarrow 2$, also $2\leftrightarrow 3$ scatterings for the gluons are possible. That is, the following processes are implemented in BAMPS:
\begin{align}
\label{bamps_processes}
	g+g &\rightarrow g+g \nonumber\\
	g+g &\rightarrow g+g+g \nonumber\\
	g+g+g &\rightarrow g+g	 \nonumber\\
	g+g &\rightarrow c +\bar{c} \umbruch
	c+ \bar{c} &\rightarrow g+g \umbruch
	g+c &\rightarrow g+c \nonumber\\
	g+\bar{c} &\rightarrow g+\bar{c}
\end{align}

Within the stochastic method, the probability for a collision of two particles during a time step $\Delta t$ in a volume element $\Delta V$ can be obtained from the collision term of the Boltzmann equation \cite{Xu:2004mz}:
\begin{align}
\label{p22}
P_{2\rightarrow 2} = 
v_{\rm rel} \, \sigma_{2\rightarrow 2}\frac{\Delta t}{\Delta V} \ ,
\end{align}
where \cite{Cleymans:1992je}
\begin{align}
\label{v_rel}
	v_{\rm rel}= \frac{\sqrt{(P_1^\mu P_{2 \mu})^2-m_1^2 m_2^2}}{E_1 E_2}
\end{align}
stands for the relative velocity, and $\sigma_{2\rightarrow 2}$ for the binary cross sections. The cross section for $gg \rightarrow gg$ as well as the treatment of the $gg\leftrightarrow ggg$ interactions are given in \cite{Xu:2004mz}.

The most dominant charm production process is gluon fusion $g+g \rightarrow c+\bar{c}$
with a differential cross section of
\begin{equation}
\label{eq:cs_dt_gg_ccb}
\frac{\mathrm{d}\sigma_{gg  \rightarrow c \bar{c}}}{\mathrm{d}t} = \frac{{|\overline{\mathcal{M}}_{gg \rightarrow c \bar{c}}|}^2}{16 \pi s^2} \ .
\end{equation} 
${|\overline{\mathcal{M}}_{gg \rightarrow c \bar{c}}|}^2$ is the averaged matrix element, that is, the averaged (summed) over color and spin of the incoming (outgoing) particles. It can be expressed in terms of the Mandelstam variables $s$, $t$ and $u$ \cite{Levai:1994dx,Combridge:1978kx}:
\begin{flalign}
\frac{{|\overline{\mathcal{M}}_{gg \rightarrow c \bar{c}}|}^2}{\pi^2 \alpha_s^2} &=   
\frac{12}{s^2}(M^2-t)(M^2-u) + \frac{8}{3}\left(\frac{M^2-u}{M^2-t} +\frac{M^2-t}{M^2-u}\right) \nonumber \\
& \quad	-\frac{16M^2}{3}\left[ \frac{M^2+t}{(M^2-t)^2}+\frac{M^2+u}{(M^2-u)^2} \right] - \frac{6}{s}(2M^2-t-u) \nonumber \\
& \quad	+  \frac{6}{s}\frac{M^2(t-u)^2}{(M^2-t)(M^2-u)} - \frac{2}{3} \frac{M^2(s-4M^2)}{(M^2-t)(M^2-u)} \ .
\end{flalign} 
$M$ denotes the mass of the charm (or bottom) quarks. If not otherwise specified, we use $M_c = 1.5\, {\rm GeV}$ for charm and $M_b = 4.8 \, {\rm GeV}$ for bottom quarks, which are widely adopted in the literature \cite{Muller:1992xn,Levai:1994dx,Gavai:1994gb,Levai:1997bi,Cacciari:2005rk,Andronic:2006ky,Rapp:2008qc,Nason:1999ta,Frixione:1997ma,Vogt:2001nh}.  For the coupling of the strong interaction we choose to take a constant value of $\alpha_s = 0.3$.

After substituting $u$ with the relation $s+t+u = 2M^2$, the total cross section is obtained by integrating \autoref{eq:cs_dt_gg_ccb},
\cite{Combridge:1978kx,Gluck:1977zm,Babcock:1977fi,Barger:1981rx,Matsui:1985eu} 
\begin{align}
\label{cs_gg_ccb}
	  \sigma_{gg \rightarrow c\bar{c}}(s) = \frac{\pi \alpha_s^2}{3s} \left[ \left(1+\frac{4M^2}{s}+\frac{M^4}{s^2} \right) \log\left(\frac{1+\chi}{1-\chi}\right)-\left( \frac{7}{4}+\frac{31M^2}{4s} \right)\chi  \right] \ ,
\end{align}
with the abbreviation
\begin{align}
\label{abk_chi}
	  \chi = \sqrt{1-\frac{4M^2}{s}}\ .
\end{align}

The angular distribution of the charm and anti-charm quarks after the collision is sampled using \autoref{eq:cs_dt_gg_ccb}.

The back reaction $c+\bar{c} \rightarrow g+g$ is negligible in heavy ion collisions at RHIC and LHC \cite{Andronic:2006ky} due to the small number of produced charm quarks compared to the number of gluons. However, we want to implement the back reaction in a box model in Section \ref{sec:box} in order to estimate the chemical equilibration time scale. The cross section of this process can be obtained through detailed balance,
\begin{align}
\label{cs_ccb_gg}
	\sigma_{ c\bar{c}\rightarrow gg} = \frac{1}{2} \, \frac{64}{9} \, \frac{1}{\chi^2} \,   \sigma_{gg \rightarrow c\bar{c}} \ . 
\end{align}
The factor $1/2$ comes into play due to gluons being identical particles. $64/9$ takes the color and spin averaging and summation into account whereas $1/\chi^2$ is a kinematical factor.

The process for charm production by light quark and anti-quark annihilation $q+ \bar{q} \rightarrow c +\bar{c} $ is not yet included in the cascade. However, we use this process to estimate the initial charm quark yield in heavy ion collisions (see Section \ref{sec:ini_charm}). The cross section for that process is given, for instance, in \cite{Combridge:1978kx,Gluck:1977zm,Babcock:1977fi,Barger:1981rx,Matsui:1985eu}.

\section{Charm production in a static medium}
\label{sec:box}

In this section, we test our implementation with BAMPS by comparing our numerical results to the analytic solution of a rate equation. To keep the problem simple and find an analytic solution, we consider a box of gluons and charm quarks, in which only charm-anti-charm production through gluon fusion and the back reaction
\begin{align}
\label{rate_prozesse}
	g+g &\rightarrow c +\bar{c} \nonumber\\
	c+ \bar{c} &\rightarrow g+g
\end{align}
are allowed, whereas all other interactions between gluons and charm quarks are forbidden. However, we checked that adding other possible processes like $gg \rightarrow gg$, $ggg \leftrightarrow gg$,  $g c \rightarrow gc$, and $g \bar{c} \rightarrow g\bar{c}$ does not have an impact on our findings.
The additional $2 \rightarrow 2$ processes ensure the kinetic equilibration and $ggg \leftrightarrow gg$ maintains the chemical equilibration of the gluons, but they do not affect the chemical equilibration time scale of the charm quarks.

For the system with only processes from \eqref{rate_prozesse}, we write down a rate equation for the evolution of the charm density \cite{Matsui:1985eu,Biro:1993qt,Levai:1994dx,Zhang:2008zzc}
\begin{align}
\label{rate_rate_glg1}
 \partial_\mu \left( n_{c} u^\mu\right) = R_{gg \rightarrow c \bar{c}} -R_{c \bar{c} \rightarrow gg} \ ,
\end{align}
where $R_{gg \rightarrow c \bar{c}}$ and $R_{c \bar{c} \rightarrow gg}$ denote the rates and $u^\mu = \gamma \;( 1, \, \vec{v} )$  ($\gamma$ being the Lorentz factor) the four-velocity of the considered volume element with charm density $n_{c}$, which is identical to the charm pair density $n_{c\bar{c}}$ since charm and anti-charm quarks are always produced in pairs.

The rates are given by \cite{Zhang:2008zzc,Biro:1993qt,Levai:1994dx}
\begin{align}
\label{rate_rategg}
	 R_{gg \rightarrow c \bar{c}} 	 &= \frac{1}{2}\left\langle \sigma_{gg \rightarrow c \bar{c}} \; v_{\rm rel} \right\rangle n_g^2 =: \frac{1}{2}\sigma_g n_g^2 
	\\
\label{rate_rateccb}
	R_{c \bar{c} \rightarrow gg} &= \left\langle \sigma_{c \bar{c} \rightarrow gg} \; v_{\rm rel} \right\rangle n_c n_{\bar{c}} =: \sigma_c n_{c\bar{c}}^2 \ .
\end{align}
In \autoref{rate_rategg}, $n_g$ stands for the gluon density and the factor $1/2$ is needed to take into account that gluons are identical particles.
$\sigma_g := \left\langle \sigma_{gg \rightarrow c \bar{c}}\; v_{\rm rel} \right\rangle$ and $\sigma_c := \left\langle \sigma_{ c \bar{c}\rightarrow  gg}\; v_{\rm rel} \right\rangle$ are the mean cross sections weighted with the relative velocity $v_{\rm rel}$ defined in \autoref{v_rel}.
In all calculations we use a constant coupling of $\alpha_s = 0.3$ for the cross sections, which are given in equations \eqref{cs_gg_ccb} and \eqref{cs_ccb_gg}.

\subsection{Analytic solution of the rate equation}

For a static box the four-velocity of each volume element is $u^\mu = ( 1, \, 0, \, 0, \,0)$ and the rate equation \eqref{rate_rate_glg1} simplifies to
\begin{align}
	\label{rate_rate_glg2}
 \partial_t  n_{c\bar{c}} = R_{gg \rightarrow c \bar{c}} -R_{c \bar{c} \rightarrow gg}
 =\frac{1}{2}\sigma_g n_g^2 
	- \sigma_c n_{c\bar{c}}^2
\end{align}
In chemical equilibrium the rates are equal, $R^{\rm eq}_{gg \rightarrow c \bar{c}} =R^{\rm eq}_{c \bar{c} \rightarrow gg}$. From that relation one can obtain the charm density in chemical equilibrium,
\begin{align}
\label{n_ccb_eq}
	n^{\rm eq}_{c\bar{c}} = \frac{1}{2} \frac{n_{\rm tot}}{\frac{1}{\sqrt{2\sigma_g/\sigma_c}}+1} \ ,
\end{align}
where $n_{\rm tot} = n_g + 2 n_{c\bar{c}}$ denotes the constant total particle density.

Initially the box may only contain gluons which are chemically and thermally equilibrated. That is, the initial gluon density for an initial temperature of $T_0$ reads 
\begin{align}
\label{gluon_dichte_temp}
	n_g (t=0) = \nu_g \frac{T_0^3}{\pi^2} 
\end{align}
(gluons are treated as Boltzmann particles, $\nu_g = 2\cdot 8 = 16$ is the degeneracy factor for gluons), whereas for charm quarks $n_{c\bar{c}} (t=0) = 0$. 

Solving the rate equation \eqref{rate_rate_glg1} provides the time evolution of the charm density, from which one can read off the chemical equilibration time scale.
Taking into account that the total particle number is constant, the charm quark density as a function of time is given by
\begin{align}
\label{rate_lsg}
	n_{c\bar{c}} (t) = \frac{1}{2} \frac{n_{\rm tot}}{1-\zeta^2}\left[1-
	\frac{\mathrm{e}^{2t/\tau}\left(\zeta+1\right)-\zeta+1}
	{\mathrm{e}^{2t/\tau}\left(\frac{1}{\zeta}+1\right)-\frac{1}{\zeta}+1}
	\right] \ ,
\end{align}
where the abbreviations
\begin{align}
	\zeta &= \frac{n^{\rm eq}_g}{2n^{\rm eq}_{c\bar{c}}} = \frac{n_{\rm tot}-2n^{\rm eq}_{c\bar{c}}}{2n^{\rm eq}_{c\bar{c}}}  \\
\label{abbr_tau}
	\tau &= \frac{2n^{\rm eq}_{c\bar{c}}}{\sigma_g n_{\rm tot} n^{\rm eq}_g}= \frac{2n^{\rm eq}_{c\bar{c}}}{\sigma_g (n_{\rm tot}^2 - 2n_{\rm tot}n^{\rm eq}_{c\bar{c}})}
\end{align}
have been introduced.

The solution is implicitly dependent on the temperature via $\sigma_c$ and $\sigma_g$ and is only valid assuming that the temperature stays constant over the whole period of time. As we will see in the next section, due to the mass creation this is, however, not exactly but in a good approximation the case.

\subsection{Comparison between analytic and numerical solution}

The numerical solution is obtained with the parton cascade BAMPS (see Section \ref{sec:bamps}).
In order to compare it to the analytic solution from the previous section only the two  processes from \ref{rate_prozesse} are allowed. 
Again, the initial charm density is  $n_{c \bar c}(t=0) = 0$ and the gluons are sampled thermally employing $n_g(t=0)$ according to \autoref{gluon_dichte_temp} with temperatures of 400\,MeV and 800\,MeV, which correspond to the expected temperatures of the quark-gluon plasma at RHIC \cite{Thews:2000rj,Soff:2000eh, Fries:2002kt,Turbide:2005fk,Rapp:2000pe,Cooper:2002td,Wang:1996yf} and LHC \cite{Fries:2002kt,Cooper:2002td,Wang:1996yf}, respectively (cf. also Figures \ref{fig:rhic_temp_fugacity_qgp} and \ref{fig:lhc_temp_fugacity_qgp}).

\subsubsection{Charm production at RHIC temperature}
\label{sec:rates_comp_ana_num_rhic}

The initial temperature of the gluons is taken as $T_g=400\, {\rm MeV}$ \cite{Thews:2000rj,Soff:2000eh, Fries:2002kt,Turbide:2005fk,Rapp:2000pe,Cooper:2002td,Wang:1996yf}. Being precisely, during the evolution of the box the temperature is ill-defined because of the gluons being not in thermal equilibrium (due to the neglect of elastic scatterings among the gluons). However, one can extend the temperature definition to this non-equilibrium domain, introducing an ``effective gluon temperature''
\begin{align}
\label{t_eff}
	T_g(t) = \frac{E_g(t)}{3 N_g(t)}= \frac{\epsilon_g(t)}{3 \, n_g(t)} \ .
\end{align}
Here $E_g$ ($N_g$) stands for the total gluon energy (number) in the box, and $\epsilon_g$ and $n_g$ for the respective densities. 

\begin{figure}
	\centering
\includegraphics[width=\gnuplotwidth]{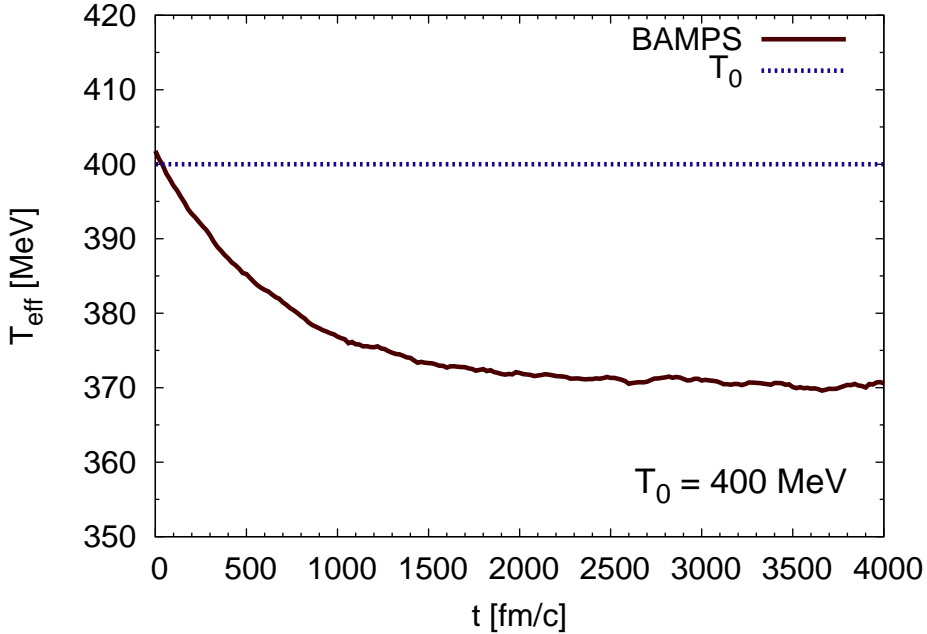}%eps
	\caption{(Color online) Time evolution of the gluon temperature in a static system with an initial temperature of $T_0=400\, {\rm MeV}$. The temperature decreases due to the creation of massive charm quarks.}
	\label{fig:rate_temp_rhic}
\end{figure}

The time evolution of the gluon temperature is illustrated in \autoref{fig:rate_temp_rhic}. 
The temperature declines to about $T_g = 370\, {\rm MeV}$ since part of the kinetic energy is used for the production of massive charm quarks. Consequently, this implies a problem in comparing the numerical results with the analytic solution, which is only valid for a constant temperature. Therefore, we will compare the analytic solutions for initial and final temperatures against the numerical results at the beginning and at the end, respectively.

\begin{figure}
	\centering
\includegraphics[width=\gnuplotwidth]{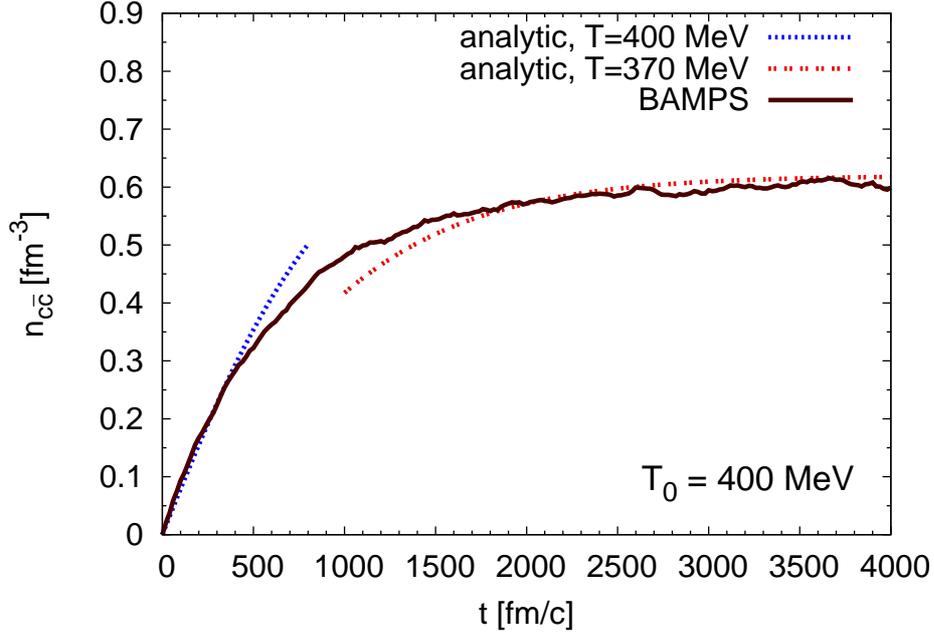}%eps
	\caption{(Color online) Time evolution of the charm pair density $n_{c\bar{c}}$ in a static medium with an initial temperature of $T_0=400\, {\rm MeV}$. For comparison, the analytic solutions from \eqref{rate_lsg} for initial and final temperature are also shown.}
	\label{fig:rate_dichte_rhic}
\end{figure}

\autoref{fig:rate_dichte_rhic} shows the time evolution of the charm pair density $n_{c\bar{c}}$. 
In addition, the analytic solutions of the rate equation for the charm density from \autoref{rate_lsg} for initial and final temperature are also plotted. In the beginning, the numerical results are in very good agreement with the analytic solution for $T=400\, {\rm MeV}$. Thereafter, the temperature of the gluon plasma decreases (cf. \autoref{fig:rate_temp_rhic}), and the numerical results must be compared with the analytic solution for the stationary final temperature of $T = 370\, {\rm MeV}$, both being in good agreement.

\begin{figure}
	\centering
\includegraphics[width=\gnuplotwidth]{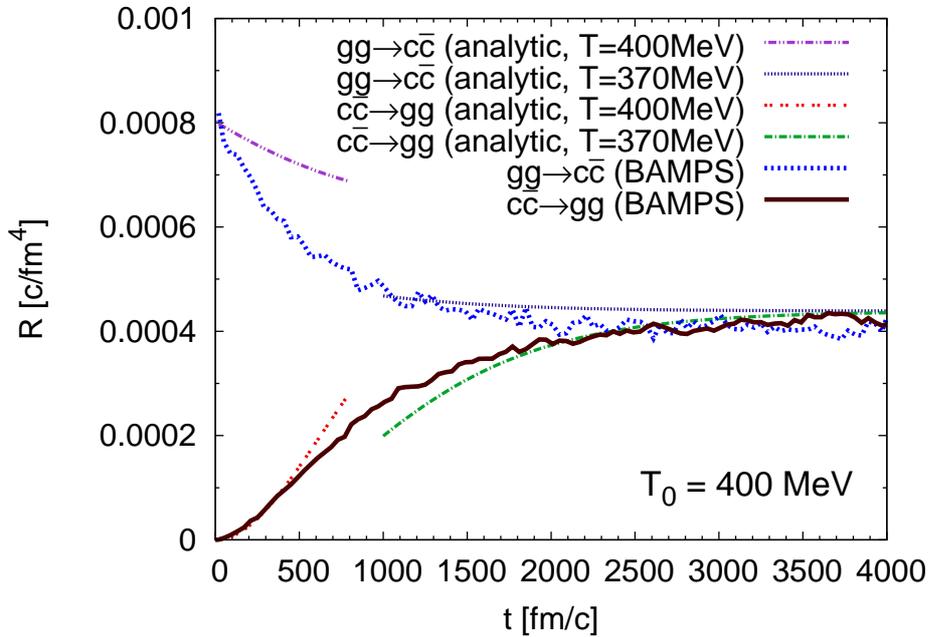}%eps
	\caption{(Color online) Time evolution of the rates in a static system with an initial temperature of $T_0=400\, {\rm MeV}$. Again, the analytic solutions for initial and final temperatures are also plotted.}
	\label{fig:rate_raten_rhic}
\end{figure}

The rates for $gg \rightarrow c\bar{c}$ and the back reaction are shown in \autoref{fig:rate_raten_rhic}. 
As one would expect from the principle of detailed balance, the rates converge to a mutual constant equilibrium value with time.
The chemical equilibration time scale may be defined as $\tau_{\rm eq}$.
By fitting the curve in \autoref{fig:rate_dichte_rhic} with 
\begin{align}
\label{fit_tau}
 n_{c\bar{c}}(t) =	n^{\rm eq}_{c\bar{c}} \,(1- {\rm e}^{-t/\tau_{\rm eq}}) \ ,
\end{align}
we obtained an approximate value of $700\, {\rm fm}/c$ for $\tau_{\rm eq}$. Of course, this result of the chemical equilibration time was estimated within a simple static model without expansion. However, the order of this result should be roughly the same as in realistic heavy ion collisions, which indicates that not very many charm quarks are produced during the QGP phase at RHIC considering this large time scale of chemical equilibration compared to the lifetime of the QGP, as we will explicitly show in Section \ref{sec:prod_qgp_rhic}.

\begin{figure}
	\centering
\includegraphics[width=\gnuplotwidth]{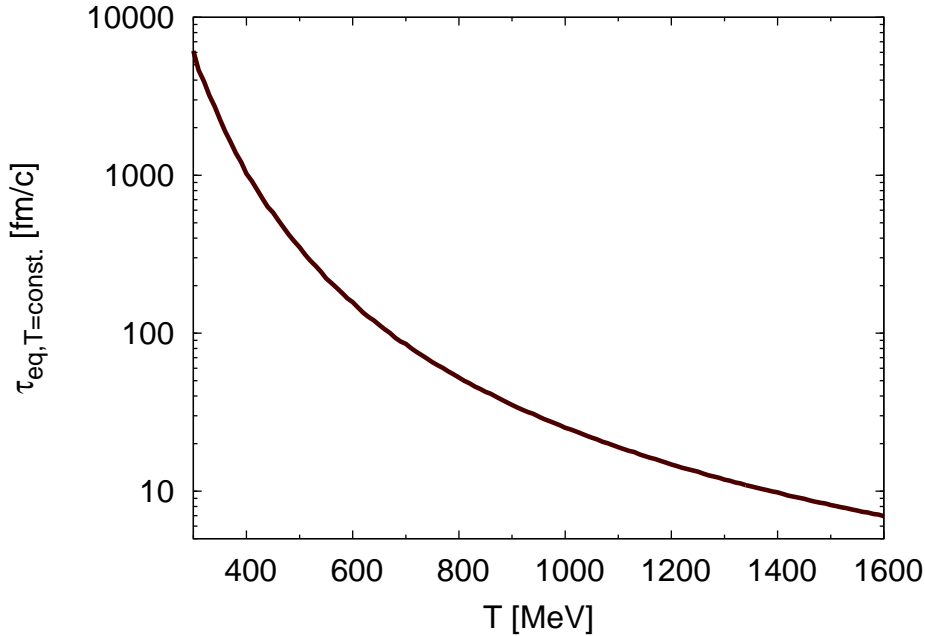}%eps
	\caption{Temperature dependence of the chemical equilibration time $\tau_{{\rm eq},T={\rm const.}}$ for charm production in a static medium of a constant temperature.}
	\label{fig:temp_tau}
\end{figure}

An obvious expression for the temperature dependence of $\tau_{\rm eq}$ can be obtained by differentiating \autoref{fit_tau} by $t$, equating with \autoref{rate_rate_glg2}, and evaluating at $t=0$:
\begin{align}
\tau_{{\rm eq}} = \frac{n^{\rm eq}_{c\bar{c}}}{R_{gg \rightarrow c \bar{c}}|_{t=0}} \ .
\end{align}
It is related to $\tau$ from \autoref{abbr_tau}, which does not have a direct physical meaning, by
\begin{align}
\tau_{{\rm eq}} = \frac{1}{1+\sqrt{2\sigma_g/\sigma_c}} \; \tau \ .
\end{align}
The temperature dependence of $\tau_{{\rm eq}}$ can be obtained analytically with the assumption of a constant temperature throughout the evolution of the system. In order to distinguish it from the $\tau_{{\rm eq}}$ obtained numerically with BAMPS, in which the temperature drop due to the mass creation is considered, we label the analytic solution as $\tau_{{\rm eq},T={\rm const.}}$.
\autoref{fig:temp_tau} depicts the temperature dependence of $\tau_{{\rm eq},T={\rm const.}}$. 
Compared to the results from BAMPS, these values for a constant temperature are slightly larger. For  high temperatures, which can be present at LHC at a very early stage of the QGP phase (cf. \autoref{fig:lhc_temp_fugacity_qgp}), the chemical equilibration time lies in the same range as the lifetime of the QGP. Consequently, a substantial charm production in the QGP at LHC can be expected, if the initial temperature of the medium is large. We will address this in more detail in Section~\ref{sec:prod_qgp_lhc}.

\begin{figure}
	\centering
\includegraphics[width=\gnuplotwidth]{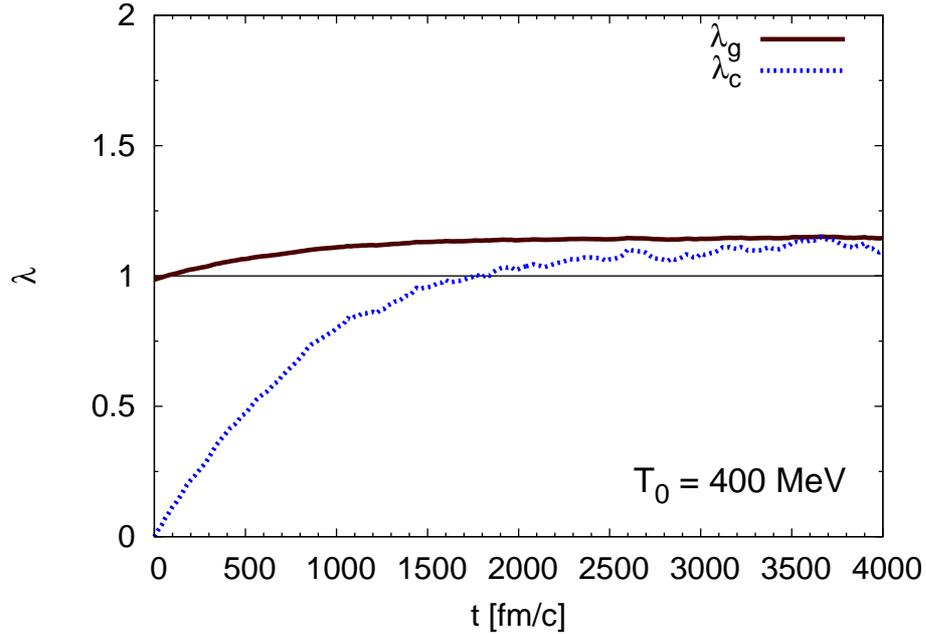}%eps
	\caption{(Color online) Evolution of gluon and charm quark fugacities in a static system. The final value is above 1 due to the fixed number of particles and the temperature drop.}
	\label{fig:fugacity_rhic}
\end{figure}

The fugacity, which is defined by 
\begin{align}
\label{fugacity}
	\lambda_i (t) = \frac{n_i (t)}{n^{\rm chem. eq}_i}
\end{align}
for particle species $i$, is an interesting variable for investigating the chemical equilibration.
\autoref{fig:fugacity_rhic} shows the fugacities of gluons and charm quarks in our simulation as a function of time.
Due to detailed balance, they adopt the same value in equilibrium, although that differs from 1, a consequence of the total particle number being constant. As a note, explicitly allowing inelastic processes ($gg \leftrightarrow ggg$) as in the full transport simulation (see Section \ref{sec:prod_qgp}) ensures chemical equilibration and leads to a final fugacity of 1 both for gluons and charm quarks.

\subsubsection{Charm production at LHC temperature}
\label{sec:rates_comp_ana_num_lhc}
\begin{figure}
	\centering
\includegraphics[width=\gnuplotwidth]{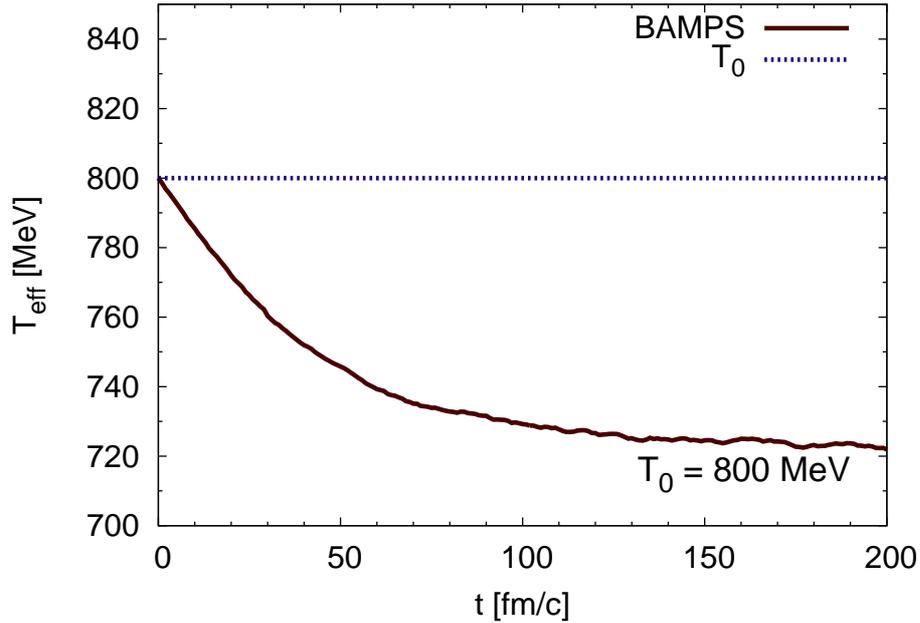}%eps
	\caption{(Color online) Evolution of the gluon temperature in a static medium with an initial temperature of  $T_0=800\, {\rm MeV}$.}
	\label{fig:rate_temp_lhc}
\end{figure}

For the initial gluon temperature, a value of $T_g=800\, {\rm MeV}$ \cite{Fries:2002kt,Cooper:2002td,Wang:1996yf} (cf. also \autoref{fig:lhc_temp_fugacity_qgp}) is chosen, which drops to about $T_g = 720\, {\rm MeV}$ as shown in \autoref{fig:rate_temp_lhc} due to the mass creation.

\begin{figure}
	\centering
\includegraphics[width=\gnuplotwidth]{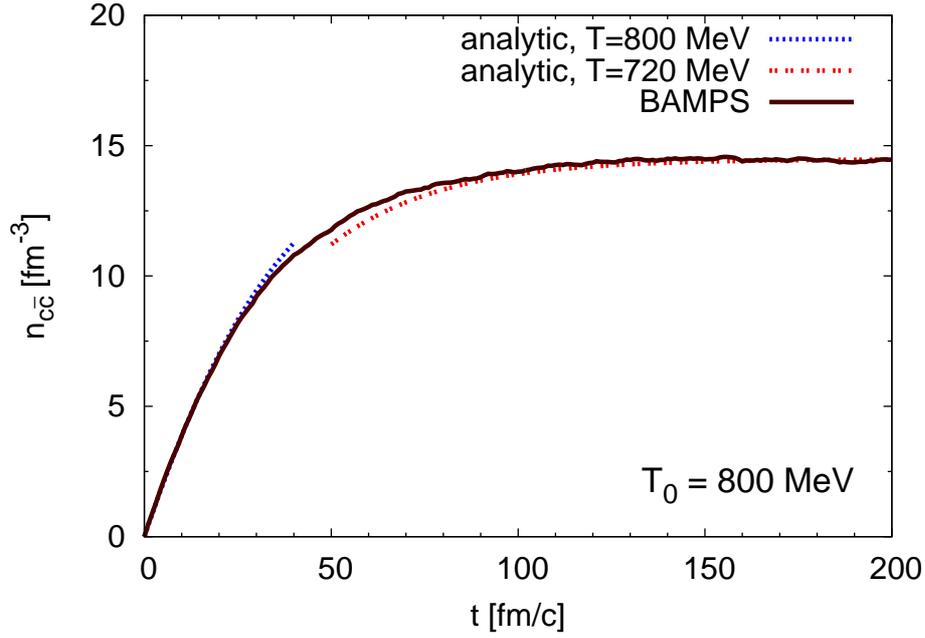}%eps
	\caption{(Color online) Charm quark density as a function of time in a static medium with an initial temperature of $T_0=800\, {\rm MeV}$. The analytic solutions for initial and final temperature are also shown.}
	\label{fig:rate_dichte_lhc}
\end{figure}

\autoref{fig:rate_dichte_lhc} depicts the numerical results of the charm pair density evolution, which is in excellent agreement with the analytic solutions taking the temperature drop into account.
From that figure, using \autoref{fit_tau}, the time scale of chemical equilibration $\tau_{\rm eq}$ can be estimated to about $30\, {\rm fm}/c$, which is much smaller than for RHIC temperature, but still sizable.

\begin{figure}
	\centering
\includegraphics[width=\gnuplotwidth]{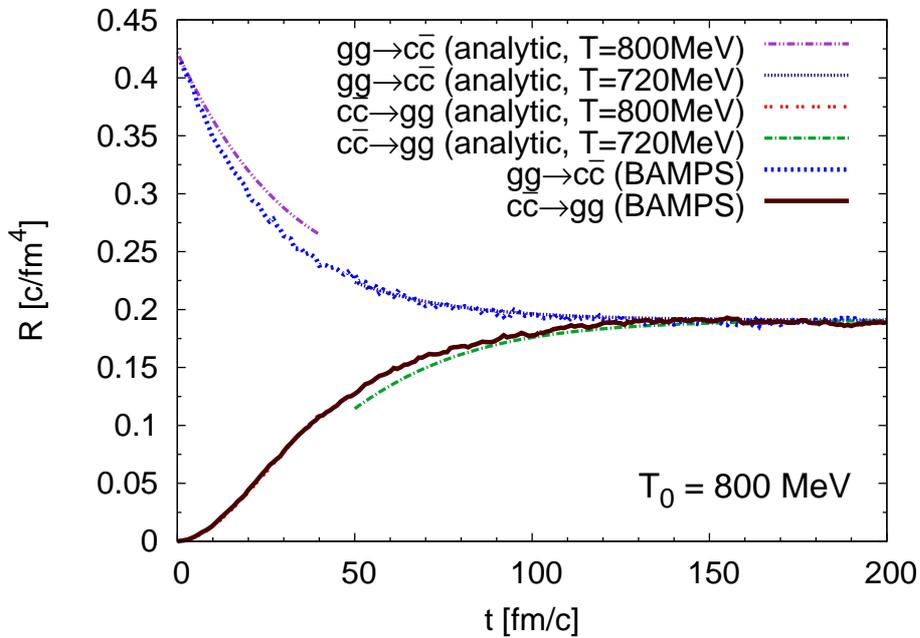}%eps
	\caption{(Color online) Evolution of the rates in a static medium with $T_0=800\, {\rm MeV}$.}
	\label{fig:rate_raten_lhc}
\end{figure}

The rates for both considered processes can be found in \autoref{fig:rate_raten_lhc}.

\section{Initial parton distribution in heavy ion collisions}
\label{sec:ini_dist}

The initial distributions of the partons play a crucial role for the dynamics of the heavy ion collision. In this section, we want to outline a prescription to describe the heavy ion collision -- according to the Glauber model -- as a superposition of nucleon-nucleon collisions, which are sampled with the event generator PYTHIA \cite{Sjostrand:2006za}. In that framework, we also study charm production and compare the yields with experimental data from nucleon-nucleon collisions.
To be able to examine the impact of the initial conditions on charm production during the QGP phase in Section \ref{sec:prod_qgp}, we also discuss other models for the initial parton distributions such as the mini-jet model and the color glass condensate.

The prescription for the position sampling of the partons according to a geometric model is described in great detail in \cite{Xu:2004mz}.

\subsection{Parton and momenta sampling with PYTHIA}
\label{sec:p_sampling_pythia}
For our simulation we use PYTHIA 6.4 \cite{Sjostrand:2006za}, allow hard and soft QCD interactions, and turn off the hadronization.
PYTHIA distinguishes between soft and hard events. Therefore, we define particles stemming from a hard (soft) event as being hard (soft), regardless of their momenta. The only exception to this rule is that beam remnants such as diquarks are always considered as soft. 

For nucleon-nucleon collisions at RHIC with a center of mass energy of $\sqrt{s_{NN}}=200 \, {\rm GeV}$, PYTHIA yields the following results. On average, 53\,\% of all processes are hard and 47\,\% soft. All particles produced in hard processes are partons. These hard partons account for about 58\,\% of all particles created in nucleon-nucleon collisions. On the contrary, in soft processes, most formed particles are non-partonic, for instance, diquarks or excited nucleons. Their fraction is about 28\,\% of all particles, whereas partons which are produced in soft events only account for 14\,\% of all particles.
The total number of produced particles in one nucleon-nucleon collision averages to 7.5, whereof 5.4 are partons. The total energy is, of course, $\sqrt{s_{NN}}=200 \, {\rm GeV}$, but all partons together possess only $30 \, {\rm GeV}$. Consequently, non-partonic particles account for  85\,\% of the total energy. The reason for that is their large energy per particle ratio, which results from the non-partonic particles either being beam remnants or stemming from elastic or diffractive hadronic processes and, therefore,  carrying a huge amount of energy.

The reason for the separation in soft and hard particles lies in the different scaling behavior from nucleon-nucleon to heavy ion collisions.  According to the Glauber model, hard processes scale with the number of binary collisions, \cite{wong} 
\begin{equation}
\label{bin_col}
N_{\rm bin}({\bf b}) \,=\sigma_{\rm p+p}\, T_{AB}({\bf b}) \ ,
\end{equation}
where ${\bf b}$ is the impact parameter, $T_{AB}({\bf b})$ the nuclear overlap function for collisions of two nuclei $A$ and $B$, and $\sigma_{\rm p+p} \approx 40 \, {\rm mb}$ for RHIC and $\sigma_{\rm p+p} \approx 60 \, {\rm mb}$ for LHC the inelastic p+p cross sections. 
For central collisions, we approximate $T_{AA}({\bf b}=0)=A^2/\pi R_A^2$ \cite{Sarcevic:1994ma, Muller:1992xn}, which leads to $T_{AA}({\bf b}=0)=30.4\, {\rm mb}^{-1}$ for Au+Au collisions at RHIC being in excellent agreement with a numerical calculation using \cite{misko_overlap}. Therefore, 
\begin{equation}
N_{\rm bin} \,=\sigma_{\rm p+p}\, T_{AA}({\bf b} =0) \approx 1200 
\end{equation}
is the number of binary collisions at RHIC for $\sqrt{s_{NN}}=200 \, {\rm GeV}$ and ${\bf b}=0$. Taking shadowing into account reduces this number to about $N_{\rm bin}  \approx 1000$ \cite{Adams:2004cb}, which we use as a scaling factor for hard partons, $C_{\rm hard}:= N_{\rm bin} $. To obtain the scaling factor $C_{\rm soft}$ for soft particles, one can make use of energy conservation,
\begin{align}
\label{scaling_energy_con}
	E_{CM} = E_{\rm pp, \, hard} \ C_{\rm hard} + E_{\rm pp, \, soft} \ C_{\rm soft} \ ,
\end{align}
where $E_{CM}$ stands for the total energy available in the heavy ion collision (at RHIC $E_{CM} = 200\, A \, {\rm GeV} = 39\, 400 \, {\rm GeV}$) and  $E_{\rm pp, \, hard (soft)}$ for the total energy of all hard (soft) particles in one nucleon-nucleon collision. Solving this equation for $C_{\rm soft}$ leads to $C_{\rm soft} \approx 100$, which is in the same order as the number of nucleons, the usually used scaling factor for soft particles.

Employing the scaling prescription obtained from \autoref{scaling_energy_con} results in the following particle yields for Au+Au collisions at RHIC with $\sqrt{s_{NN}}=200 \, {\rm GeV}$: Partons from hard processes account on average for about 93\,\% of the produced particles. 4.5\,\% of all particles are non-partonic particles from soft processes, whereas 2.5\,\% are soft partons. The total number of produced particles is about 4600, of which 4400 are partons. 
However, for the energy distribution the picture looks a bit different. Hard partons carry just 53\,\% of the total energy and non-partonic particles with 44\,\% nearly the same amount. Soft partons only account for 3\,\% of the total energy. Consequently, the whole energy deposited by partons, which is available for the parton cascade, is about 56\,\% of the total energy.

\begin{figure}
	\centering
\begin{minipage}{0.02\linewidth}
(a)
\end{minipage}
\begin{minipage}{0.97\linewidth}
\includegraphics[width=\gnuplotwidth]{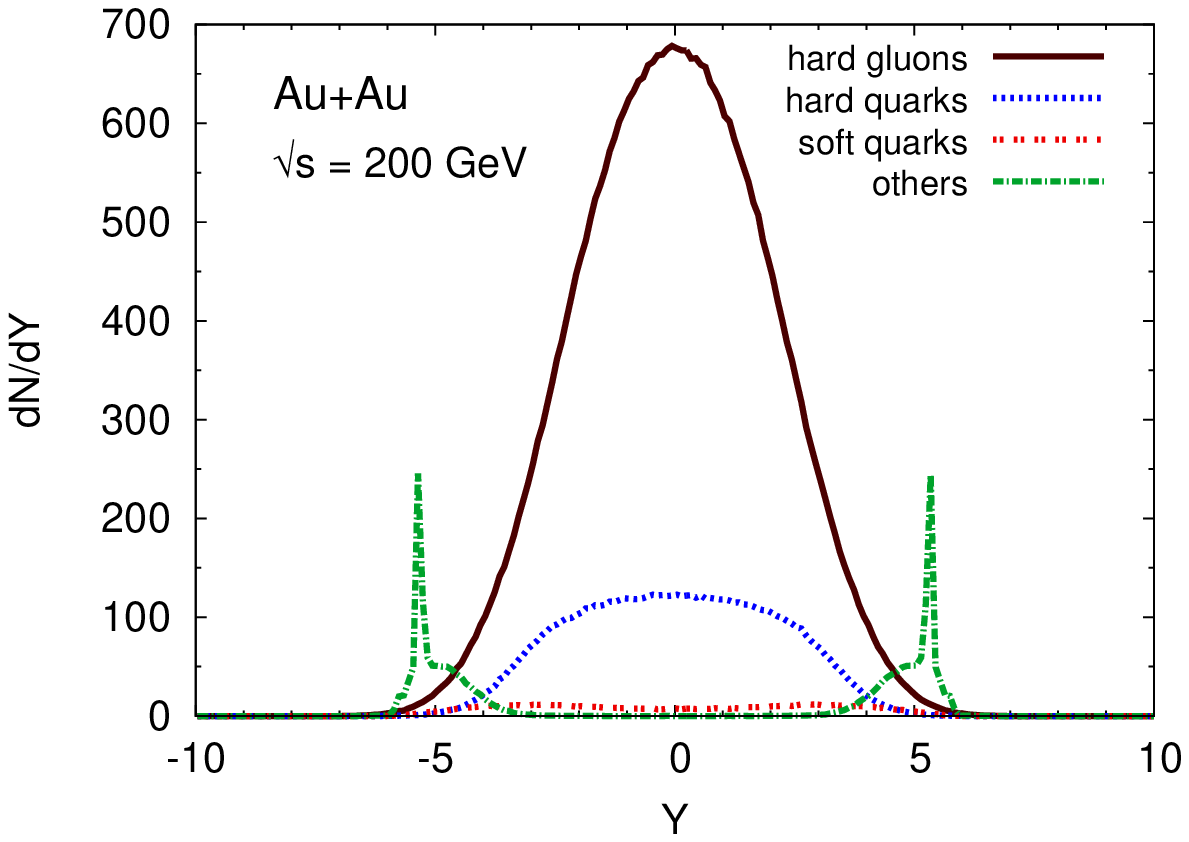}%eps
\end{minipage}

\begin{minipage}{0.02\linewidth}
(b)
\end{minipage}
\begin{minipage}{0.97\linewidth}
\includegraphics[width=\gnuplotwidth]{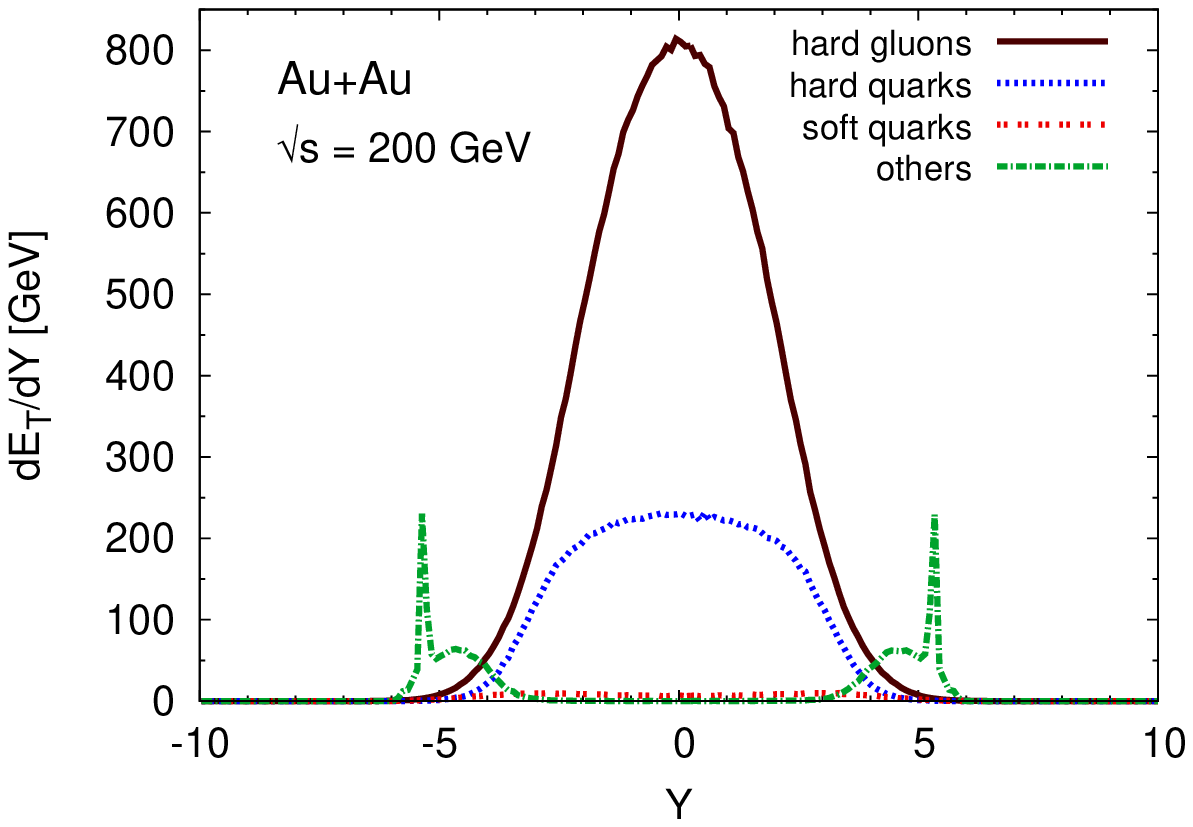}%eps
\end{minipage}
	\caption{(Color online) Rapidity distribution of particle number $\mathrm{d}N/\mathrm{d}y$ (a) and transverse energy $\mathrm{d}E_T/\mathrm{d}y$ (b) in central Au+Au collisions at RHIC. As explained in the text, the distributions are obtained by simulating nucleon-nucleon collisions with PYTHIA and scaling them to heavy ion collisions. CTEQ6l is used for the parton distribution functions. ``\emph{others}'' denotes non-partonic particles such as diquarks or protons from soft processes. The distribution of soft gluons is almost zero and is, therefore, not shown here or in the following figures.}
	\label{pythia_hi_dndy_dedy}
\end{figure}

\autoref{pythia_hi_dndy_dedy} depicts the rapidity distribution of the particle number and of their transverse energy in a Au+Au collisions at RHIC.
Gluons from hard processes dominate the spectrum at mid-rapidity in both distributions. In addition, hard quarks take a  considerable fraction of the transverse energy. Partons from soft processes have lost much of their significance compared to unscaled nucleon-nucleon collisions due to the smaller scaling factor of soft particles.

\begin{figure}
	\centering
\includegraphics[width=\gnuplotwidth]{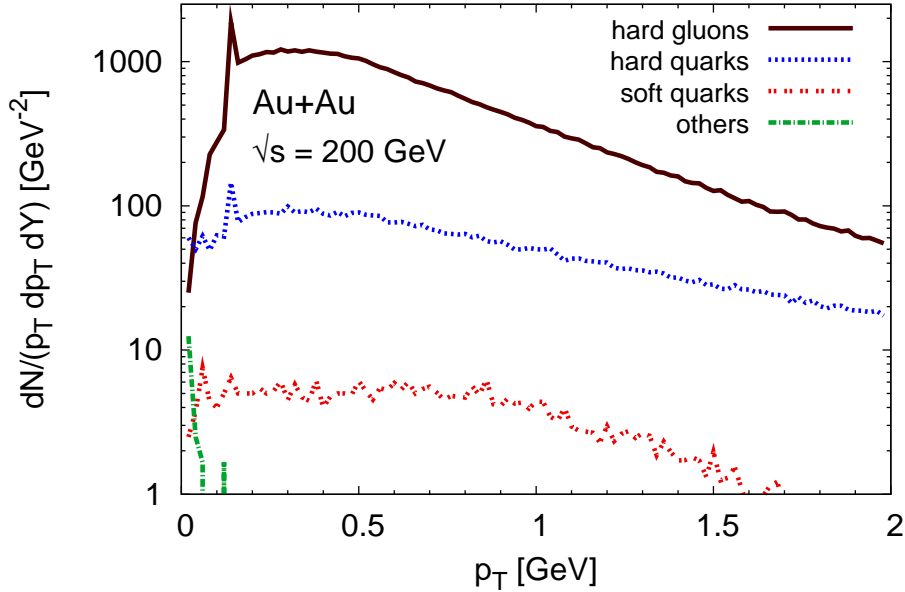}%eps
	\caption{(Color online) Transverse momentum spectra $\mathrm{d}N/(p_T\mathrm{d}p_T\mathrm{d}y)$ of particles at rapidity $y \in [-0.5,0.5]$ in central Au+Au collisions at RHIC.}
	\label{pythia_hi_dndpt}
\end{figure}

In \autoref{pythia_hi_dndpt}, the transverse momentum spectra are shown.
Of course, hard partons also dominate these spectra. The transverse momentum of soft quarks is always smaller than 2\,GeV because of a momentum cut-off in PYTHIA. For partons from hard events, we did not introduce a cut-off. However, PYTHIA has an internal cut-off for so called semi-hard scatterings, which we also consider as hard events. The cut-off is dependent on $\sqrt{s}$ and lies for RHIC energy at 141\,MeV causing a jump at this value in the $p_T$ spectra of hard partons.

\begin{figure}
	\centering
\begin{minipage}{0.02\linewidth}
(a)
\end{minipage}
\begin{minipage}{0.97\linewidth}
\includegraphics[width=\gnuplotwidth]{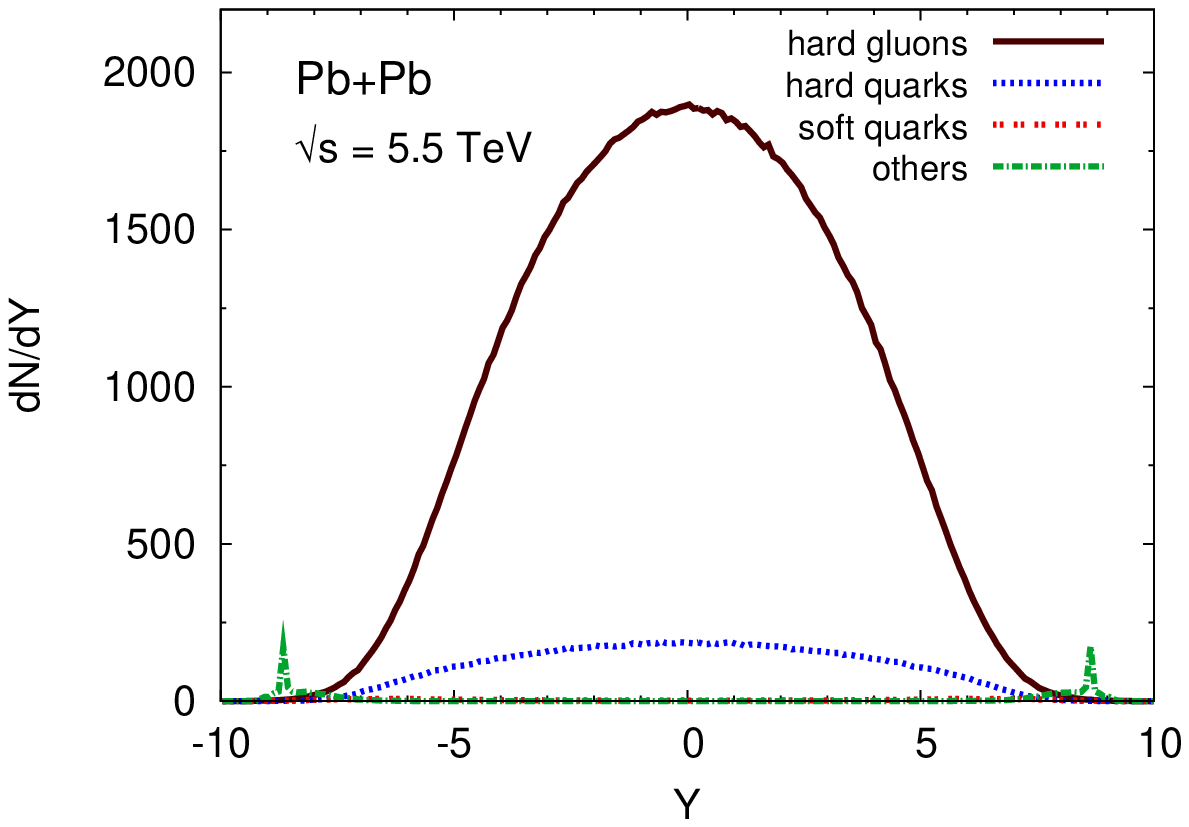}%eps
\end{minipage}

\begin{minipage}{0.02\linewidth}
(b)
\end{minipage}
\begin{minipage}{0.97\linewidth}
\includegraphics[width=\gnuplotwidth]{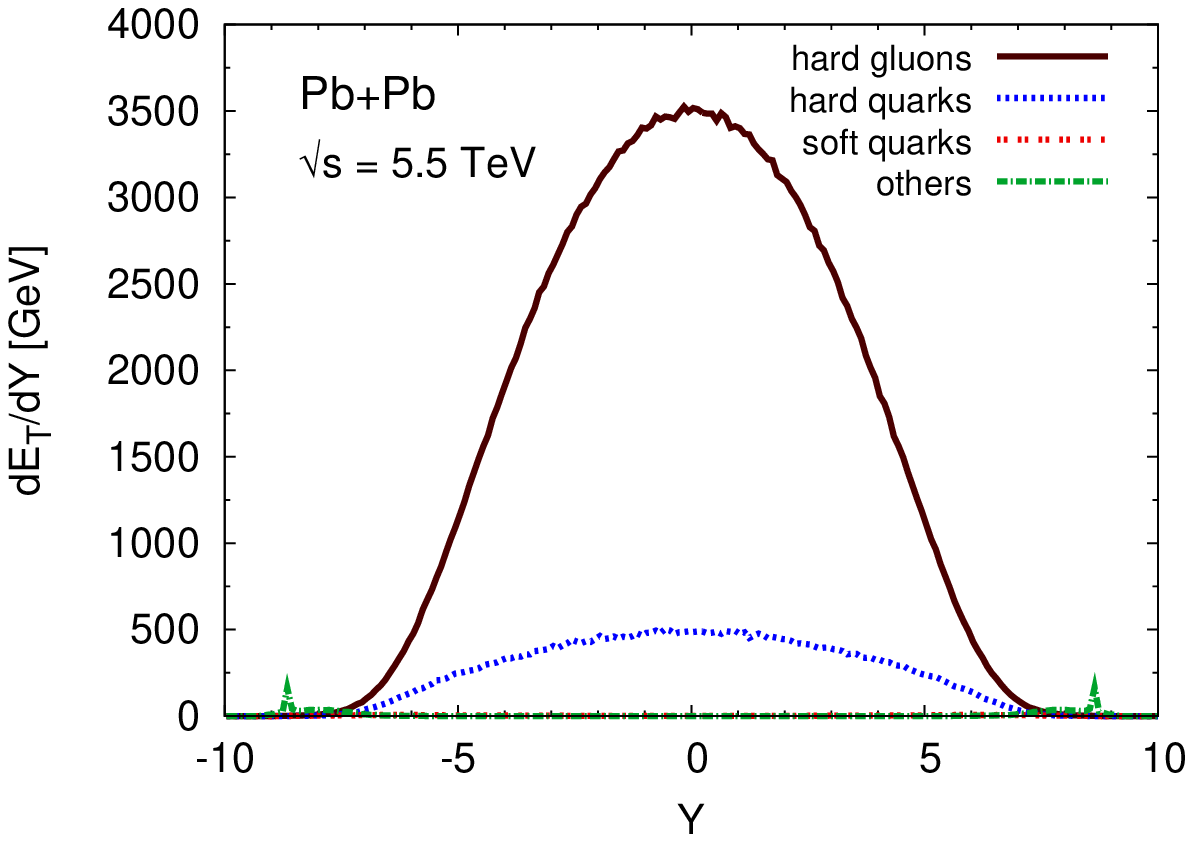}%eps
\end{minipage}
	\caption{(Color online) As in \autoref{pythia_hi_dndy_dedy}, but for central Pb+Pb collisions at LHC ($\sqrt{s_{NN}} = 5.5 \, {\rm TeV}$).}
	\label{pythia_hi_dndy_dedy_lhc}
\end{figure}

For LHC, the initial parton distribution is also sampled with PYTHIA and scaled by using the same prescription. For central Pb+Pb collisions at $\sqrt{s_{NN}} = 5.5 \, {\rm TeV}$, the overlap function is $T_{AA}({\bf b}=0)\approx 32.7\, {\rm mb}^{-1}$ and proton-proton cross section $\sigma_{\rm p+p} \approx 60 \, {\rm mb}$, which determines the number of binary collisions to be about 2000. Actually, we reduce this value to $N_{\rm bin}\approx 1500$ to take also shadowing into account \cite{Emel'yanov:1999bn,Accardi:2004be}.
\autoref{pythia_hi_dndy_dedy_lhc} depicts the rapidity distributions of the particle number and their transverse energy for heavy ion collisions at LHC.

% Since gluons dominate the early phase of the QGP and give the biggest contribution for charm production, BAMPS currently neglects light quarks. Therefore, in order to conserve energy and particle number, light quarks from PYTHIA are converted to gluons before we evolve the system with the parton cascade. However, the implementation of light quarks in BAMPS is scheduled for the near future.
Currently, BAMPS does not include light quarks. Therefore, to conserve energy and particle number, light quarks from PYTHIA are converted to gluons before we evolve the system with the parton cascade. This conversion does not account for effects such as the faster thermalization of gluons compared to light quarks and the larger cross section of $gg \rightarrow c\bar{c}$ than of $q\bar{q} \rightarrow c\bar{c}$. However, since gluons dominate the early phase of the QGP, where most of the heavy quark production takes place, we do not expect an impact on the production of heavy quarks. Nevertheless, this will be checked in the near future after light quarks are included in BAMPS.

\subsection{Other models for the initial parton distribution}
\label{sec:other_ini_models}

Two other models for the initial parton distribution were already used in the parton cascade BAMPS: the mini-jet model \cite{Xu:2004mz,Xu:2007aa} and a color glass condensate (CGC) inspired model \cite{El:2007vg,Xu:2008zi}.

In the mini-jet model \cite{Kajantie:1987pd,Eskola:1988yh}, the initial parton distribution is given by several independent 2-jet events, which are sampled according to \cite{Wang:1991hta}
\begin{equation}
\label{cs_minijet}
\frac{\mathrm{d}\sigma_{\rm jet}}{\mathrm{d}p_T^2\mathrm{d}y_1\mathrm{d}y_2} = K \sum_{a,b}
x_1f_a(x_1,p_T^2)x_2f_b(x_2,p_T^2) \frac{\mathrm{d}\sigma_{ab}}{\mathrm{d} t} \ ,
\end{equation} 
where $p_T$ denotes the transverse momentum, $y$ the rapidity and $x$ the Bjorken variable. The cross section  $\sigma_{ab}$ is calculated in LO pQCD and a $K=2$ factor is introduced to account for higher orders.

To avoid problems at low momenta where pQCD is not valid anymore, we employ a momentum cut-off, which is set to $p_0= 1.4 \, {\rm GeV}$ for RHIC in order to get the final transverse energy distribution of the gluons in agreement with data \cite{Xu:2005wv}. For LHC we choose $p_0= 3.3 \, {\rm GeV}$ to get a gluon yield, which is comparable to the PYTHIA result.

The number of produced partons is given by
\begin{equation}
N_{{\rm partons}}(b) =\sigma_{\rm jet}\, T_{AB}({\bf b}) \ .
\end{equation}

To generate the CGC initial conditions we use the model from \cite{Drescher:2006pi,Drescher:2006ca}. In that model gluons are sampled using the $k_T$ factorization ansatz \cite{Gribov:1984tu}
\begin{eqnarray}
\label{cgc_sampling}
  \frac{{\rm d}N_g}{{\rm d}^2 r_{T}{\rm d}y}=
  \frac{4N_c}{N_c^2-1} \int^{p_T^\mathrm{max}}\frac{{\rm d}^2p_T}{p^2_T}
  \int^{p_T} \frac{{\rm d}^2 k_T}{4} \;\alpha_s   \,
     \phi_A\left(x_1, \frac{({\bf p}_T+{\bf k}_T)^2}{4}\right) \,
              \phi_B\left(x_2, \frac{({\bf p}_T{-}{\bf k}_T)^2}{4}\right) \ ,
\end{eqnarray}
where  $N_c=3$ stands for the number of colors, $x_{1,2} = p_T\exp(\pm y)/\sqrt{s}$ for the light cone momentum fractions and $\sqrt{s}$ for the center of mass energy.

According to the KLN (Kharzeev-Levin-Nardi) ansatz \cite{Kharzeev:2000ph,Kharzeev:2002ei}, the unintegrated gluon distribution function $\phi(x,k_T^2)$ is related to the saturation scale $Q_{\rm sat}$ via
 \begin{equation}
\label{cgc_q_sat}
  \phi(x,k_T^2;{\bf r}_T)\sim
  \frac{1}{\alpha_s(Q^2_{\rm sat})}\frac{Q_{\rm sat}^2}
   {{\rm max}(Q_{\rm sat}^2,k_T^2)} \ .
\end{equation}
The normalization of \autoref{cgc_q_sat} and therefore also \eqref{cgc_sampling} is determined by the experimentally measured particle multiplicity at mid-rapidity in central collisions at RHIC. For LHC it is normalized in such a way that it is in agreement with the yield from PYTHIA.

\begin{figure}
\begin{minipage}[t]{0.49\textwidth}
\centering
\includegraphics[width=1.0\textwidth]{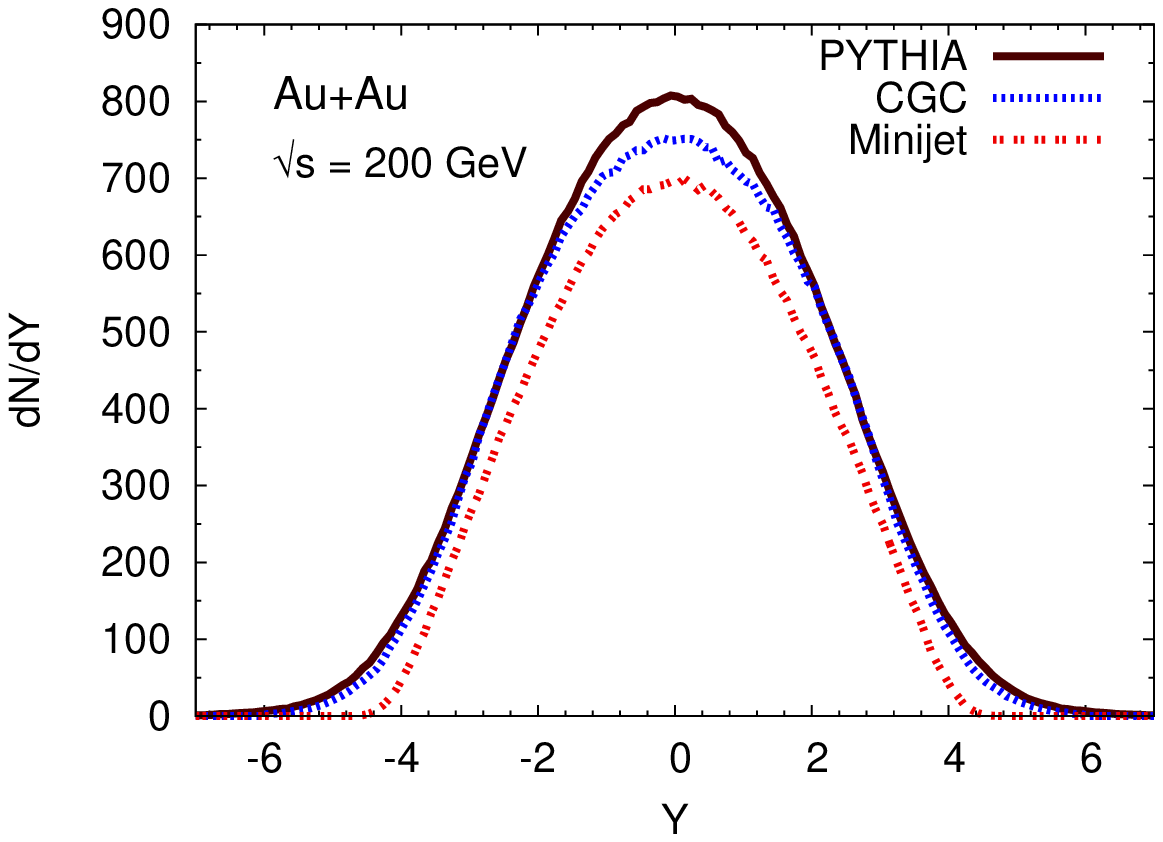}%eps

(a)
\end{minipage}
\hfill
\begin{minipage}[t]{0.49\textwidth}
\centering
\includegraphics[width=1.0\textwidth]{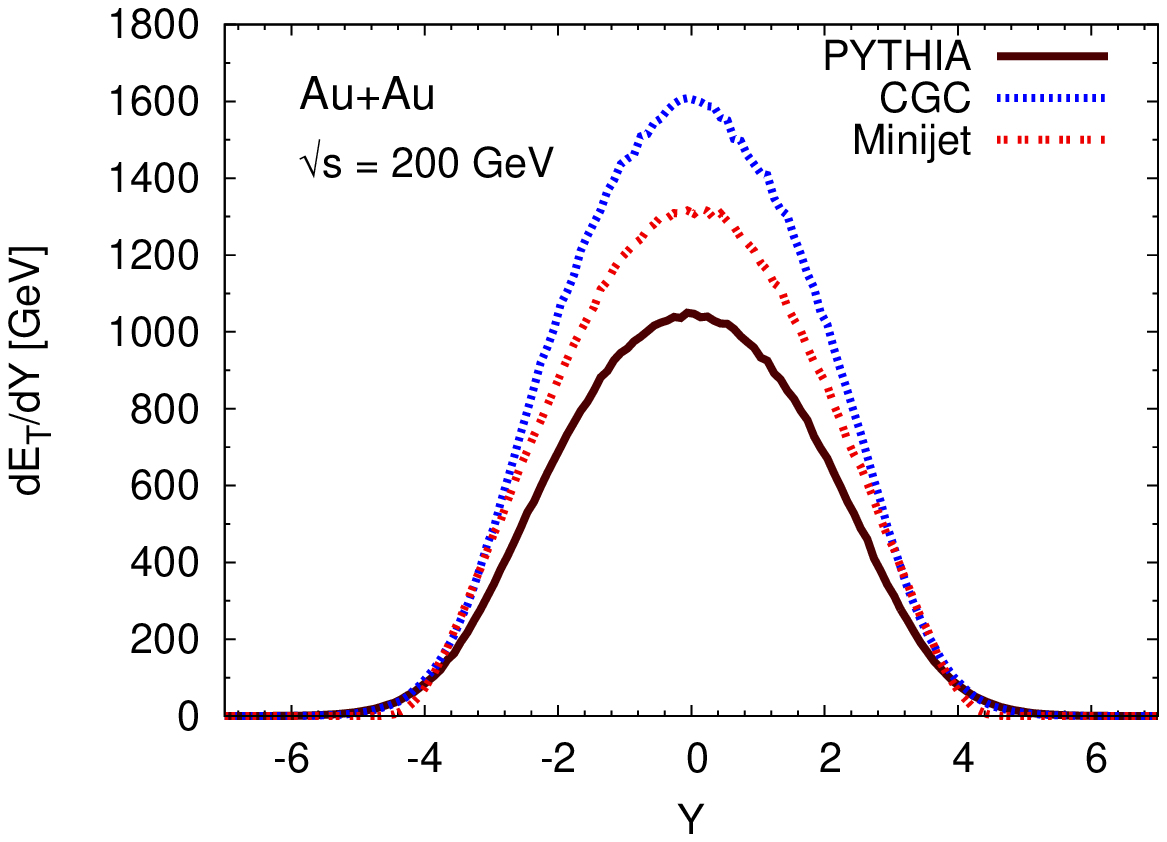}%eps

(b)
\end{minipage}
\caption{(Color online) Rapidity distribution of parton number $\mathrm{d}N/\mathrm{d}y$ (a) and their transverse energy $\mathrm{d}E_T/\mathrm{d}y$ (b) for PYTHIA, mini-jet and CGC initial conditions in a central Au+Au collision at RHIC.}
\label{fig:vergleich_ini_models}
\end{figure}

\autoref{fig:vergleich_ini_models} compares the initial particle and energy distributions of all partons for the three used models as a function of rapidity.
Although the particle number at mid-rapidity in  all three models is nearly the same, the energy distributions differ considerably. The small value from PYTHIA is a consequence of omitting beam remnants like diquarks, since BAMPS works on a partonic level. A future project will be to include string fragmentation of diquarks and partons in order to transport part of their energy into the partonic phase.

\begin{figure}
\begin{minipage}[t]{0.49\textwidth}
\centering
\includegraphics[width=1.0\textwidth]{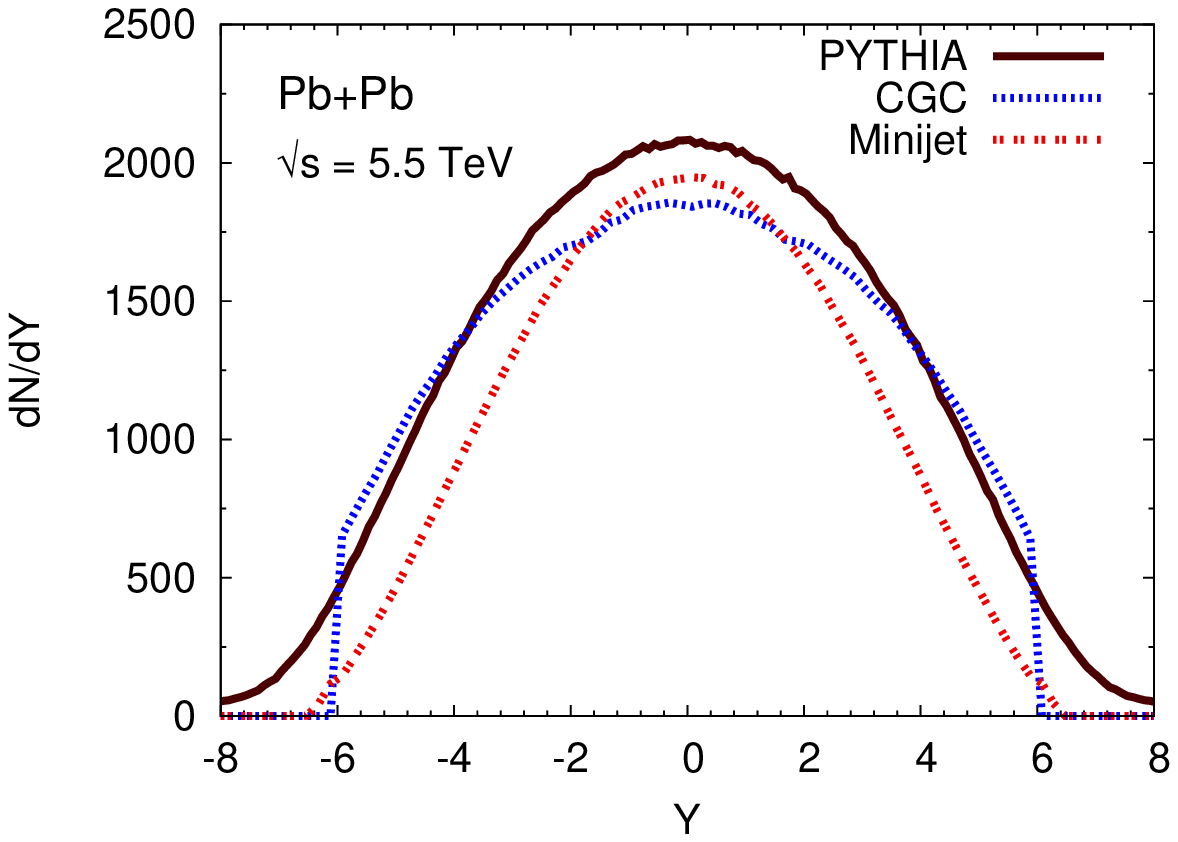}%eps

(a)
\end{minipage}
\hfill
\begin{minipage}[t]{0.49\textwidth}
\centering
\includegraphics[width=1.0\textwidth]{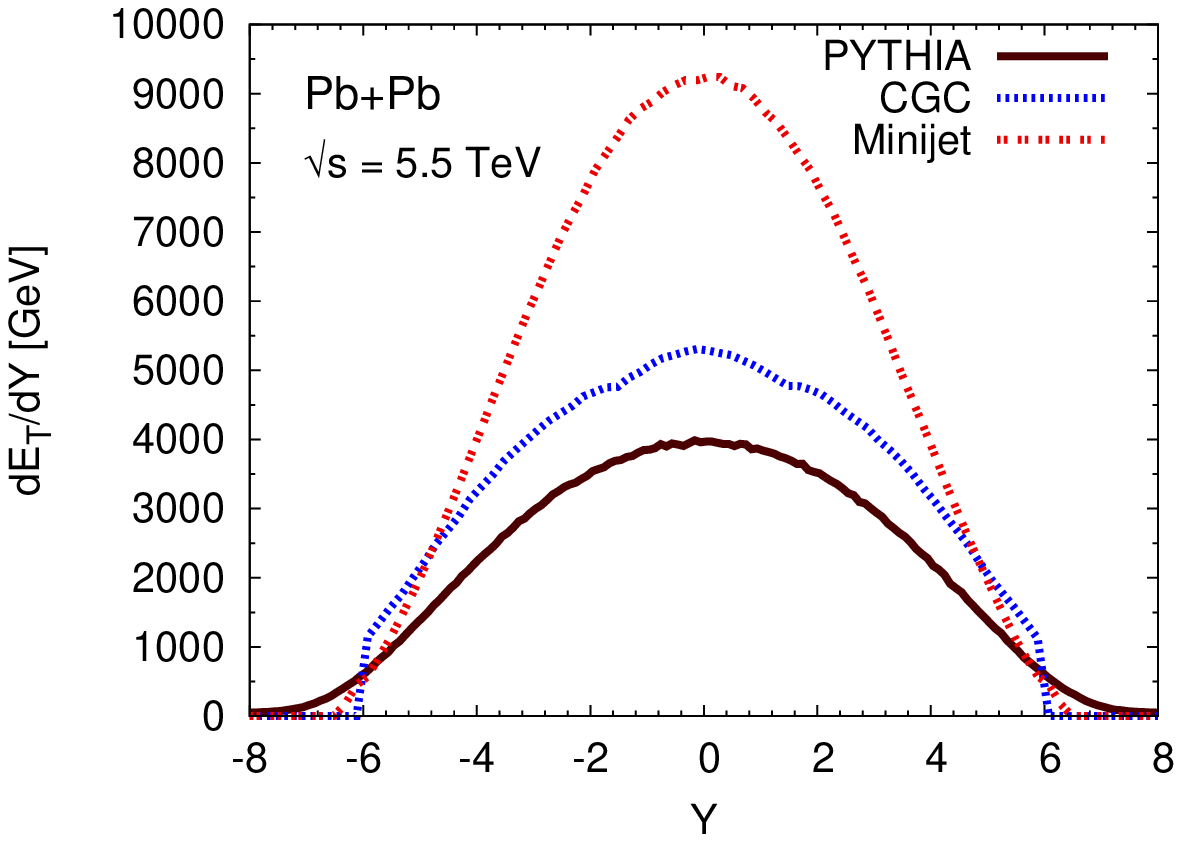}%eps

(b)
\end{minipage}
\caption{(Color online) The same as in \autoref{fig:vergleich_ini_models} for  central Pb+Pb collisions at LHC.}
\label{fig:vergleich_ini_models_lhc}
\end{figure}

\autoref{fig:vergleich_ini_models_lhc} shows the same distributions for LHC.
The model for the CGC only samples gluons with rapidity smaller than 6, which is, of course, not a physical effect. However, since we are interested in the mid-rapidity region that is not a problem.
PYTHIA yields also for LHC the smallest initial energy distribution.
The particle number distributions are comparable with various other models \cite{Armesto:2008fj}. However, there are huge uncertainties due to shadowing of $30-40$\,\% \cite{Emel'yanov:1999bn,Accardi:2004be}, which also affects secondary charm production in the QGP due to the unclear initial properties of the medium. On the other hand primary charm production in initial hard scatterings is influenced by shadowing as well (cf. next section).

\subsection{Charm production in hard processes}
\label{sec:ini_charm}

The most general expression for the double differential cross section for charm pair production at a collision of the hadrons $A$ and $B$ is given by the partonic cross section for the process $\widehat{\sigma}_{ij \rightarrow c \bar{c}}$ convoluted with the parton distribution functions (PDF) $f_i$ in the hadrons \cite{Gavai:1994gb}:
\begin{align}
\label{cs_pp_ccb_allgemein}
	E_c E_{\bar{c}} \frac{\mathrm{d}\sigma_{c\bar{c}}^{AB}}{\mathrm{d}^3p_c \mathrm{d}^3p_{\bar{c}}}
	= \sum_{i,j}\int \mathrm{d}x_1\, \mathrm{d}x_2 f_i^A(x_1,\mu_F)f_j^B(x_2,\mu_F)
	E_c E_{\bar{c}} \frac{\mathrm{d}\widehat{\sigma}_{ij \rightarrow c \bar{c}}
	(x_1P_A,x_2P_B,M_c,\mu_R)}{\mathrm{d}^3p_c \mathrm{d}^3p_{\bar{c}}} \ .
\end{align}
The indices $i$ and $j$ denote the partons from $A$ and $B$. $\mu_F$ and $\mu_R$ are the factorization and renormalization scale, respectively, which are mostly chosen to be equal, $\mu=\mu_F=\mu_R$, where $\mu = 2 M_c$ \cite{Gavai:1994gb,Smith:1996sb,Eskola:2003fk} or $\mu = \sqrt{p_T^2+M_c^2}$ \cite{Cacciari:2005rk,Vogt:2007aw,Eskola:2003fk} are common choices.

\subsubsection{Leading order charm production within the mini-jet model}
\label{sec:charm_lo}

In leading order charm quarks are produced in the two processes
\begin{align}
\label{charm_prozesse}
	g+g &\rightarrow c +\bar{c} \nonumber \\
	q+ \bar{q} &\rightarrow c +\bar{c}  \ .
\end{align}

Experimentally, due to the confinement one cannot measure single charm quarks $c$ or $\bar{c}$ but only mesons with charm content $H(c \bar{q})$, $\overline{H}(\bar{c} q)$ or $K(c\bar{c})$. The invariant cross section for an open charm production process $A + B \rightarrow H + \overline{H} $ with two charmed mesons in the final state reads \cite{Gavai:1994gb}
\begin{equation}
\label{pp_cs_dist_frag}
\begin{split}
 	E_H E_{\overline{H}} \frac{\mathrm{d}
	\sigma_{H\overline{H}}^{AB}}{\mathrm{d}^3p_H \mathrm{d}^3p_{\overline{H}}} & =  \int \frac{\hat{s}}{2 \pi} \, \mathrm{d}x_1
	\mathrm{d}x_2 \mathrm{d}z_c \mathrm{d}z_{\bar{c}} \,
	C(x_1,x_2) 
	\frac{E_H E_{\overline{H}}}{E_c
	E_{\bar{c}}} 
	\\ 	&   \qquad 
\frac{D_{H/c}(z_c)}{z_c^3}
	\frac{D_{\overline{H}/\bar{c}}(z_{\bar{c}})}{z_{\bar{c}}^3}
	\delta^4 (P_1 + P_2 - P_c - P_{\bar{c}}) \  , 
\end{split}
\end{equation}
with
\begin{align}
\label{pp_cs_dist_frag_c}
	C(x_1,x_2)  = 
	f^A_g(x_1) \, f^B_g (x_2) \, \frac{\mathrm{d}\hat{\sigma}_{gg \rightarrow c \bar{c}}}{\mathrm{d}\hat{t}} + 
	\sum_q \left[ f^A_q(x_1) \, f^B_{\bar{q}} (x_2) + f^A_{\bar{q}}(x_1) \, f^B_q (x_2) \right] \frac{\mathrm{d}\hat{\sigma}_{q\bar{q} \rightarrow c \bar{c}}}{\mathrm{d}\hat{t}}  \ .
\end{align}
Here $\sqrt{\hat{s}}$ is the center of mass energy of the partons, which is related to the center of mass energy of the hadrons $\sqrt{s}$ via $\hat{s} = x_1 x_2 s$.
The partonic cross sections are given in Section \ref{sec:bamps}. 

The fragmentation function $D_{H/c}(z)$ describes the hadronization of the charm quarks, with $z = |\vec{p}_H|/|\vec{p}_c|$. Since fragmentation only affects the momentum distribution but not the total $c\bar{c}$ cross section \cite{Gavai:1994gb}, in which we are interested, the momentum of $H$ is assumed to be equal to the charm quark momentum, that is, $z = 1$ and $D_{H/c}(z) = \delta (1-z)$. Thus, taking energy conservation into account, \autoref{pp_cs_dist_frag} can be written as
\beq
\label{pp_cs_dist}
\frac{\mathrm{d} \sigma_{c\bar{c}}^{AB}}{\mathrm{d}p_T^2 \mathrm{d}y_c \mathrm{d}y_{\bar{c}}} = x_1 x_2
C(x_1,x_2) \ ,
\eeq
$y$ being the rapidity. In the center-of-mass frame, the charm and anti-charm quarks hold the same transverse momentum $p_T$. 

Integration of the previous equation leads to the total cross section
\begin{align}
\label{ges_pp_cs_ccb_diff}
	\sigma_{c\bar{c}} = \frac{1}{2} \int_0^{s/4-M_c^2}\mathrm{d}p_T^2 \int_{y_c^{\text{min}}}^{y_c^{\text{max}}} \mathrm{d}y_c 
\int_{y_{\bar{c}}^{\text{min}}}^{y_{\bar{c}}^{\text{max}}} \mathrm{d}y_{\bar{c}} \ 
\frac{\mathrm{d} \sigma_{c\bar{c}}^{AB}}{\mathrm{d}p_T^2 \mathrm{d}y_c \mathrm{d}y_{\bar{c}}} \ .
\end{align}
The factor $1/2$ results from the method of charm number counting, which is widely adopted in the literature. The charm quark number corresponding to the cross section should not be the total number of particles (charm plus anti-charm quarks) but the number of charm pairs. 
The number of produced charm pairs and the cross section is simply related via \cite{Vogt:2001nh,Andronic:2006ky}
\begin{align}
	N_{c \bar{c}} = \sigma_{c \bar{c}}^{NN} \, T_{AB}({\bf b}) \ .
\end{align}

The limits of the integrals over the rapidities in \autoref{ges_pp_cs_ccb_diff} can be obtained from the definition of the rapidity and some kinematic considerations,
\begin{align}
\label{grenzen_ges_pp_cs_ccb_diff}
	y_c^{\text{max/min}} & = \pm \ln \left(\frac{1}{\chi_T} + \sqrt{\frac{1}{\chi_T^2}-1}\right) 
	\umbruch
	y_{\bar{c}}^{\text{max/min}} & = \pm \ln \left(\frac{2}{\chi_T} - \mathrm{e}^{\pm y_c}\right) \ ,
\end{align}
with $\chi_T=2 M_T / \sqrt{s}$.

From equations \eqref{pp_cs_dist_frag} and \eqref{pp_cs_dist_frag_c} it is obvious that the charm production cross section is fundamentally dependent on the PDFs. Therefore, using different PDFs provided by the \emph{Les Houches Accord Parton Density Function} (LHAPDF) group \cite{Whalley:2005nh} we explore this dependency in detail, as listed in \autoref{tab:charm_minijet_rhic}.
\begin{table*}
	\centering
\begin{tabular}{c|c|c|c|c}
PDF &  Scale $\mu_F = \mu_R$ & $M_{c} \,[ \rm{GeV}]$ & $\sigma \,[ \rm{\mu b}]$ & $\rm{d}\sigma/\rm{d}y |_{y=0} \,[ \rm{\mu b}]$\tabularnewline
\hline
\hline

 \multirow{5}{*}{CTEQ6m} & \multirow{2}{*}{$2M_c$} & 1.2 &       160   &    38      \tabularnewline
\cline{3-5}  &  & 1.5  &      72   &    19      \tabularnewline
\cline{2-5} & \multirow{2}{*}{$\sqrt{p_T^2+M_c^2}$} & 1.2 &      140   &    36       \tabularnewline
\cline{3-5} &  & 1.5 &       79   &    20       \tabularnewline
\cline{2-5}
&  \multicolumn{2}{c}{PYTHIA} \vline  &  540  &  130   \tabularnewline

\cline{2-5}
\hline 
\hline 

 \multirow{5}{*}{ CTEQ6l} & \multirow{2}{*}{$2M_c$} & 1.2 &       230   &    57      \tabularnewline

\cline{3-5}  &  & 1.5  &      90   &    25       \tabularnewline

\cline{2-5} & \multirow{2}{*}{$\sqrt{p_T^2+M_c^2}$} & 1.2 &       280   &    68       \tabularnewline

\cline{3-5} &  & 1.5 &       120   &    31       \tabularnewline
\cline{2-5}
&  \multicolumn{2}{c}{PYTHIA} \vline  & 370   &   91  \tabularnewline

\cline{2-5}
\hline 
\hline

 \multirow{5}{*}{ MRST2007lomod} & \multirow{2}{*}{$2M_c$} & 1.2 &      290   &    65      \tabularnewline

\cline{3-5}  &  & 1.5  &       120   &    30       \tabularnewline

\cline{2-5} & \multirow{2}{*}{$\sqrt{p_T^2+M_c^2}$} & 1.2 &      320   &    66      \tabularnewline

\cline{3-5} &  & 1.5 &       150   &    35      \tabularnewline
\cline{2-5}
&  \multicolumn{2}{c}{PYTHIA} \vline  &  370  &   89  \tabularnewline
\cline{2-5}
\hline 
\hline 

 \multirow{5}{*}{GRV98lo}  & \multirow{2}{*}{$2M_c$} & 1.2 &       190   &    38      \tabularnewline

\cline{3-5}  &  & 1.5 &      78   &    17      \tabularnewline

\cline{2-5} & \multirow{2}{*}{$\sqrt{p_T^2+M_c^2}$} & 1.2 &      220   &    43       \tabularnewline

\cline{3-5} &  & 1.5 &       97   &    20     \tabularnewline
\cline{2-5}
&  \multicolumn{2}{c}{PYTHIA} \vline  &  120  &   30  \tabularnewline
\cline{2-5}

\hline 
\hline 
\multicolumn{3}{c}{PHENIX} \vline & $544 \pm 381$ & $123\pm 47$ \tabularnewline
\hline 
\hline 
\multicolumn{3}{c}{STAR} \vline & $1400 \pm 600$  & $300\pm 130$ \tabularnewline
\hline 
\hline 
\end{tabular}
\caption{LO pQCD cross sections for charm pair production in nucleon-nucleon collisions at $\sqrt{s}=200 \, {\rm GeV}$ with a running $\alpha_s$, $N_f = 3$, $\lambda_{\rm QCD}=346 \, \rm{MeV}$ \cite{Bethke:2006ac} and a $K$ factor of $K=1$ for various parton distribution functions, renormalization ($\mu_R$) and factorization ($\mu_F$) scales  and charm masses $M_c$. Results from PYTHIA and experimental data \cite{:2008asa_PHENIX,Adare:2006hc_PHENIX_dsigmadY,Adams:2004fc_STAR_dcsdY_cstot} are also listed. In PYTHIA $M_c= 1.5 \, {\rm GeV}$ is used.}
\label{tab:charm_minijet_rhic}
\end{table*}

For that study we also vary the charm mass and the factorization and renormalization scale. The calculations are done with a running coupling $\alpha_s$. In contrast, a fixed coupling of  $\alpha_s=0.3$ yields much smaller results. Additionally, results from PYTHIA (see next section) and experimental data \cite{:2008asa_PHENIX,Adare:2006hc_PHENIX_dsigmadY,Adams:2004fc_STAR_dcsdY_cstot} are shown in the table.
The reason for the strong deviation between both experiments is currently under intensive investigation.

Our findings are in good agreement with the literature. In \cite{Levai:1994dx}, the charm production cross section in LO is calculated to $\sigma_{c \bar{c}}^{NN} = 160 \, {\rm \mu b}$ using the HIJING model and taking shadowing into account. \cite{Vogt:2001nh} finds a value between 133 and $153 \, {\rm \mu b}$ depending on the choice of the parton distribution functions. In \cite{Pop:2009sd}, the cross section lies between 300 and $750  \, {\rm \mu b}$, according to whether shadowing and/or strong color electric fields are incorporated.

Our results in \autoref{tab:charm_minijet_rhic} as well as those from the previously mentioned LO models are strongly dependent on various parameters. Mainly, the dependence on the scales shows that LO calculations are not sufficient to describe these processes. Indeed, compared to experimental data, the LO results are much too small. In contrast, next-to-leading order (NLO) calculations are closer to the data \cite{Cacciari:2005rk}. However, errors due to uncertainties in PDFs, scales and charm mass are still significantly high.

\autoref{tab:charm_minijet_lhc} lists our predictions for the charm production cross section of nucleon-nucleon collisions at LHC energy. Again, there is a considerable dependency on the aforementioned parameters.
\begin{table*}
	\centering
\begin{tabular}{c|c|c|c|c}
PDF &  Scale $\mu_F = \mu_R$ & $M_{c} \,[ \rm{GeV}]$ & $\sigma \,[ \rm{\mu b}]$ & $\rm{d}\sigma/\rm{d}y |_{y=0} \,[ \rm{\mu b}]$\tabularnewline
\hline
\hline 

 \multirow{5}{*}{ CTEQ6m} & \multirow{2}{*}{$2M_c$} & 1.2 &   1600   &   170   \tabularnewline
\cline{3-5}  &  & 1.5  &    1100   &   130    \tabularnewline
\cline{2-5} & \multirow{2}{*}{$\sqrt{p_T^2+M_c^2}$} & 1.2 &   770    &  83     \tabularnewline
\cline{3-5} &  & 1.5 &   690    &   78    \tabularnewline
\cline{2-5}
&  \multicolumn{2}{c}{PYTHIA} \vline  &  2600  &  300   \tabularnewline
\cline{2-5}
\hline 
\hline 

  \multirow{5}{*}{CTEQ6l} & \multirow{2}{*}{$2M_c$} & 1.2 &    5300     &    640     \tabularnewline

\cline{3-5}  &  & 1.5  &   3200    &    400   \tabularnewline

\cline{2-5} & \multirow{2}{*}{$\sqrt{p_T^2+M_c^2}$} & 1.2 &  3500     &  420     \tabularnewline

\cline{3-5} &  & 1.5 &   2400    &   310    \tabularnewline
\cline{2-5}
&  \multicolumn{2}{c}{PYTHIA} \vline  & 2500   &   310  \tabularnewline
\cline{2-5}
\hline 
\hline 

  \multirow{5}{*}{MRST2007lomod} & \multirow{2}{*}{$2M_c$} & 1.2 &       6600   &    730      \tabularnewline

\cline{3-5}  &  & 1.5  &       3900   &    460       \tabularnewline

\cline{2-5} & \multirow{2}{*}{$\sqrt{p_T^2+M_c^2}$} & 1.2 &       4700   &    480       \tabularnewline

\cline{3-5} &  & 1.5 &       3100   &    350       \tabularnewline
\cline{2-5}
&  \multicolumn{2}{c}{PYTHIA} \vline  &  2700  &   320  \tabularnewline
\cline{2-5}
\hline 
\hline 

 \multirow{5}{*}{GRV98lo}  & \multirow{2}{*}{$2M_c$} & 1.2 &       7600   &    890      \tabularnewline

\cline{3-5}  &  & 1.5 &       4100   &    500      \tabularnewline

\cline{2-5} & \multirow{2}{*}{$\sqrt{p_T^2+M_c^2}$} & 1.2 &   -    &   -    \tabularnewline

\cline{3-5} &  & 1.5 &   -    &  -     \tabularnewline
\cline{2-5}
&  \multicolumn{2}{c}{PYTHIA} \vline  &  -  &   -  \tabularnewline
\cline{2-5}

\hline 
\hline 
\end{tabular}
\caption{As in \autoref{tab:charm_minijet_rhic}, but at LHC energy of $\sqrt{s} = 5.5 \, {\rm TeV}$. The results for GRV98lo with a factorization scale of  $\mu_F = \sqrt{p_T^2+M_c^2}$ and for PYTHIA cannot be calculated, because these PDFs are not designed for such large scales.}
\label{tab:charm_minijet_lhc}
\end{table*} 
\cite{Levai:1994dx} calculates the cross section for LHC to $\sigma_{c \bar{c}}^{NN} = 5750 \, {\rm \mu b}$ and \cite{Pop:2009sd} to $6400 \, {\rm \mu b}$. In \cite{Vogt:2001nh} it varies between 2000 and $7000 \, {\rm \mu b}$, depending on the PDFs chosen.

\subsubsection{Charm production with PYTHIA}
\label{sec:ini_charm_pythia}
We also used PYTHIA to estimate the number of produced charm quarks in a nucleon-nucleon collision. The corresponding cross sections are listed in Tables \ref{tab:charm_minijet_rhic} and \ref{tab:charm_minijet_lhc}. 
Applying the scaling prescription introduced in Section \ref{sec:p_sampling_pythia}, the number of initial charm quarks in a heavy ion collision can be computed from this value. Since they are produced in hard processes, their number scales with the number of binary collisions.\footnote{Actually, the scaling factor is a bit smaller if one takes shadowing into account as is done in this calculation (cf. Section \ref{sec:p_sampling_pythia}).}
\autoref{tab:pythia_charm_yield} lists the number of initially produced charm quarks in Au+Au collisions at RHIC according to PYTHIA for various parton distribution functions.
\begin{table}
	\centering
		\begin{tabular}{l|c|D{!}{.}{5.4}}
			PDF & Reference &  
			\multicolumn{1}{c}{Number of charm pairs}			\\ \hline
			\hline
CTEQ5l (LO) & \cite{Lai:1999wy} & 8!9 \\ \hline
CTEQ6l (LO) & \cite{Pumplin:2002vw_CTEQ6} & 9!2 \\ \hline
CTEQ6m ($\overline{MS}$) & \cite{Pumplin:2002vw_CTEQ6} & 13!6 \\ \hline
MRST2001LO & \cite{Martin:2002dr} & 9!6 \\ \hline
MRST2007LOmod & \cite{Sherstnev:2007nd} & 9!2\\ \hline
HERAPDF01 & 
\cite{Nagano:2008ip} 
 & 12!3 \\ \hline
GJR08 (FF LO) & \cite{Gluck:2007ck,Gluck:2008gs} & 3!0 \\ \hline
GRV98 (LO) & \cite{Gluck:1998xa} & 3!0
		\end{tabular}
	\caption{Number of charm pairs produced in primary hard scatterings in central Au+Au collisions at RHIC for some parton distribution functions by sampling nucleon-nucleon collisions with PYTHIA and scaling to Au+Au collisions (cf. Section \ref{sec:p_sampling_pythia}).}
	\label{tab:pythia_charm_yield}
\end{table}
Again, the uncertainties of using different PDFs are reflected. The mean value lies at about 9 pairs.

According to \cite{Duraes:2004zt} $2-6.5$~charm quarks are produced in initial hard collisions at RHIC in LO, dependent on the PDFs, charm mass and shadowing.
\cite{Muller:1992xn} estimates this number to 2 using the mini-jet model and a phenomenological factor of $K=2$.
\cite{Gavin:1996bx} extrapolates from p+p collisions in NLO calculations 8.7~produced pairs. The authors also calculate the number of charm quarks at mid-rapidity to about 3 pairs, which is in good agreement with our results of about 2 for CTEQ6l and 3 for CTEQ6m.
In another NLO calculation \cite{Vogt:2002vx} the total number of charm quarks varies between 8 and 13.
Thus, recent results from the literature are in good agreement with our results from PYTHIA.

\begin{figure}
	\centering
\includegraphics[width=\gnuplotwidth]{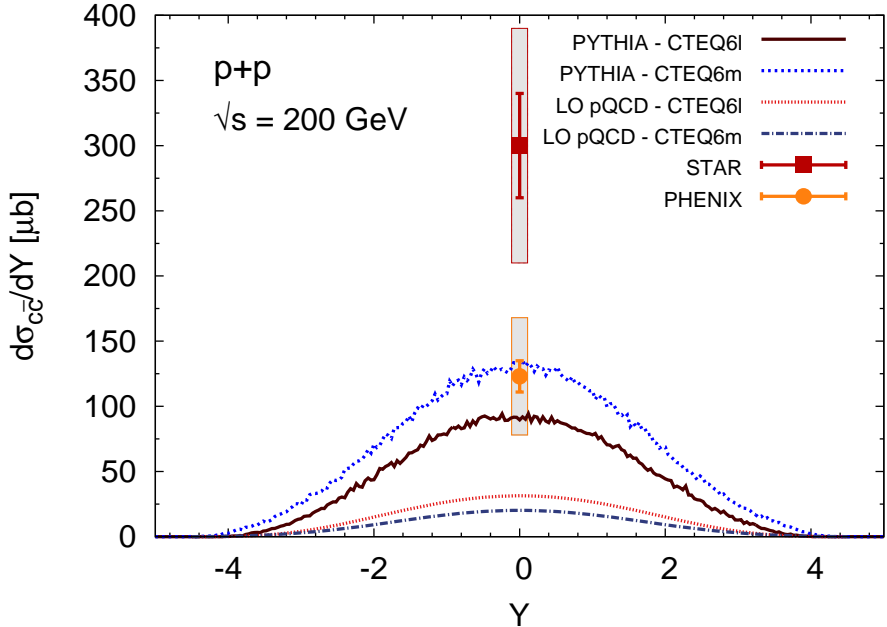}%eps
	\caption{(Color online) Charm production cross section $\mathrm{d}\sigma_{c \bar{c}}^{NN}/\mathrm{d}y$ as a function of rapidity $y$ in a nucleon-nucleon collision at RHIC energy simulated with PYTHIA and pQCD, respectively, for the PDFs CTEQ6l and CTEQ6m together with experimental data \cite{Adams:2004fc_STAR_dcsdY_cstot,Adare:2006hc_PHENIX_dsigmadY}. The pQCD calculation is done in LO with $\mu_F = \mu_R=\sqrt{p_T^2+M_c^2}$, $N_f=3$, $M_c= 1.5\, {\rm GeV}$, $\lambda_{\rm QCD}=346 \, \rm{MeV}$ \cite{Bethke:2006ac} and $K=1$.}
	\label{fig:ini_charm_dn_dy_exp}
\end{figure}

Experimental data for the differential charm production cross section at mid-rapidity in a nucleon-nucleon collision at $\sqrt{s}=200 \, {\rm GeV}$ $\mathrm{d}\sigma_{c \bar{c}}^{NN}/\mathrm{d}y$ is available both from STAR and PHENIX. The former one measured a value of $\mathrm{d}\sigma_{c \bar{c}}^{NN}/\mathrm{d}y = 300\pm 130 \, {\rm \mu b}$ \cite{Adams:2004fc_STAR_dcsdY_cstot} and the latter of $123\pm 47 \, {\rm \mu b}$ \cite{Adare:2006hc_PHENIX_dsigmadY}, again showing big deviations.
The rapidity distribution of the charm production cross section simulated with PYTHIA is depicted in \autoref{fig:ini_charm_dn_dy_exp} together with the experimental data points and the pQCD calculations from Section \ref{sec:charm_lo}.
The charm distribution from the LO pQCD calculation lies far below the experimental data. Hence, higher order corrections should be taken into account or the introduction of a phenomenological factor of $K > 2$ is necessary. The results from PYTHIA agree well with the PHENIX data point, which is a bit peculiar, since PYTHIA is based on LO pQCD cross sections. However, PYTHIA is tuned with a running coupling and $K$ factors in order to describe experimental data well.

Although CTEQ6m reproduces the data better than CTEQ6l, we will use in the following the latter PDF set, because it is designed for LO event generators such as PYTHIA \cite{Pumplin:2002vw_CTEQ6}. Therefore, we can use PYTHIA also for a reliable sampling of the initial gluon distribution.

\begin{table}
	\centering
		\begin{tabular}{l|c|c|c} 
		&& \multicolumn{2}{c}{Heavy quark pairs}\\ 
			PDF & Reference & 
			Charm  & Bottom			\\ \hline
			\hline
CTEQ6l (LO) & \cite{Pumplin:2002vw_CTEQ6} & 62 & 7.2 \\ \hline
CTEQ6m ($\overline{MS}$) & \cite{Pumplin:2002vw_CTEQ6} & 66 & 6.9 \\ \hline
MRST2007LOmod & \cite{Sherstnev:2007nd} & 67 & 8.9
		\end{tabular}
	\caption{As in \autoref{tab:pythia_charm_yield}, but for central Pb+Pb collisions at LHC.}
	\label{tab:pythia_charm_yield_lhc}
\end{table}

\autoref{tab:pythia_charm_yield_lhc} lists the number of charm and bottom quarks produced during initial nucleon scatterings in Pb+Pb collisions at LHC according to PYTHIA. The results from different PDFs deviate not as much as for collisions at RHIC. Nevertheless, there are still big systematic errors due to uncertainties in mass, shadowing, factorization and renormalization scale, although these are not reflected in the table.

\cite{Muller:1992xn} predicts 34 produced charm pairs in Pb+Pb collisions at LHC within the mini-jet model. The NLO value from \cite{Gavin:1996bx} was estimated to 450, but reduced to $67-150$ after taking newer PDFs and shadowing into account \cite{Vogt:2002vx}. \cite{Vogt:2002vx} also makes a prediction for bottom quarks, of which about 5 pairs should be produced at LHC.

\begin{figure}
	\centering
\includegraphics[width=\gnuplotwidth]{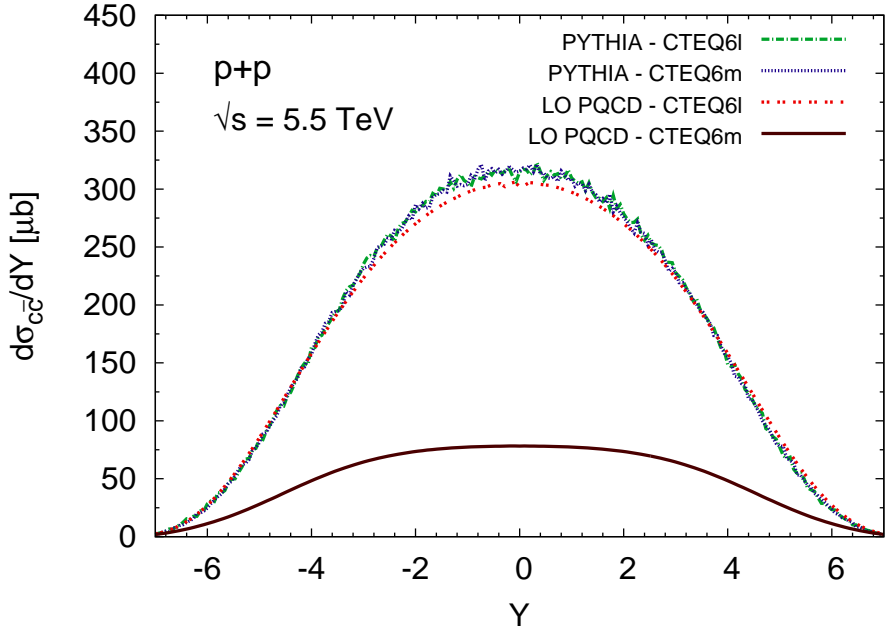}%eps
	\caption{(Color online) As in \autoref{fig:ini_charm_dn_dy_exp}, but for LHC.}
	\label{fig:ini_charm_dn_dy_lhc}
\end{figure}

The charm production cross sections $\mathrm{d}\sigma_{c \bar{c}}^{NN}/\mathrm{d}y$ as a function of rapidity $y$ for nucleon-nucleon collision at LHC energy simulated with PYTHIA and LO pQCD, respectively, with different PDFs  are shown in \autoref{fig:ini_charm_dn_dy_lhc}.
It is surprising that the curves from PYTHIA and LO pQCD  do not differ very much for CTEQ6L. For CTEQ6M, however, the curve from PYTHIA is nearly identical to that from CTEQ6L, but it differs by more than a factor of 3 from the LO pQCD result for the same PDF. From this surprising result, which can also be observed in \autoref{tab:pythia_charm_yield_lhc}, one sees again the big uncertainties due to different parameters, models, PDFs and shadowing. The NLO calculation including shadowing from \cite{Vogt:2001nh} lies about a factor of 2 above the PYTHIA yield with CTEQ6L.

\section{Heavy quark production in the QGP}
\label{sec:prod_qgp}

\subsection{Charm production at RHIC}
\label{sec:prod_qgp_rhic}

In this section, charm production during the quark gluon plasma phase in central Au+Au collisions at RHIC will be investigated within the framework of the parton cascade BAMPS (cf. Section \ref{sec:bamps}). In contrast to Section~\ref{sec:box}, where we limited ourselves to a static medium with just two processes, we consider now the full BAMPS simulation of the expanding fireball with all interactions from \ref{bamps_processes}. In that framework, because of the $gg \leftrightarrow ggg$ processes, rapid thermalization \cite{Xu:2007aa}, for instance, or the build-up of the elliptic flow \cite{Xu:2008av} of the gluonic medium can be explained. 

The initial distribution of the charm quarks is simulated using PYTHIA with the parton distribution function CTEQ6l and scaled to heavy ion collisions as described in Section \ref{sec:p_sampling_pythia}. Since light quarks are not implemented in BAMPS yet, they are treated as massless gluons to take conservation of energy and particle number into account. In this gluonic medium charm quarks are produced by gluon fusion $g+g\rightarrow c+\bar c$ and interact with gluons in elastic scatterings within our model. For both processes,  LO cross sections (cf. Section \ref{sec:bamps}) and a constant coupling of $\alpha_s=0.3$ are used. First, we study charm production without multiplying the cross section by any $K$ factor, although we will occasionally employ $K=2$ later.  Charm annihilation $c+\bar c \rightarrow g+g$ is very unlikely due to the small number of charm quarks which are produced and can be neglected \cite{Andronic:2006ky}.

\begin{figure}
	\centering
\includegraphics[width=\gnuplotwidth]{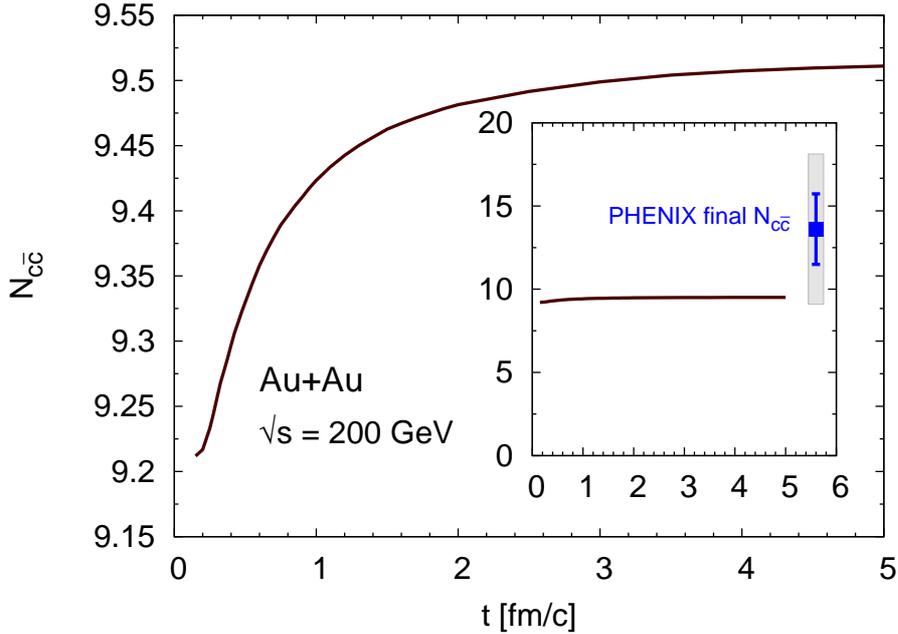}%eps
	\caption{(Color online) Number of charm quark pairs produced in a central Au+Au collision at RHIC according to BAMPS. The initial parton distribution is obtained with PYTHIA. Note that only a small interval of the $y$-axis is shown. Therefore, we also depict the full range in the inset. In addition, the experimental value for the number of final charm pairs is plotted \cite{Adler:2004ta}.}
	\label{fig:charm_yield_rhic_central_pythia}
\end{figure}

The primary sources of produced charm quarks at RHIC are initial hard parton scatterings during nucleon-nucleon collisions. With PYTHIA and CTEQ6l the initial yield  amounts to 9.2 charm quark pairs as was shown in the previous section. The time evolution of the number of charm quarks during the quark gluon plasma  phase within our BAMPS simulation is depicted in \autoref{fig:charm_yield_rhic_central_pythia}.
The number of charm pairs increases slightly in the first $1-2\,{\rm fm}/c$ by 0.3 pairs on average due to the high initial temperature of the plasma.
Subsequently, the number saturates at a value of 9.5 charm pairs after $5\,{\rm fm}/c$, which indicates a marginal charm production of only 3\,\% of the total final charm quarks in the QGP. To illustrate this, the full range of the $y$-axis is shown in the plot in the inset.
Additionally, the experimental value of charm pairs, which is obtained from analyzing $D$ meson decays, is also shown. However, we calculated $N_{c\bar{c}}$ and its errors from $N_{c\bar{c}}/T_{AA}$ and  $T_{AA}$  since these are the only values given in \cite{Adler:2004ta}.
The number of charm pairs estimated with BAMPS and PYTHIA is below the experimental value, but still within the errors. This deviation results from the initial charm number obtained with PYTHIA and CTEQ6l, which is also smaller than the experimental value in proton-proton collisions (cf. \autoref{fig:ini_charm_dn_dy_exp}).

In addition, experimental data show that the number of produced charm quarks at RHIC scales with the number of binary collisions \cite{Adler:2004ta,Adams:2004fc_STAR_dcsdY_cstot,:2008hja}. That is, charm quarks are mainly created in initial hard parton scatterings and not in the QGP, which is in good agreement with our findings.

The small charm yield in the QGP at RHIC confirms our results from Section \ref{sec:rates_comp_ana_num_rhic}, where we estimated a huge chemical equilibration time scale for charm quarks in a gluonic box at RHIC temperatures.

\begin{figure}
	\centering
\includegraphics[width=\gnuplotwidth]{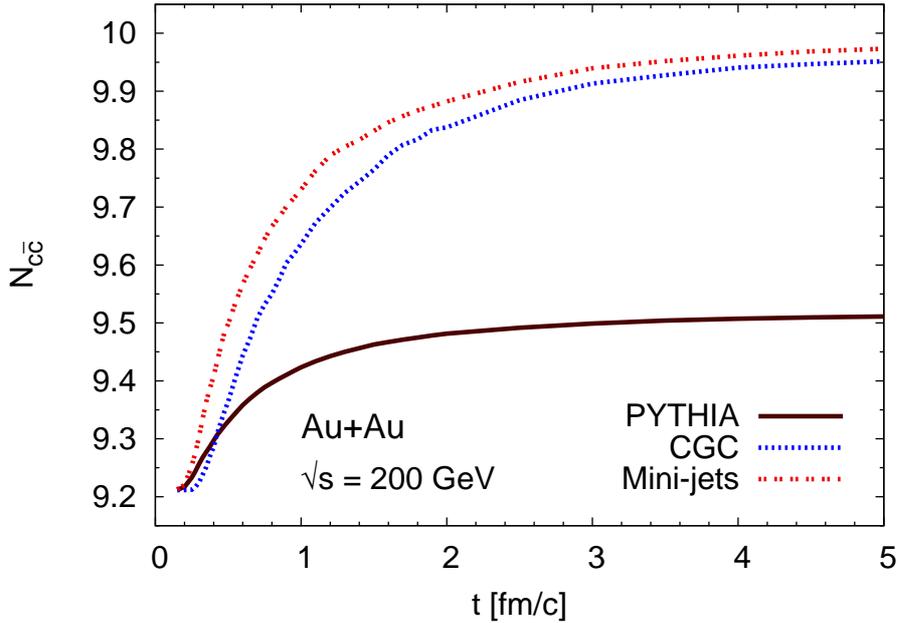}%eps
	\caption{(Color online) As in \autoref{fig:charm_yield_rhic_central_pythia}, but with PYTHIA, CGC, and mini-jet initial conditions for gluons. For better comparison, the initial charm distribution from PYTHIA is used for all models.}
	\label{fig:charm_yield_rhic_central_initial}
\end{figure}

The number of charm quarks produced in the QGP is very sensitive to the initial gluon distribution. Therefore, we employed in addition to PYTHIA the color glass condensate (CGC) and the mini-jet model for the initial gluon distribution (cf. Section \ref{sec:other_ini_models}). Both models lead to an increase of a factor of 2.5 in the number of charm pairs produced in the QGP compared to PYTHIA, as is shown in \autoref{fig:charm_yield_rhic_central_initial}.
This larger yield is a result of the larger initial gluon energy density of both models compared to PYTHIA, as discussed in Section \ref{sec:other_ini_models}. 

\begin{figure}
	\centering
\includegraphics[width=\gnuplotwidth]{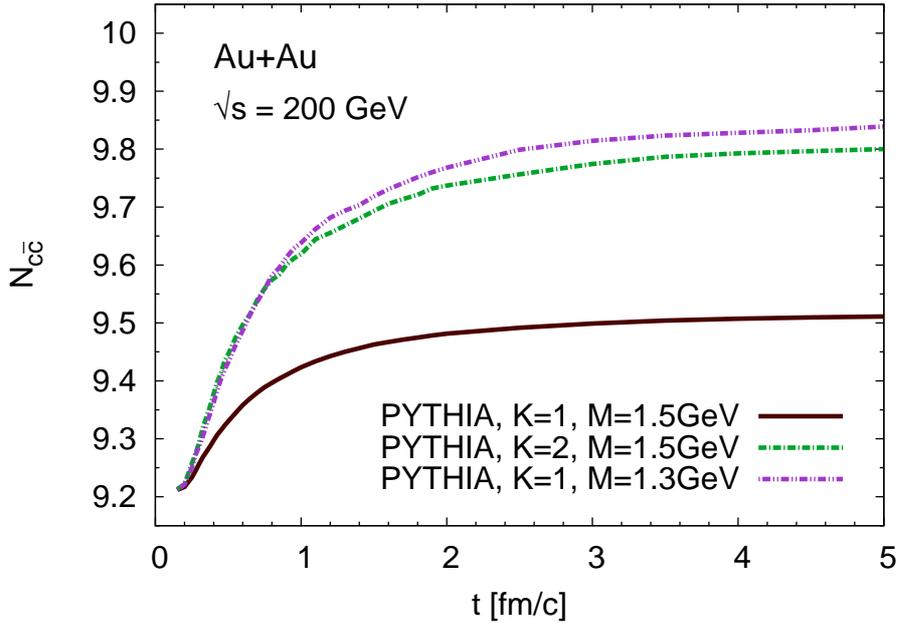}%eps
	\caption{(Color online) As in \autoref{fig:charm_yield_rhic_central_pythia}, but varying the charm mass $M$ and the $K$ factor for the cross section of $gg\rightarrow c\bar c$. }
	\label{fig:charm_yield_rhic_central_k_M}
\end{figure}

Often, the charm production cross section is multiplied by a $K$ factor of 2, which leads (naturally) to twice as many charm quarks. Nearly the same effect is attained by altering the charm mass from 1.5\,GeV to 1.3\,GeV, as depicted in \autoref{fig:charm_yield_rhic_central_k_M}.

Either changing the initial gluon distribution, choosing a smaller charm mass, or employing a $K$ factor can enhance the charm production by about a factor of 2, but the ratio of charm quarks produced in the QGP and total charm yield is still just 6\,\%. A combination of all these parameter changes (minijet initial conditions with high initial energy density, small charm mass of $M=1.3 \, {\rm GeV}$ and $K$ factor of 2) yields 3.4 charm quarks which are produced during the QGP phase. This corresponds to 27\,\% of the final charm quarks, a value being the upper limit of charm production in our model.

In \autoref{tab:qgp_charm_yield_rhic_midrap} our results are summarized. 
\begin{table}
	\centering
		\begin{tabular}{c|c|c||D{!}{.}{2.3}|D{!}{.}{3.2}} 
IC & $M_c$ & $K$  & \multicolumn{1}{c|}{$\frac{{\rm d}N_{c \bar c}}{{\rm d}y}$}& \multicolumn{1}{c}{$N_{c \bar c}$}
  \\ \hline \hline
PYTHIA & 1.5  & 1  & 1!68 & 9!5  \\ \hline
Mini-jets& 1.5 & 1  & 1!74 & 10!0  \\ \hline
CGC & 1.5  &  1  & 1!74  &  9!9 \\ \hline
PYTHIA & 1.5  &  2  & 1!74 &  9!8  \\ \hline
PYTHIA & 1.3  &  1  & 1!71 &9!8    \\ \hline
Mini-jets  & 1.3  &  2  & 2!16&  12!6 
		\end{tabular}
	\caption{Comparison of the total charm pair yield $N_{c\bar c}$ and the number of charm pairs at mid-rapidity ($y\in [-0.5,0.5]$) $\frac{{\rm d}N_{c \bar c}}{{\rm d}y}$  after the QGP phase at central Au+Au collisions at RHIC simulated with BAMPS. Masses $M_c$ are in units of GeV. 
$K$ stands for  $K$ factor, and IC for initial conditions.}
	\label{tab:qgp_charm_yield_rhic_midrap}
\end{table}
In addition to the total charm numbers, we list the numbers at mid-rapidity. 
\autoref{tab:qgp_charm_yield_rhic} gives an overview about charm production at RHIC in other models. 
\begin{table*}
	\centering
% 		\begin{tabular}{c||D{!}{.}{3.2}||c|c|c|m{5.0cm}} 
		\begin{tabular}{c||D{!}{.}{3.2}||c|c|c|l} 
Reference & \multicolumn{1}{c||}{$N_{c \bar{c},{\rm QGP}}$}  & $M_c$ & $K$ & IC & Comments \\ \hline \hline
Present work & 0!32 & 1.5  & 1 &  PYTHIA & Gluonic medium, $\alpha_s=0.3$ \\ \hline
Present work & 0!80 & 1.5  &  1 & Mini-jets & Gluonic medium, $\alpha_s=0.3$ \\ \hline
Present work & 0!77 & 1.5  &  1 & CGC & Gluonic medium, $\alpha_s=0.3$ \\ \hline
Present work & 0!63 & 1.5  &  2 & PYTHIA & Gluonic medium, $\alpha_s=0.3$ \\ \hline
Present work & 0!61 & 1.3  &  1 & PYTHIA & Gluonic medium, $\alpha_s=0.3$ \\ \hline
Present work & 3!4 & 1.3  &  2 & Mini-jets & Gluonic medium, $\alpha_s=0.3$ \\ \hline
\hline
\cite{Levai:1997bi} & 0!97 & 1.5 & 1 & HIJING & Gluonic medium \\ \hline
\cite{Levai:1997bi} & 1!38 & 1.5 & 1 & HIJING & Quarks and gluons with thermal masses \\ \hline
\cite{Levai:1997bi} & 3!2 & 1.2 & 1 & HIJING & Gluonic medium \\ \hline
\cite{Levai:1997bi} & 4!9 & 1.2 & 1 & HIJING &  Quarks and gluons with thermal masses \\ \hline 
\cite{Levai:1997bi} & 0!01 & 1.5 & 1 & Minijets & Gluonic medium \\ \hline 

\cite{Gavin:1996bx} & 1 & 1.2 & 1 & Hydro & \\ \hline 

\cite{Duraes:2004zt} & 3!8 & 1.5 & 1 & Therm. & running coupling $\alpha_s$ \\ \hline 
\cite{Duraes:2004zt} & 39 & 1.2 & 2 & Therm. & running $\alpha_s$ \\ \hline 
\cite{Duraes:2004zt} & 11 & 1.5 & 1 & Therm. & constant $\alpha_s(M_c^2)=0.37$ \\ \hline 
\cite{Duraes:2004zt} & 120 & 1.2 & 2 & Therm. & constant $\alpha_s(M_c^2)=0.42$ \\ \hline

\cite{Muller:1992xn} &  \sim 2 & 1.5 & 2 & HIJING &  \\ \hline 
\cite{Levai:1994dx} & \sim 0!2 & 1.5 & 2 & HIJING &  \\ 
	
		\end{tabular}
	\caption{Comparison of charm pair yield $N_{c \bar{c},{\rm QGP}}$ in the QGP at RHIC from various models. Masses $M_c$ are given in GeV. 
$K$ stands for  $K$ factor and IC for initial conditions.}
	\label{tab:qgp_charm_yield_rhic}
\end{table*}
The big deviations of the various models result from the influence of many parameters, which are often chosen differently in different models. For instance, there is disagreement regarding the temperature of the QGP, thermalization time scale, constant or running $\alpha_s$, thermal masses of quarks and gluons, volume, energy density, $K$ factor or charm mass. Within our results we also saw the sensitivity of the production rates concerning some of these parameters.

\begin{figure}
	\centering
\begin{minipage}{0.02\linewidth}
(a)
\end{minipage}
\begin{minipage}{0.97\linewidth}
\includegraphics[width=\gnuplotwidth]{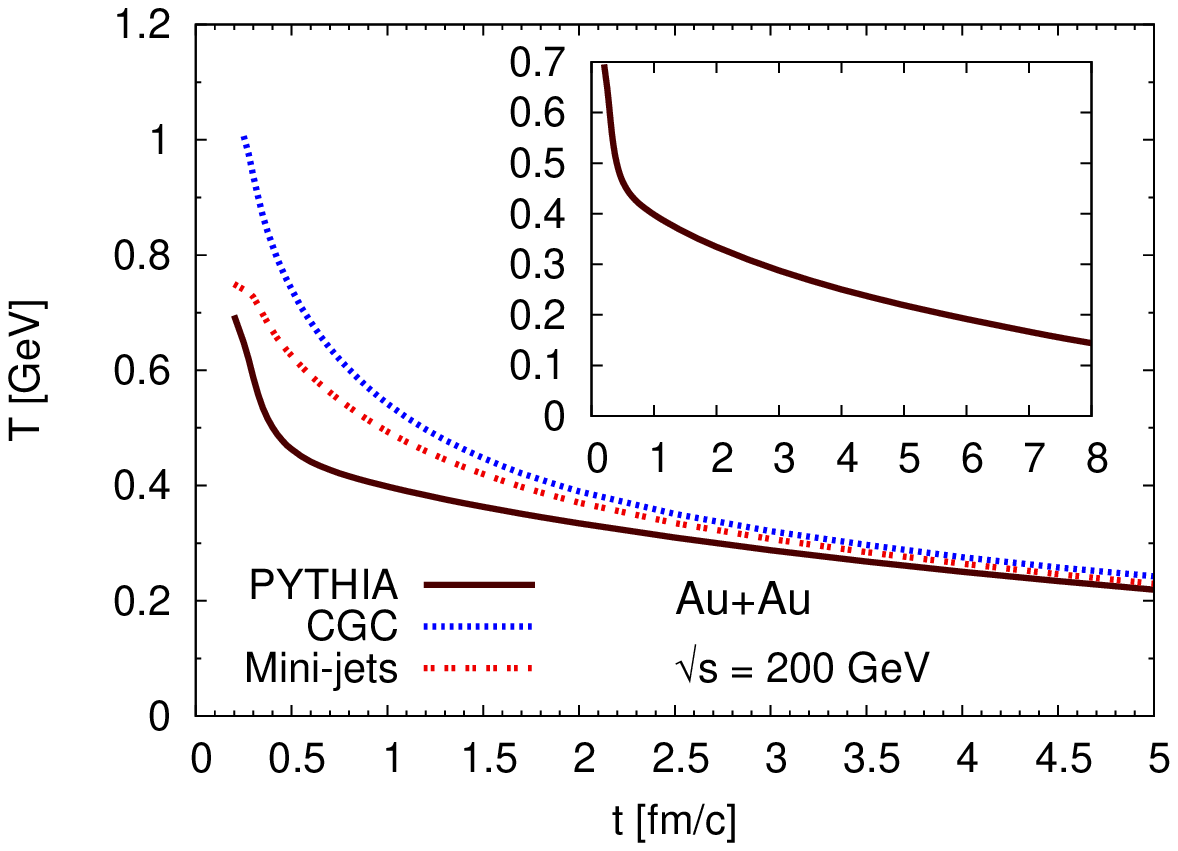}%eps
\end{minipage}

\begin{minipage}{0.02\linewidth}
(b)
\end{minipage}
\begin{minipage}{0.97\linewidth}
\includegraphics[width=\gnuplotwidth]{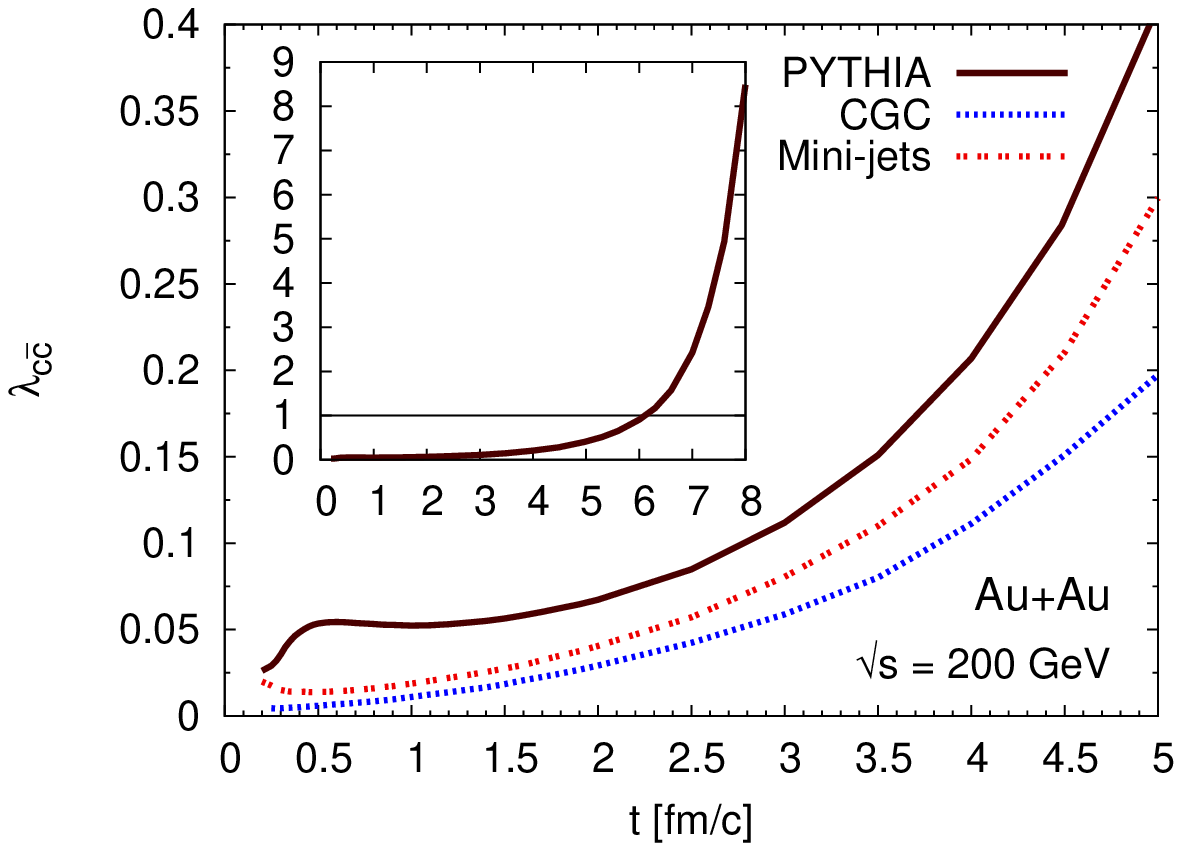}%eps
\end{minipage}
	\caption{(Color online) Evolution of the effective gluon temperature (a) and the charm quark fugacity (b) in the center of the collision ($r\leq2 \,{\rm fm}$ and $ \eta \in[-0.5,0.5]$) at RHIC for different initial models. The insets show the evolution of both quantities for PYTHIA initial conditions until $8 \,{\rm fm}$ when the temperature drops below the phase transition temperature of about $150-200\,{\rm MeV}$ \cite{Cheng:2006qk,Aoki:2006br,Aoki:2009sc}.}
	\label{fig:rhic_temp_fugacity_qgp}
\end{figure}

Another interesting topic we want to address is the question of whether charm quarks are chemically equilibrated. For that, we investigate the fugacity defined in \autoref{fugacity} for charm quarks which are located in central tubes with radius $r=2 \,{\rm fm}$ in transverse direction and longitudinal boundaries at space-time rapidity $ \eta = \pm 0.5$. The equilibrium value for the charm quark number $n^{\rm chem. eq}_{c\bar{c}}$ is computed with the effective temperature of the gluonic medium. Both variables are shown in \autoref{fig:rhic_temp_fugacity_qgp} for all three initial models.
The temperature is quite large for RHIC energies, because we consider only a gluonic medium. If we took also light quarks into account, the temperature of the medium would be smaller.

The number of charm quarks is below the equilibrium value and increases with time. As we saw in Section \ref{sec:rates_comp_ana_num_rhic}, however, their equilibration time scale is by far too large to see a significant rise before hadronization. In addition, the production rate decreases with time due to the accompanied temperature decline \cite{Zhang:2008zzc}. 
If the system evolves for a longer time, as is shown in the insets in \autoref{fig:rhic_temp_fugacity_qgp}, the fugacity increases dramatically because of the decreasing temperature. At the temperature of $161\,{\rm MeV}$ 
the fugacity is about 4, which is of the same order as the result from \cite{Andronic:2006ky,Andronic:2007ff} for the same temperature. However, in \cite{Andronic:2006ky,Andronic:2007ff} the fugacity is computed on a hadronic level from the equilibrium number of charmed hadrons, whereas in our approach the fugacity is calculated on a quark level from the equilibrium number of charm quarks. Taking this and the difference of the considered volume into account, it is understandable that the values of the fugacity are not exactly the same. 

At this low effective temperature of  $161\,{\rm MeV}$, the energy density in the considered volume is about $0.3\,{\rm GeV/fm}^3$ (in a totally equilibrated gluonic medium it would be $0.43\,{\rm GeV/fm}^3$). As a note, if the energy density in a cell in BAMPS drops below the critical energy density of $1\,{\rm GeV/fm}^3$ the particles stream freely and do not interact anymore. In the present simulation this energy density in the considered volume corresponds to $T \approx 220\,{\rm MeV}$ (cf. $T \approx 200\,{\rm MeV}$ in an equilibrated medium).

\subsection{Charm production at LHC}
\label{sec:prod_qgp_lhc}

Due to the higher center of mass energy at LHC than at RHIC, more charm quarks are produced during the QGP phase and the yield from initial nucleon-nucleon scatterings is also much higher  at LHC. However, since the charm production in the QGP increases almost exponentially with the QGP temperature but the initial charm yield only logarithmically with the collision energy \cite{Zhang:2008zzc}, the importance of the former rises significantly.

\begin{figure}
	\centering
\includegraphics[width=\gnuplotwidth]{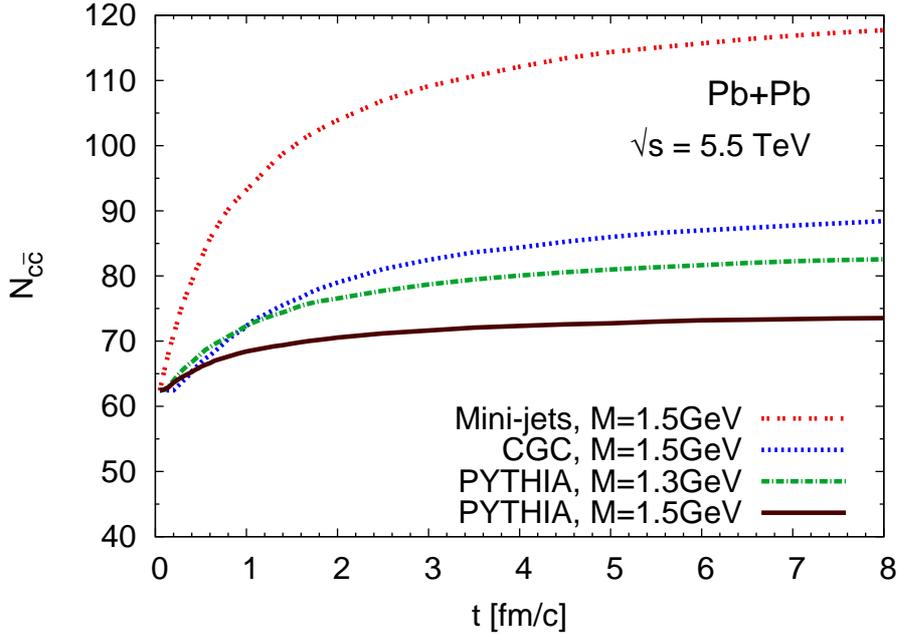}%eps
	\caption{(Color online) Number of charm pairs in a central Pb+Pb collision at LHC simulated with BAMPS with PYTHIA, CGC, and mini-jet initial conditions for gluons.}
	\label{fig:charm_yield_lhc_central}
\end{figure}

For the initial gluon distribution we use again PYTHIA, the CGC, and the mini-jet model.
In Section \ref{sec:ini_charm_pythia} we estimated with PYTHIA and CTEQ6l about 62 charm pairs for the initial yield at LHC, which evolve in the QGP as a function of time as shown in \autoref{fig:charm_yield_lhc_central}. 
With initial conditions from PYTHIA about 11 charm pairs are produced, which corresponds to about 15\,\% of the total final charm quarks. In contrast to RHIC, where nearly all charm quarks are produced in initial hard parton scatterings, at LHC a sizable fraction is created in the QGP. For initial gluon distributions from the CGC or the mini-jet model this fraction is even higher. In the latter framework the charm yield in the QGP is actually comparable to the initial yield (47\,\% of the total charm comes from the QGP).
The introduction of a $K$ factor of 2 or the lowering of the charm mass to $M_c = 1.3 \, {\rm GeV}$ increases these values again by a factor of 2.

\autoref{tab:qgp_charm_yield_lhc_midrap} summarizes our results and \autoref{tab:qgp_charm_yield_lhc} compares them to other models.
\begin{table}
	\centering
		\begin{tabular}{c|c|c||D{!}{.}{3.2}|D{!}{.}{4.1}} 
IC & $M_c$ & $K$  & \multicolumn{1}{c|}{$\frac{{\rm d}N_{c \bar c}}{{\rm d}y}$} & \multicolumn{1}{c}{$N_{c \bar c}$}
  \\ \hline \hline
PYTHIA & 1.5  & 1  & 7!8 &   74   \\ \hline
Mini-jets  & 1.5 & 1 & 14!7 & 118 \\ \hline
CGC & 1.5  &  1  & 10!6&  88  \\ \hline
PYTHIA & 1.5  &  2   & 9!0 & 85 \\ \hline
PYTHIA & 1.3  &  1  &  8!7 &  83
		\end{tabular}
	\caption{As in \autoref{tab:qgp_charm_yield_rhic_midrap}, but for central Pb+Pb collisions at LHC.}
	\label{tab:qgp_charm_yield_lhc_midrap}
\end{table}
\begin{table*}
	\centering
			\begin{tabular}{c||c||c|c|c|m{7.0cm}} 
% 			\begin{tabular}{c||c||c|c|c|l} 
Reference & $N_{c \bar{c},{\rm QGP}}$  & $M_c$ & $K$ & IC & Comments \\ \hline \hline
Present work & 11 & 1.5  & 1 &  PYTHIA & Gluonic medium, $\alpha_s=0.3$ \\ \hline
Present work & 55 & 1.5  &  1 & Mini-jets & Gluonic medium, $\alpha_s=0.3$ \\ \hline
Present work & 26 & 1.5  &  1 & CGC & Gluonic medium, $\alpha_s=0.3$ \\ \hline
Present work & 23 & 1.5  & 2 &  PYTHIA & Gluonic medium, $\alpha_s=0.3$ \\ \hline
Present work & 20 & 1.3  & 1 &  PYTHIA & Gluonic medium, $\alpha_s=0.3$ \\ \hline
Present work & 38 & 1.5  & 2 &  PYTHIA & Gluonic medium, $\alpha_s=0.3$ \\ \hline
\hline
\cite{Levai:1997bi} & 43 & 1.5 & 1 & HIJING & Gluonic medium \\ \hline
\cite{Levai:1997bi} & 94 & 1.5 & 1 & HIJING & Quarks and gluons with thermal masses \\ \hline
\cite{Levai:1997bi} & 102 & 1.2 & 1 & HIJING & Gluonic medium \\ \hline
\cite{Levai:1997bi} & 245 & 1.2 & 1 & HIJING & Quarks and gluons with thermal masses \\ \hline 
\cite{Levai:1997bi} & 21 & 1.5 & 1 & Minijets & Gluonic medium \\ \hline 

\cite{Gavin:1996bx} & 23 & 1.2 & 1 & Hydro. & \\ \hline 

\cite{BraunMunzinger:2000dv} & 5 & 1.5 & 1 & SSPC \& HIJING & $\tau_0 = 0.25 \, {\rm fm}/c$, $T_0 = 1.02 \, {\rm GeV}$, $\lambda_0^g = 0.43$, $\lambda_0^q = 0.082$ \\ \hline 

\cite{Zhang:2008zzc} & (7) & 1.3 & 1 & Therm. & $T_0=700  \, {\rm MeV}$, NLO,  at mid-rapidity \\ \hline 
\cite{Zhang:2008zzc} &  (33) & 1.3 & 1 & Therm. & $T_0=750  \, {\rm MeV}$, NLO, at mid-rapidity \\ 

		\end{tabular}
	\caption{As in \autoref{tab:qgp_charm_yield_rhic}, but for central Pb+Pb collisions at LHC.}
	\label{tab:qgp_charm_yield_lhc}
\end{table*}
As for RHIC (cf. previous section), most of these predictions from other models are only for thermally produced charm quarks, whereas in our calculations the prethermal charm production within the first  $\approx 0.5 \, {\rm fm}/c$ plays a crucial role.

\begin{figure}
	\centering
\begin{minipage}{0.02\linewidth}
(a)
\end{minipage}
\begin{minipage}{0.97\linewidth}
\includegraphics[width=\gnuplotwidth]{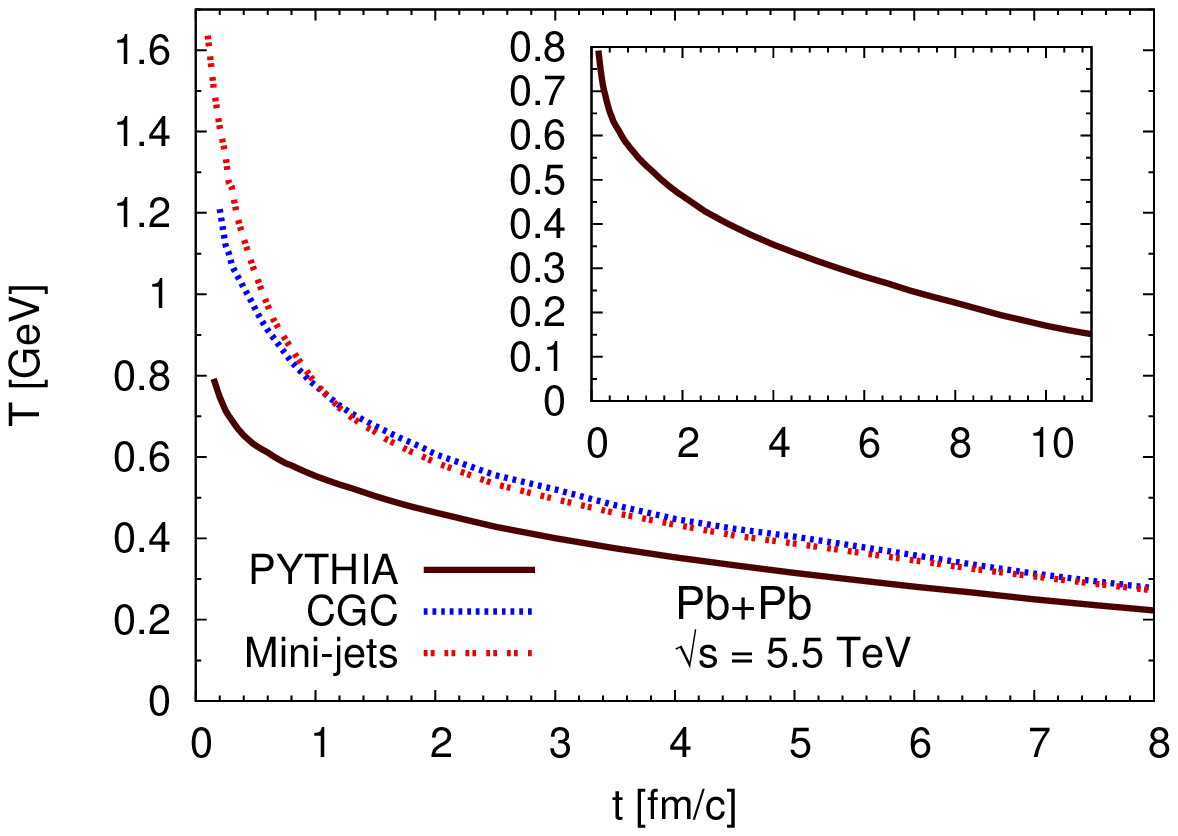}%eps
\end{minipage}

\begin{minipage}{0.02\linewidth}
(b)
\end{minipage}
\begin{minipage}{0.97\linewidth}
\includegraphics[width=\gnuplotwidth]{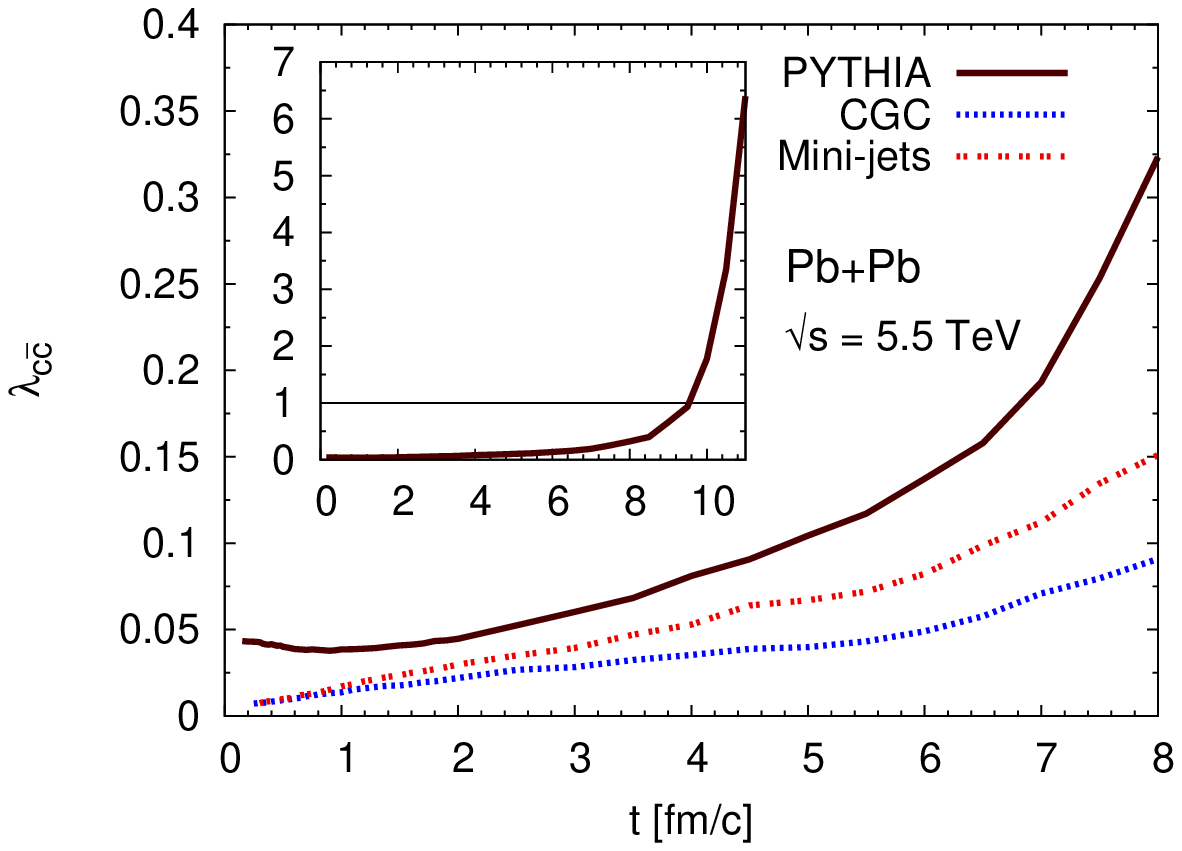}%eps
\end{minipage}
	\caption{(Color online) As in \autoref{fig:rhic_temp_fugacity_qgp}, but for LHC.}
	\label{fig:lhc_temp_fugacity_qgp}
\end{figure}

\autoref{fig:lhc_temp_fugacity_qgp} shows the evolution of the plasma temperature and the charm quark fugacity at LHC.
The charm quark fugacity being below 1 and the smaller chemical equilibration time at this temperature (cf. Section \ref{sec:rates_comp_ana_num_lhc}) are the reasons for the significant charm production during the QGP phase at LHC.
The inlays show that the fugacity increases dramatically, when the temperature approaches its value at the phase transition. At a temperature of $161\,{\rm MeV}$ the charm quark fugacity is about 4, which is considerably smaller than the fugacity from \cite{Andronic:2006ky} based on charmed hadrons.

\subsection{Bottom production at LHC}
\label{sec:lhc_bottom}

\begin{figure}
	\centering
\includegraphics[width=\gnuplotwidth]{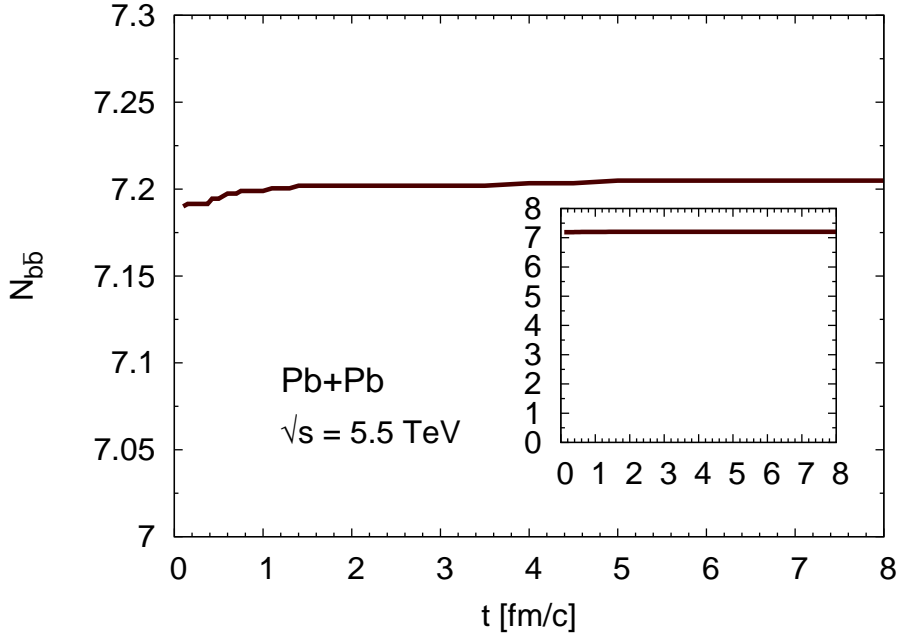}%eps
	\caption{(Color online) Number of bottom pairs produced in central Pb+Pb collisions at LHC simulated with BAMPS. The initial parton distribution is sampled with PYTHIA.}
	\label{fig:bottom_yield_lhc_central}
\end{figure}

In Section \ref{sec:ini_charm_pythia} we estimated the initial bottom yield at LHC with PYTHIA and CTEQ6l to about 7.2 pairs. 
\autoref{fig:bottom_yield_lhc_central} shows the evolution of the number of bottom pairs, which is only slightly dependent on time.
According to our simulation only 0.01 bottom pairs are produced on average in the QGP at LHC, which corresponds to about 0.2\,\% of the total final bottom quarks. Consequently, bottom production in the QGP phase is negligible. This is a result of the huge bottom mass $M_b \gg T$, which is much larger than the temperature of the medium \cite{Grandchamp:2005yw}.

Therefore, bottom quarks are a promising probe at the LHC. On the one hand their number is large enough to be measurable; on the other hand one knows that bottom quarks are produced initially. Due to this information one will be able to draw conclusions about the early phase of the QGP from observables like the elliptic flow or the nuclear modification factor.

\section{Conclusions}
\label{sec:conclusion}

We have studied charm and bottom production in heavy ion collisions at RHIC and LHC using PYTHIA and the mini-jet model for the primary yield and the partonic transport model BAMPS for secondary production in the QGP. For that, we implemented the LO $2\rightarrow 2$ processes for heavy quark production, their annihilation and their interaction with gluons. The charm production in the QGP at RHIC can be neglected, but at LHC we expect that open charm will no longer scale with the number of binary collisions due to a large charm production in the QGP.

The BAMPS simulation of charm production in a static gluonic medium was compared with the analytic solution of the corresponding rate equation, where an excellent agreement was found. From these calculations we estimated the order of magnitude of the chemical equilibration time scale for charm production. For a medium with an initial temperature of 400\,MeV -- the approximate temperature of the QGP at RHIC -- we found a time scale of about $700\,{\rm fm}/c$, which is much larger than the lifetime of the QGP hinting of a small charm production at RHIC. For an LHC temperature of 800\,MeV, we estimated the chemical equilibration time scale to be $30\,{\rm fm}/c$. For slightly higher temperatures, it is even comparable to the lifetime of the QGP at LHC. Consequently, one can expect a significant charm production in the QGP at LHC, which is, however, still strongly dependent on the initial conditions.

The production of heavy quarks in initial hard parton scatterings during nucleon-nucleon collisions was estimated with PYTHIA and the mini-jet model in LO pQCD, revealing a strong sensitivity to the chosen parton distribution functions, heavy quark mass, factorization, and renormalization scale. Charm production within the mini-jet model lay about a factor of 5 below the PHENIX data, while PYTHIA's value was close to the experimental data.

Based on this initial heavy quark yield, we studied the production of charm quarks in central heavy ion collisions at RHIC and LHC and bottom quarks at LHC in a full space-time simulation of the QGP with BAMPS. In addition to the 9.2 initially produced charm pairs with PYTHIA and CTEQ6l, between 0.3 and 3.4 pairs were created in the QGP, depending on the initial gluon distribution (from PYTHIA, CGC, or mini-jets), charm mass, and $K$ factor. Because of this small production, the charm quark number during the QGP phase is below its equilibrium value and only rises above this value when the temperature of the medium drops below 200\,MeV.

For Pb+Pb collisions at LHC, the number of initially produced charm pairs is 62 according to PYTHIA with CTEQ6l. Subsequently, in the QGP, the yield is significant: between 11 and 55 pairs are created, depending on the model for the initial gluon distribution. 
%This leads to the prediction that open charm will not scale with the number of binary collisions at LHC.
For the LHC, we estimated the initial bottom yield as 7.2 pairs, whereas their production in the QGP can be neglected.

In conclusion, heavy quarks are an interesting probe for the early stage of heavy ion collisions due to their strong sensitivity on the initial conditions and the properties of the QGP within the first ${\rm fm}/c$. However, it is crucial to disentangle the origins of heavy quarks: whether they were produced in primary or in secondary production. If one knows the ratio of both, the measured total yield can reveal information about important variables in the QGP: The primary yield is strongly influenced by the parton distribution functions, mass, scales and also shadowing; the secondary production depends on the temperature of the medium, which is again dependent on the initial conditions including shadowing. Therefore, both yields are correlated and can be disentangled by comparing the effects of using different models for initial conditions and shadowing on the total heavy quark yield within an accurate description of the QGP. In the present article, we have studied the impact of different initial conditions but considered shadowing just by lowering the number of binary collisions. In a future investigation, we will take more sophisticated models for shadowing into account, which will also affect primary heavy quark production.

Other observables, that can give great insight into the dynamical properties of the QGP, are the elliptic flow and the nuclear modification factor $R_{AA}$ of heavy quarks \cite{Djordjevic:2003zk,Djordjevic:2005db,Wicks:2005gt,Moore:2004tg,vanHees:2005wb,Armesto:2005mz,Gossiaux:2008jv}. Experimentally, a strong elliptic flow of heavy flavor electrons and a small $R_{AA}$ have been observed \cite{Adare:2006nq}, which could not be described within the leading order regime of pQCD \cite{Zhang:2005ni}. However, next-to-leading order calculations \cite{Djordjevic:2003zk,Wicks:2005gt} that also include radiative energy loss are very promising for explaining the data. A future task will be to investigate the energy loss and these observables with our model, implementing also light quarks, higher order corrections, a running coupling, and an improved Debye screening \cite{Peshier:2008bg,Peigne:2008nd,Gossiaux:2008jv,Gossiaux:2009mk}. For that, BAMPS is a well suited model, since the framework for  $2 \leftrightarrow 3$ interactions is already implemented \cite{Xu:2004mz}. Extending recent BAMPS studies concerning $v_2$ and $R_{AA}$ of gluons \cite{Xu:2007jv,Fochler:2008ts,Xu:2008av,Xu:2010cq} to the heavy flavor sector will reveal further information about interactions in the QGP. Moreover, we want to study the effect of hadronization on the elliptic flow or $R_{AA}$, which could be done, for instance, in a fragmentation, coalescence or statistical hadronization model \cite{Greco:2003vf,Lin:2003jy,Zhang:2005ni,Andronic:2006ky}. In this framework $J/\psi$ creation can be investigated as well. If necessary, we will also consider other schemes such as  AdS/CFT correspondence \cite{CasalderreySolana:2006rq,Herzog:2006gh} in addition to pQCD.

\section*{Acknowledgements}
J. U. would like to thank Steffen Bass and Berndt M\"uller for fruitful discussions and the kind hospitality at Duke University, where part of this work has been done. We are also thankful to Andr\'e Peshier, Anton Andronic, Andr\'e Mischke, and Hendrik van Hees for helpful discussions.

The BAMPS simulations were performed at the Center for Scientific Computing of the Goethe University. This work was supported by the Helmholtz International Center for FAIR within the framework of the LOEWE program (Landes-Offensive zur Entwicklung Wissenschaftlich-\"okonomischer Exzellenz) launched by the State of Hesse.

\bibliography{charmprod}

%Merlin.mbs v4.21 2009-07-09.
\begin{thebibliography}{100}%
\makeatletter
\providecommand \@ifxundefined [1]{%
 \ifx #1\undefined \expandafter \@firstoftwo
 \else \expandafter \@secondoftwo
\fi
}%
\providecommand \@ifnum [1]{%
 \ifnum #1\expandafter \@firstoftwo
 \else \expandafter \@secondoftwo
\fi
}%
\providecommand \enquote [1]{``#1''}%
\providecommand \bibnamefont  [1]{#1}%
\providecommand \bibfnamefont [1]{#1}%
\providecommand \citenamefont [1]{#1}%
\providecommand\href[0]{\@sanitize\@href}%
\providecommand\@href[1]{\endgroup\@@startlink{#1}\endgroup\@@href}%
\providecommand\@@href[1]{#1\@@endlink}%
\providecommand \@sanitize [0]{\begingroup\catcode`\&12\catcode`\#12\relax}%
\@ifxundefined \pdfoutput {\@firstoftwo}{%
 \@ifnum{\z@=\pdfoutput}{\@firstoftwo}{\@secondoftwo}%
}{%
 \providecommand\@@startlink[1]{\leavevmode\special{html:<a href="#1">}}%
 \providecommand\@@endlink[0]{\special{html:</a>}}%
}{%
 \providecommand\@@startlink[1]{%
  \leavevmode
  \pdfstartlink
   attr{/Border[0 0 1 ]/H/I/C[0 1 1]}%
   user{/Subtype/Link/A<</Type/Action/S/URI/URI(#1)>>}%
  \relax
 }%
 \providecommand\@@endlink[0]{\pdfendlink}%
}%
\providecommand \url  [0]{\begingroup\@sanitize \@url }%
\providecommand \@url [1]{\endgroup\@href {#1}{\urlprefix}}%
\providecommand \urlprefix [0]{URL }%
\providecommand \Eprint[0]{\href }%
\@ifxundefined \urlstyle {%
  \providecommand \doi [1]{doi:\discretionary{}{}{}#1}%
}{%
  \providecommand \doi [0]{doi:\discretionary{}{}{}\begingroup
  \urlstyle{rm}\Url }%
}%
\providecommand \doibase [0]{http://dx.doi.org/}%
\providecommand \Doi[1]{\href{\doibase#1}}%
\providecommand \bibAnnote [3]{%
  \BibitemShut{#1}%
  \begin{quotation}\noindent
    \textsc{Key:}\ #2\\\textsc{Annotation:}\ #3%
  \end{quotation}%
}%
\providecommand \bibAnnoteFile [2]{%
  \IfFileExists{#2}{\bibAnnote {#1} {#2} {\input{#2}}}{}%
}%
\providecommand \typeout [0]{\immediate \write \m@ne }%
\providecommand \selectlanguage [0]{\@gobble}%
\providecommand \bibinfo [0]{\@secondoftwo}%
\providecommand \bibfield [0]{\@secondoftwo}%
\providecommand \translation [1]{[#1]}%
\providecommand \BibitemOpen[0]{}%
\providecommand \bibitemStop [0]{}%
\providecommand \bibitemNoStop [0]{.\EOS\space}%
\providecommand \EOS [0]{\spacefactor3000\relax}%
\providecommand \BibitemShut [1]{\csname bibitem#1\endcsname}%
%</preamble>
\bibitem{Adams:2005dq}%
  \BibitemOpen
  \bibfield{author}{%
  \bibinfo {author} {\bibfnamefont{J.}~\bibnamefont{Adams}} \emph{et~al.}
  (\bibinfo {collaboration} {STAR}),\ }%
  \bibfield{journal}{%
  \bibinfo {journal} {Nucl. Phys.}\ }%
  \textbf{\bibinfo {volume} {A757}},\ \bibinfo {pages} {102} (\bibinfo {year}
  {2005}),\ \Eprint{http://arxiv.org/abs/nucl-ex/0501009}{nucl-ex/0501009}%
  \bibAnnoteFile{NoStop}{Adams:2005dq}%
%%CITATION = NUCL-EX 0501009;%%
\bibitem{Adcox:2004mh}%
  \BibitemOpen
  \bibfield{author}{%
  \bibinfo {author} {\bibfnamefont{K.}~\bibnamefont{Adcox}} \emph{et~al.}
  (\bibinfo {collaboration} {PHENIX}),\ }%
  \bibfield{journal}{%
  \bibinfo {journal} {Nucl. Phys.}\ }%
  \textbf{\bibinfo {volume} {A757}},\ \bibinfo {pages} {184} (\bibinfo {year}
  {2005}),\ \Eprint{http://arxiv.org/abs/nucl-ex/0410003}{nucl-ex/0410003}%
  \bibAnnoteFile{NoStop}{Adcox:2004mh}%
%%CITATION = NUCL-EX 0410003;%%
\bibitem{Arsene:2004fa}%
  \BibitemOpen
  \bibfield{author}{%
  \bibinfo {author} {\bibfnamefont{I.}~\bibnamefont{Arsene}} \emph{et~al.}
  (\bibinfo {collaboration} {BRAHMS}),\ }%
  \bibfield{journal}{%
  \bibinfo {journal} {Nucl. Phys.}\ }%
  \textbf{\bibinfo {volume} {A757}},\ \bibinfo {pages} {1} (\bibinfo {year}
  {2005}),\ \Eprint{http://arxiv.org/abs/nucl-ex/0410020}{nucl-ex/0410020}%
  \bibAnnoteFile{NoStop}{Arsene:2004fa}%
%%CITATION = NUCL-EX 0410020;%%
\bibitem{Back:2004je}%
  \BibitemOpen
  \bibfield{author}{%
  \bibinfo {author} {\bibfnamefont{B.~B.}\ \bibnamefont{Back}} \emph{et~al.}
  (\bibinfo {collaboration} {PHOBOS}),\ }%
  \bibfield{journal}{%
  \bibinfo {journal} {Nucl. Phys.}\ }%
  \textbf{\bibinfo {volume} {A757}},\ \bibinfo {pages} {28} (\bibinfo {year}
  {2005}),\ \Eprint{http://arxiv.org/abs/nucl-ex/0410022}{nucl-ex/0410022}%
  \bibAnnoteFile{NoStop}{Back:2004je}%
%%CITATION = NUCL-EX 0410022;%%
\bibitem{Kolb:2000sd}%
  \BibitemOpen
  \bibfield{author}{%
  \bibinfo {author} {\bibfnamefont{P.~F.}\ \bibnamefont{Kolb}}, \bibinfo
  {author} {\bibfnamefont{J.}~\bibnamefont{Sollfrank}},\ and\ \bibinfo {author}
  {\bibfnamefont{U.~W.}\ \bibnamefont{Heinz}},\ }%
  \bibfield{journal}{%
  \bibinfo {journal} {Phys. Rev.}\ }%
  \textbf{\bibinfo {volume} {C62}},\ \bibinfo {pages} {054909} (\bibinfo {year}
  {2000}),\ \Eprint{http://arxiv.org/abs/hep-ph/0006129}{hep-ph/0006129}%
  \bibAnnoteFile{NoStop}{Kolb:2000sd}%
%%CITATION = HEP-PH 0006129;%%
\bibitem{Heinz:2001xi}%
  \BibitemOpen
  \bibfield{author}{%
  \bibinfo {author} {\bibfnamefont{U.~W.}\ \bibnamefont{Heinz}}\ and\ \bibinfo
  {author} {\bibfnamefont{P.~F.}\ \bibnamefont{Kolb}},\ }%
  \bibfield{journal}{%
  \bibinfo {journal} {Nucl. Phys.}\ }%
  \textbf{\bibinfo {volume} {A702}},\ \bibinfo {pages} {269} (\bibinfo {year}
  {2002}),\ \Eprint{http://arxiv.org/abs/hep-ph/0111075}{hep-ph/0111075}%
  \bibAnnoteFile{NoStop}{Heinz:2001xi}%
%%CITATION = HEP-PH 0111075;%%
\bibitem{Huovinen:2001cy}%
  \BibitemOpen
  \bibfield{author}{%
  \bibinfo {author} {\bibfnamefont{P.}~\bibnamefont{Huovinen}}, \bibinfo
  {author} {\bibfnamefont{P.~F.}\ \bibnamefont{Kolb}}, \bibinfo {author}
  {\bibfnamefont{U.~W.}\ \bibnamefont{Heinz}}, \bibinfo {author}
  {\bibfnamefont{P.~V.}\ \bibnamefont{Ruuskanen}},\ and\ \bibinfo {author}
  {\bibfnamefont{S.~A.}\ \bibnamefont{Voloshin}},\ }%
  \bibfield{journal}{%
  \Doi{10.1016/S0370-2693(01)00219-2}{\bibinfo {journal} {Phys. Lett.}}\ }%
  \textbf{\bibinfo {volume} {B503}},\ \bibinfo {pages} {58} (\bibinfo {year}
  {2001}),\ \Eprint{http://arxiv.org/abs/hep-ph/0101136}{arXiv:hep-ph/0101136}%
  \bibAnnoteFile{NoStop}{Huovinen:2001cy}%
%%CITATION = HEP-PH/0101136;%%
\bibitem{Adams:2003am}%
  \BibitemOpen
  \bibfield{author}{%
  \bibinfo {author} {\bibfnamefont{J.}~\bibnamefont{Adams}} \emph{et~al.}
  (\bibinfo {collaboration} {STAR}),\ }%
  \bibfield{journal}{%
  \Doi{10.1103/PhysRevLett.92.052302}{\bibinfo {journal} {Phys. Rev. Lett.}}\
  }%
  \textbf{\bibinfo {volume} {92}},\ \bibinfo {pages} {052302} (\bibinfo {year}
  {2004}),\
  \Eprint{http://arxiv.org/abs/nucl-ex/0306007}{arXiv:nucl-ex/0306007}%
  \bibAnnoteFile{NoStop}{Adams:2003am}%
%%CITATION = NUCL-EX/0306007;%%
\bibitem{Adler:2003kt}%
  \BibitemOpen
  \bibfield{author}{%
  \bibinfo {author} {\bibfnamefont{S.~S.}\ \bibnamefont{Adler}} \emph{et~al.}
  (\bibinfo {collaboration} {PHENIX}),\ }%
  \bibfield{journal}{%
  \Doi{10.1103/PhysRevLett.91.182301}{\bibinfo {journal} {Phys. Rev. Lett.}}\
  }%
  \textbf{\bibinfo {volume} {91}},\ \bibinfo {pages} {182301} (\bibinfo {year}
  {2003}),\
  \Eprint{http://arxiv.org/abs/nucl-ex/0305013}{arXiv:nucl-ex/0305013}%
  \bibAnnoteFile{NoStop}{Adler:2003kt}%
%%CITATION = NUCL-EX/0305013;%%
\bibitem{Romatschke:2007mq}%
  \BibitemOpen
  \bibfield{author}{%
  \bibinfo {author} {\bibfnamefont{P.}~\bibnamefont{Romatschke}}\ and\ \bibinfo
  {author} {\bibfnamefont{U.}~\bibnamefont{Romatschke}},\ }%
  \bibfield{journal}{%
  \Doi{10.1103/PhysRevLett.99.172301}{\bibinfo {journal} {Phys. Rev. Lett.}}\
  }%
  \textbf{\bibinfo {volume} {99}},\ \bibinfo {pages} {172301} (\bibinfo {year}
  {2007}),\ \Eprint{http://arxiv.org/abs/0706.1522}{arXiv:0706.1522 [nucl-th]}%
  \bibAnnoteFile{NoStop}{Romatschke:2007mq}%
%%CITATION = 0706.1522;%%
\bibitem{Xu:2007jv}%
  \BibitemOpen
  \bibfield{author}{%
  \bibinfo {author} {\bibfnamefont{Z.}~\bibnamefont{Xu}}, \bibinfo {author}
  {\bibfnamefont{C.}~\bibnamefont{Greiner}},\ and\ \bibinfo {author}
  {\bibfnamefont{H.}~\bibnamefont{St\"ocker}},\ }%
  \bibfield{journal}{%
  \Doi{10.1103/PhysRevLett.101.082302}{\bibinfo {journal} {Phys. Rev. Lett.}}\
  }%
  \textbf{\bibinfo {volume} {101}},\ \bibinfo {pages} {082302} (\bibinfo {year}
  {2008}),\ \Eprint{http://arxiv.org/abs/0711.0961}{arXiv:0711.0961 [nucl-th]}%
  \bibAnnoteFile{NoStop}{Xu:2007jv}%
%%CITATION = 0711.0961;%%
\bibitem{Bethke:2006ac}%
  \BibitemOpen
  \bibfield{author}{%
  \bibinfo {author} {\bibfnamefont{S.}~\bibnamefont{Bethke}},\ }%
  \bibfield{journal}{%
  \Doi{10.1016/j.ppnp.2006.06.001}{\bibinfo {journal} {Prog. Part. Nucl.
  Phys.}}\ }%
  \textbf{\bibinfo {volume} {58}},\ \bibinfo {pages} {351} (\bibinfo {year}
  {2007}),\ \Eprint{http://arxiv.org/abs/hep-ex/0606035}{arXiv:hep-ex/0606035}%
  \bibAnnoteFile{NoStop}{Bethke:2006ac}%
%%CITATION = HEP-EX/0606035;%%
\bibitem{Nason:1987xz}%
  \BibitemOpen
  \bibfield{author}{%
  \bibinfo {author} {\bibfnamefont{P.}~\bibnamefont{Nason}}, \bibinfo {author}
  {\bibfnamefont{S.}~\bibnamefont{Dawson}},\ and\ \bibinfo {author}
  {\bibfnamefont{R.~K.}\ \bibnamefont{Ellis}},\ }%
  \bibfield{journal}{%
  \Doi{10.1016/0550-3213(88)90422-1}{\bibinfo {journal} {Nucl. Phys.}}\ }%
  \textbf{\bibinfo {volume} {B303}},\ \bibinfo {pages} {607} (\bibinfo {year}
  {1988})%
  \bibAnnoteFile{NoStop}{Nason:1987xz}%
%%CITATION = NUPHA,B303,607;%%
\bibitem{Nason:1989zy}%
  \BibitemOpen
  \bibfield{author}{%
  \bibinfo {author} {\bibfnamefont{P.}~\bibnamefont{Nason}}, \bibinfo {author}
  {\bibfnamefont{S.}~\bibnamefont{Dawson}},\ and\ \bibinfo {author}
  {\bibfnamefont{R.~K.}\ \bibnamefont{Ellis}},\ }%
  \bibfield{journal}{%
  \Doi{10.1016/0550-3213(89)90286-1}{\bibinfo {journal} {Nucl. Phys.}}\ }%
  \textbf{\bibinfo {volume} {B327}},\ \bibinfo {pages} {49} (\bibinfo {year}
  {1989})%
  \bibAnnoteFile{NoStop}{Nason:1989zy}%
%%CITATION = NUPHA,B327,49;%%
\bibitem{Dokshitzer:2001zm}%
  \BibitemOpen
  \bibfield{author}{%
  \bibinfo {author} {\bibfnamefont{Y.~L.}\ \bibnamefont{Dokshitzer}}\ and\
  \bibinfo {author} {\bibfnamefont{D.~E.}\ \bibnamefont{Kharzeev}},\ }%
  \bibfield{journal}{%
  \Doi{10.1016/S0370-2693(01)01130-3}{\bibinfo {journal} {Phys. Lett.}}\ }%
  \textbf{\bibinfo {volume} {B519}},\ \bibinfo {pages} {199} (\bibinfo {year}
  {2001}),\ \Eprint{http://arxiv.org/abs/hep-ph/0106202}{arXiv:hep-ph/0106202}%
  \bibAnnoteFile{NoStop}{Dokshitzer:2001zm}%
%%CITATION = HEP-PH/0106202;%%
\bibitem{Zhang:2003wk}%
  \BibitemOpen
  \bibfield{author}{%
  \bibinfo {author} {\bibfnamefont{B.-W.}\ \bibnamefont{Zhang}}, \bibinfo
  {author} {\bibfnamefont{E.}~\bibnamefont{Wang}},\ and\ \bibinfo {author}
  {\bibfnamefont{X.-N.}\ \bibnamefont{Wang}},\ }%
  \bibfield{journal}{%
  \Doi{10.1103/PhysRevLett.93.072301}{\bibinfo {journal} {Phys. Rev. Lett.}}\
  }%
  \textbf{\bibinfo {volume} {93}},\ \bibinfo {pages} {072301} (\bibinfo {year}
  {2004}),\
  \Eprint{http://arxiv.org/abs/nucl-th/0309040}{arXiv:nucl-th/0309040}%
  \bibAnnoteFile{NoStop}{Zhang:2003wk}%
%%CITATION = NUCL-TH/0309040;%%
\bibitem{Rapp:2008qc}%
  \BibitemOpen
  \bibfield{author}{%
  \bibinfo {author} {\bibfnamefont{R.}~\bibnamefont{Rapp}}\ and\ \bibinfo
  {author} {\bibfnamefont{H.}~\bibnamefont{van Hees}}}%
   (\bibinfo {year} {2008}),\
  \Eprint{http://arxiv.org/abs/0803.0901}{arXiv:0803.0901 [hep-ph]}%
  \bibAnnoteFile{NoStop}{Rapp:2008qc}%
%%CITATION = 0803.0901;%%
\bibitem{Abelev:2006db}%
  \BibitemOpen
  \bibfield{author}{%
  \bibinfo {author} {\bibfnamefont{B.~I.}\ \bibnamefont{Abelev}} \emph{et~al.}
  (\bibinfo {collaboration} {STAR}),\ }%
  \bibfield{journal}{%
  \Doi{10.1103/PhysRevLett.98.192301}{\bibinfo {journal} {Phys. Rev. Lett.}}\
  }%
  \textbf{\bibinfo {volume} {98}},\ \bibinfo {pages} {192301} (\bibinfo {year}
  {2007}),\
  \Eprint{http://arxiv.org/abs/nucl-ex/0607012}{arXiv:nucl-ex/0607012}%
  \bibAnnoteFile{NoStop}{Abelev:2006db}%
%%CITATION = NUCL-EX/0607012;%%
\bibitem{Adare:2006nq}%
  \BibitemOpen
  \bibfield{author}{%
  \bibinfo {author} {\bibfnamefont{A.}~\bibnamefont{Adare}} \emph{et~al.}
  (\bibinfo {collaboration} {PHENIX}),\ }%
  \bibfield{journal}{%
  \Doi{10.1103/PhysRevLett.98.172301}{\bibinfo {journal} {Phys. Rev. Lett.}}\
  }%
  \textbf{\bibinfo {volume} {98}},\ \bibinfo {pages} {172301} (\bibinfo {year}
  {2007}),\
  \Eprint{http://arxiv.org/abs/nucl-ex/0611018}{arXiv:nucl-ex/0611018}%
  \bibAnnoteFile{NoStop}{Adare:2006nq}%
%%CITATION = NUCL-EX/0611018;%%
\bibitem{Armesto:2005mz}%
  \BibitemOpen
  \bibfield{author}{%
  \bibinfo {author} {\bibfnamefont{N.}~\bibnamefont{Armesto}}, \bibinfo
  {author} {\bibfnamefont{M.}~\bibnamefont{Cacciari}}, \bibinfo {author}
  {\bibfnamefont{A.}~\bibnamefont{Dainese}}, \bibinfo {author}
  {\bibfnamefont{C.~A.}\ \bibnamefont{Salgado}},\ and\ \bibinfo {author}
  {\bibfnamefont{U.~A.}\ \bibnamefont{Wiedemann}},\ }%
  \bibfield{journal}{%
  \Doi{10.1016/j.physletb.2005.12.073}{\bibinfo {journal} {Phys. Lett.}}\ }%
  \textbf{\bibinfo {volume} {B637}},\ \bibinfo {pages} {362} (\bibinfo {year}
  {2006}),\ \Eprint{http://arxiv.org/abs/hep-ph/0511257}{arXiv:hep-ph/0511257}%
  \bibAnnoteFile{NoStop}{Armesto:2005mz}%
%%CITATION = HEP-PH/0511257;%%
\bibitem{vanHees:2005wb}%
  \BibitemOpen
  \bibfield{author}{%
  \bibinfo {author} {\bibfnamefont{H.}~\bibnamefont{van Hees}}, \bibinfo
  {author} {\bibfnamefont{V.}~\bibnamefont{Greco}},\ and\ \bibinfo {author}
  {\bibfnamefont{R.}~\bibnamefont{Rapp}},\ }%
  \bibfield{journal}{%
  \Doi{10.1103/PhysRevC.73.034913}{\bibinfo {journal} {Phys. Rev.}}\ }%
  \textbf{\bibinfo {volume} {C73}},\ \bibinfo {pages} {034913} (\bibinfo {year}
  {2006}),\
  \Eprint{http://arxiv.org/abs/nucl-th/0508055}{arXiv:nucl-th/0508055}%
  \bibAnnoteFile{NoStop}{vanHees:2005wb}%
%%CITATION = NUCL-TH/0508055;%%
\bibitem{Moore:2004tg}%
  \BibitemOpen
  \bibfield{author}{%
  \bibinfo {author} {\bibfnamefont{G.~D.}\ \bibnamefont{Moore}}\ and\ \bibinfo
  {author} {\bibfnamefont{D.}~\bibnamefont{Teaney}},\ }%
  \bibfield{journal}{%
  \Doi{10.1103/PhysRevC.71.064904}{\bibinfo {journal} {Phys. Rev.}}\ }%
  \textbf{\bibinfo {volume} {C71}},\ \bibinfo {pages} {064904} (\bibinfo {year}
  {2005}),\ \Eprint{http://arxiv.org/abs/hep-ph/0412346}{arXiv:hep-ph/0412346}%
  \bibAnnoteFile{NoStop}{Moore:2004tg}%
%%CITATION = HEP-PH/0412346;%%
\bibitem{Wicks:2005gt}%
  \BibitemOpen
  \bibfield{author}{%
  \bibinfo {author} {\bibfnamefont{S.}~\bibnamefont{Wicks}}, \bibinfo {author}
  {\bibfnamefont{W.}~\bibnamefont{Horowitz}}, \bibinfo {author}
  {\bibfnamefont{M.}~\bibnamefont{Djordjevic}},\ and\ \bibinfo {author}
  {\bibfnamefont{M.}~\bibnamefont{Gyulassy}},\ }%
  \bibfield{journal}{%
  \Doi{10.1016/j.nuclphysa.2006.12.048}{\bibinfo {journal} {Nucl. Phys.}}\ }%
  \textbf{\bibinfo {volume} {A784}},\ \bibinfo {pages} {426} (\bibinfo {year}
  {2007}),\
  \Eprint{http://arxiv.org/abs/nucl-th/0512076}{arXiv:nucl-th/0512076}%
  \bibAnnoteFile{NoStop}{Wicks:2005gt}%
%%CITATION = NUCL-TH/0512076;%%
\bibitem{Peigne:2008nd}%
  \BibitemOpen
  \bibfield{author}{%
  \bibinfo {author} {\bibfnamefont{S.}~\bibnamefont{Peigne}}\ and\ \bibinfo
  {author} {\bibfnamefont{A.}~\bibnamefont{Peshier}},\ }%
  \bibfield{journal}{%
  \Doi{10.1103/PhysRevD.77.114017}{\bibinfo {journal} {Phys. Rev.}}\ }%
  \textbf{\bibinfo {volume} {D77}},\ \bibinfo {pages} {114017} (\bibinfo {year}
  {2008}),\ \Eprint{http://arxiv.org/abs/0802.4364}{arXiv:0802.4364 [hep-ph]}%
  \bibAnnoteFile{NoStop}{Peigne:2008nd}%
%%CITATION = 0802.4364;%%
\bibitem{Gossiaux:2008jv}%
  \BibitemOpen
  \bibfield{author}{%
  \bibinfo {author} {\bibfnamefont{P.~B.}\ \bibnamefont{Gossiaux}}\ and\
  \bibinfo {author} {\bibfnamefont{J.}~\bibnamefont{Aichelin}},\ }%
  \bibfield{journal}{%
  \Doi{10.1103/PhysRevC.78.014904}{\bibinfo {journal} {Phys. Rev.}}\ }%
  \textbf{\bibinfo {volume} {C78}},\ \bibinfo {pages} {014904} (\bibinfo {year}
  {2008}),\ \Eprint{http://arxiv.org/abs/0802.2525}{arXiv:0802.2525 [hep-ph]}%
  \bibAnnoteFile{NoStop}{Gossiaux:2008jv}%
%%CITATION = 0802.2525;%%
\bibitem{Matsui:1986dk}%
  \BibitemOpen
  \bibfield{author}{%
  \bibinfo {author} {\bibfnamefont{T.}~\bibnamefont{Matsui}}\ and\ \bibinfo
  {author} {\bibfnamefont{H.}~\bibnamefont{Satz}},\ }%
  \bibfield{journal}{%
  \Doi{10.1016/0370-2693(86)91404-8}{\bibinfo {journal} {Phys. Lett.}}\ }%
  \textbf{\bibinfo {volume} {B178}},\ \bibinfo {pages} {416} (\bibinfo {year}
  {1986})%
  \bibAnnoteFile{NoStop}{Matsui:1986dk}%
%%CITATION = PHLTA,B178,416;%%
\bibitem{Andronic:2006ky}%
  \BibitemOpen
  \bibfield{author}{%
  \bibinfo {author} {\bibfnamefont{A.}~\bibnamefont{Andronic}}, \bibinfo
  {author} {\bibfnamefont{P.}~\bibnamefont{Braun-Munzinger}}, \bibinfo {author}
  {\bibfnamefont{K.}~\bibnamefont{Redlich}},\ and\ \bibinfo {author}
  {\bibfnamefont{J.}~\bibnamefont{Stachel}},\ }%
  \bibfield{journal}{%
  \Doi{10.1016/j.nuclphysa.2007.02.013}{\bibinfo {journal} {Nucl. Phys.}}\ }%
  \textbf{\bibinfo {volume} {A789}},\ \bibinfo {pages} {334} (\bibinfo {year}
  {2007}),\
  \Eprint{http://arxiv.org/abs/nucl-th/0611023}{arXiv:nucl-th/0611023}%
  \bibAnnoteFile{NoStop}{Andronic:2006ky}%
%%CITATION = NUCL-TH/0611023;%%
\bibitem{Sjostrand:2006za}%
  \BibitemOpen
  \bibfield{author}{%
  \bibinfo {author} {\bibfnamefont{T.}~\bibnamefont{Sjostrand}}, \bibinfo
  {author} {\bibfnamefont{S.}~\bibnamefont{Mrenna}},\ and\ \bibinfo {author}
  {\bibfnamefont{P.}~\bibnamefont{Skands}},\ }%
  \bibfield{journal}{%
  \bibinfo {journal} {JHEP}\ }%
  \textbf{\bibinfo {volume} {05}},\ \bibinfo {pages} {026} (\bibinfo {year}
  {2006}),\ \Eprint{http://arxiv.org/abs/hep-ph/0603175}{arXiv:hep-ph/0603175}%
  \bibAnnoteFile{NoStop}{Sjostrand:2006za}%
%%CITATION = HEP-PH/0603175;%%
\bibitem{Kajantie:1987pd}%
  \BibitemOpen
  \bibfield{author}{%
  \bibinfo {author} {\bibfnamefont{K.}~\bibnamefont{Kajantie}}, \bibinfo
  {author} {\bibfnamefont{P.~V.}\ \bibnamefont{Landshoff}},\ and\ \bibinfo
  {author} {\bibfnamefont{J.}~\bibnamefont{Lindfors}},\ }%
  \bibfield{journal}{%
  \Doi{10.1103/PhysRevLett.59.2527}{\bibinfo {journal} {Phys. Rev. Lett.}}\ }%
  \textbf{\bibinfo {volume} {59}},\ \bibinfo {pages} {2527} (\bibinfo {year}
  {1987})%
  \bibAnnoteFile{NoStop}{Kajantie:1987pd}%
%%CITATION = PRLTA,59,2527;%%
\bibitem{Eskola:1988yh}%
  \BibitemOpen
  \bibfield{author}{%
  \bibinfo {author} {\bibfnamefont{K.~J.}\ \bibnamefont{Eskola}}, \bibinfo
  {author} {\bibfnamefont{K.}~\bibnamefont{Kajantie}},\ and\ \bibinfo {author}
  {\bibfnamefont{J.}~\bibnamefont{Lindfors}},\ }%
  \bibfield{journal}{%
  \Doi{10.1016/0550-3213(89)90586-5}{\bibinfo {journal} {Nucl. Phys.}}\ }%
  \textbf{\bibinfo {volume} {B323}},\ \bibinfo {pages} {37} (\bibinfo {year}
  {1989})%
  \bibAnnoteFile{NoStop}{Eskola:1988yh}%
%%CITATION = NUPHA,B323,37;%%
\bibitem{Iancu:2003xm}%
  \BibitemOpen
  \bibfield{author}{%
  \bibinfo {author} {\bibfnamefont{E.}~\bibnamefont{Iancu}}\ and\ \bibinfo
  {author} {\bibfnamefont{R.}~\bibnamefont{Venugopalan}}}%
   (\bibinfo {year} {2003}),\
  \Eprint{http://arxiv.org/abs/hep-ph/0303204}{arXiv:hep-ph/0303204}%
  \bibAnnoteFile{NoStop}{Iancu:2003xm}%
%%CITATION = HEP-PH/0303204;%%
\bibitem{Drescher:2006ca}%
  \BibitemOpen
  \bibfield{author}{%
  \bibinfo {author} {\bibfnamefont{H.~J.}\ \bibnamefont{Drescher}}\ and\
  \bibinfo {author} {\bibfnamefont{Y.}~\bibnamefont{Nara}},\ }%
  \bibfield{journal}{%
  \Doi{10.1103/PhysRevC.75.034905}{\bibinfo {journal} {Phys. Rev.}}\ }%
  \textbf{\bibinfo {volume} {C75}},\ \bibinfo {pages} {034905} (\bibinfo {year}
  {2007}),\
  \Eprint{http://arxiv.org/abs/nucl-th/0611017}{arXiv:nucl-th/0611017}%
  \bibAnnoteFile{NoStop}{Drescher:2006ca}%
%%CITATION = NUCL-TH/0611017;%%
\bibitem{Xu:2004mz}%
  \BibitemOpen
  \bibfield{author}{%
  \bibinfo {author} {\bibfnamefont{Z.}~\bibnamefont{Xu}}\ and\ \bibinfo
  {author} {\bibfnamefont{C.}~\bibnamefont{Greiner}},\ }%
  \bibfield{journal}{%
  \Doi{10.1103/PhysRevC.71.064901}{\bibinfo {journal} {Phys. Rev.}}\ }%
  \textbf{\bibinfo {volume} {C71}},\ \bibinfo {pages} {064901} (\bibinfo {year}
  {2005}),\ \Eprint{http://arxiv.org/abs/hep-ph/0406278}{arXiv:hep-ph/0406278}%
  \bibAnnoteFile{NoStop}{Xu:2004mz}%
%%CITATION = HEP-PH/0406278;%%
\bibitem{Xu:2007aa}%
  \BibitemOpen
  \bibfield{author}{%
  \bibinfo {author} {\bibfnamefont{Z.}~\bibnamefont{Xu}}\ and\ \bibinfo
  {author} {\bibfnamefont{C.}~\bibnamefont{Greiner}},\ }%
  \bibfield{journal}{%
  \Doi{10.1103/PhysRevC.76.024911}{\bibinfo {journal} {Phys. Rev.}}\ }%
  \textbf{\bibinfo {volume} {C76}},\ \bibinfo {pages} {024911} (\bibinfo {year}
  {2007}),\ \Eprint{http://arxiv.org/abs/hep-ph/0703233}{arXiv:hep-ph/0703233}%
  \bibAnnoteFile{NoStop}{Xu:2007aa}%
%%CITATION = HEP-PH/0703233;%%
\bibitem{Cleymans:1992je}%
  \BibitemOpen
  \bibfield{author}{%
  \bibinfo {author} {\bibfnamefont{J.}~\bibnamefont{Cleymans}}, \bibinfo
  {author} {\bibfnamefont{V.~V.}\ \bibnamefont{Goloviznin}},\ and\ \bibinfo
  {author} {\bibfnamefont{K.}~\bibnamefont{Redlich}},\ }%
  \bibfield{journal}{%
  \Doi{10.1103/PhysRevD.47.989}{\bibinfo {journal} {Phys. Rev.}}\ }%
  \textbf{\bibinfo {volume} {D47}},\ \bibinfo {pages} {989} (\bibinfo {year}
  {1993})%
  \bibAnnoteFile{NoStop}{Cleymans:1992je}%
%%CITATION = PHRVA,D47,989;%%
\bibitem{Levai:1994dx}%
  \BibitemOpen
  \bibfield{author}{%
  \bibinfo {author} {\bibfnamefont{P.}~\bibnamefont{Levai}}, \bibinfo {author}
  {\bibfnamefont{B.}~\bibnamefont{M\"uller}},\ and\ \bibinfo {author}
  {\bibfnamefont{X.-N.}\ \bibnamefont{Wang}},\ }%
  \bibfield{journal}{%
  \Doi{10.1103/PhysRevC.51.3326}{\bibinfo {journal} {Phys. Rev.}}\ }%
  \textbf{\bibinfo {volume} {C51}},\ \bibinfo {pages} {3326} (\bibinfo {year}
  {1995}),\ \Eprint{http://arxiv.org/abs/hep-ph/9412352}{arXiv:hep-ph/9412352}%
  \bibAnnoteFile{NoStop}{Levai:1994dx}%
%%CITATION = HEP-PH/9412352;%%
\bibitem{Combridge:1978kx}%
  \BibitemOpen
  \bibfield{author}{%
  \bibinfo {author} {\bibfnamefont{B.~L.}\ \bibnamefont{Combridge}},\ }%
  \bibfield{journal}{%
  \Doi{10.1016/0550-3213(79)90449-8}{\bibinfo {journal} {Nucl. Phys.}}\ }%
  \textbf{\bibinfo {volume} {B151}},\ \bibinfo {pages} {429} (\bibinfo {year}
  {1979})%
  \bibAnnoteFile{NoStop}{Combridge:1978kx}%
%%CITATION = NUPHA,B151,429;%%
\bibitem{Muller:1992xn}%
  \BibitemOpen
  \bibfield{author}{%
  \bibinfo {author} {\bibfnamefont{B.}~\bibnamefont{M\"uller}}\ and\ \bibinfo
  {author} {\bibfnamefont{X.-N.}\ \bibnamefont{Wang}},\ }%
  \bibfield{journal}{%
  \Doi{10.1103/PhysRevLett.68.2437}{\bibinfo {journal} {Phys. Rev. Lett.}}\ }%
  \textbf{\bibinfo {volume} {68}},\ \bibinfo {pages} {2437} (\bibinfo {year}
  {1992})%
  \bibAnnoteFile{NoStop}{Muller:1992xn}%
%%CITATION = PRLTA,68,2437;%%
\bibitem{Gavai:1994gb}%
  \BibitemOpen
  \bibfield{author}{%
  \bibinfo {author} {\bibfnamefont{R.~V.}\ \bibnamefont{Gavai}} \emph{et~al.},\
  }%
  \bibfield{journal}{%
  \bibinfo {journal} {Int. J. Mod. Phys.}\ }%
  \textbf{\bibinfo {volume} {A10}},\ \bibinfo {pages} {2999} (\bibinfo {year}
  {1995}),\ \Eprint{http://arxiv.org/abs/hep-ph/9411438}{arXiv:hep-ph/9411438}%
  \bibAnnoteFile{NoStop}{Gavai:1994gb}%
%%CITATION = HEP-PH/9411438;%%
\bibitem{Levai:1997bi}%
  \BibitemOpen
  \bibfield{author}{%
  \bibinfo {author} {\bibfnamefont{P.}~\bibnamefont{Levai}}\ and\ \bibinfo
  {author} {\bibfnamefont{R.}~\bibnamefont{Vogt}},\ }%
  \bibfield{journal}{%
  \Doi{10.1103/PhysRevC.56.2707}{\bibinfo {journal} {Phys. Rev.}}\ }%
  \textbf{\bibinfo {volume} {C56}},\ \bibinfo {pages} {2707} (\bibinfo {year}
  {1997}),\ \Eprint{http://arxiv.org/abs/hep-ph/9704360}{arXiv:hep-ph/9704360}%
  \bibAnnoteFile{NoStop}{Levai:1997bi}%
%%CITATION = HEP-PH/9704360;%%
\bibitem{Cacciari:2005rk}%
  \BibitemOpen
  \bibfield{author}{%
  \bibinfo {author} {\bibfnamefont{M.}~\bibnamefont{Cacciari}}, \bibinfo
  {author} {\bibfnamefont{P.}~\bibnamefont{Nason}},\ and\ \bibinfo {author}
  {\bibfnamefont{R.}~\bibnamefont{Vogt}},\ }%
  \bibfield{journal}{%
  \Doi{10.1103/PhysRevLett.95.122001}{\bibinfo {journal} {Phys. Rev. Lett.}}\
  }%
  \textbf{\bibinfo {volume} {95}},\ \bibinfo {pages} {122001} (\bibinfo {year}
  {2005}),\ \Eprint{http://arxiv.org/abs/hep-ph/0502203}{arXiv:hep-ph/0502203}%
  \bibAnnoteFile{NoStop}{Cacciari:2005rk}%
%%CITATION = HEP-PH/0502203;%%
\bibitem{Nason:1999ta}%
  \BibitemOpen
  \bibfield{author}{%
  \bibinfo {author} {\bibfnamefont{P.}~\bibnamefont{Nason}} \emph{et~al.}}%
   (\bibinfo {year} {1999}),\
  \Eprint{http://arxiv.org/abs/hep-ph/0003142}{arXiv:hep-ph/0003142}%
  \bibAnnoteFile{NoStop}{Nason:1999ta}%
%%CITATION = HEP-PH/0003142;%%
\bibitem{Frixione:1997ma}%
  \BibitemOpen
  \bibfield{author}{%
  \bibinfo {author} {\bibfnamefont{S.}~\bibnamefont{Frixione}}, \bibinfo
  {author} {\bibfnamefont{M.~L.}\ \bibnamefont{Mangano}}, \bibinfo {author}
  {\bibfnamefont{P.}~\bibnamefont{Nason}},\ and\ \bibinfo {author}
  {\bibfnamefont{G.}~\bibnamefont{Ridolfi}},\ }%
  \bibfield{journal}{%
  \bibinfo {journal} {Adv. Ser. Direct. High Energy Phys.}\ }%
  \textbf{\bibinfo {volume} {15}},\ \bibinfo {pages} {609} (\bibinfo {year}
  {1998}),\ \Eprint{http://arxiv.org/abs/hep-ph/9702287}{arXiv:hep-ph/9702287}%
  \bibAnnoteFile{NoStop}{Frixione:1997ma}%
%%CITATION = HEP-PH/9702287;%%
\bibitem{Vogt:2001nh}%
  \BibitemOpen
  \bibfield{author}{%
  \bibinfo {author} {\bibfnamefont{R.}~\bibnamefont{Vogt}} (\bibinfo
  {collaboration} {Hard Probe}),\ }%
  \bibfield{journal}{%
  \Doi{10.1142/S0218301303001272}{\bibinfo {journal} {Int. J. Mod. Phys.}}\ }%
  \textbf{\bibinfo {volume} {E12}},\ \bibinfo {pages} {211} (\bibinfo {year}
  {2003}),\ \Eprint{http://arxiv.org/abs/hep-ph/0111271}{arXiv:hep-ph/0111271}%
  \bibAnnoteFile{NoStop}{Vogt:2001nh}%
%%CITATION = HEP-PH/0111271;%%
\bibitem{Gluck:1977zm}%
  \BibitemOpen
  \bibfield{author}{%
  \bibinfo {author} {\bibfnamefont{M.}~\bibnamefont{Gl\"uck}}, \bibinfo
  {author} {\bibfnamefont{J.~F.}\ \bibnamefont{Owens}},\ and\ \bibinfo {author}
  {\bibfnamefont{E.}~\bibnamefont{Reya}},\ }%
  \bibfield{journal}{%
  \Doi{10.1103/PhysRevD.17.2324}{\bibinfo {journal} {Phys. Rev.}}\ }%
  \textbf{\bibinfo {volume} {D17}},\ \bibinfo {pages} {2324} (\bibinfo {year}
  {1978})%
  \bibAnnoteFile{NoStop}{Gluck:1977zm}%
%%CITATION = PHRVA,D17,2324;%%
\bibitem{Babcock:1977fi}%
  \BibitemOpen
  \bibfield{author}{%
  \bibinfo {author} {\bibfnamefont{J.}~\bibnamefont{Babcock}}, \bibinfo
  {author} {\bibfnamefont{D.~W.}\ \bibnamefont{Sivers}},\ and\ \bibinfo
  {author} {\bibfnamefont{S.}~\bibnamefont{Wolfram}},\ }%
  \bibfield{journal}{%
  \Doi{10.1103/PhysRevD.18.162}{\bibinfo {journal} {Phys. Rev.}}\ }%
  \textbf{\bibinfo {volume} {D18}},\ \bibinfo {pages} {162} (\bibinfo {year}
  {1978})%
  \bibAnnoteFile{NoStop}{Babcock:1977fi}%
%%CITATION = PHRVA,D18,162;%%
\bibitem{Barger:1981rx}%
  \BibitemOpen
  \bibfield{author}{%
  \bibinfo {author} {\bibfnamefont{V.~D.}\ \bibnamefont{Barger}}, \bibinfo
  {author} {\bibfnamefont{F.}~\bibnamefont{Halzen}},\ and\ \bibinfo {author}
  {\bibfnamefont{W.~Y.}\ \bibnamefont{Keung}},\ }%
  \bibfield{journal}{%
  \Doi{10.1103/PhysRevD.25.112}{\bibinfo {journal} {Phys. Rev.}}\ }%
  \textbf{\bibinfo {volume} {D25}},\ \bibinfo {pages} {112} (\bibinfo {year}
  {1982})%
  \bibAnnoteFile{NoStop}{Barger:1981rx}%
%%CITATION = PHRVA,D25,112;%%
\bibitem{Matsui:1985eu}%
  \BibitemOpen
  \bibfield{author}{%
  \bibinfo {author} {\bibfnamefont{T.}~\bibnamefont{Matsui}}, \bibinfo {author}
  {\bibfnamefont{B.}~\bibnamefont{Svetitsky}},\ and\ \bibinfo {author}
  {\bibfnamefont{L.~D.}\ \bibnamefont{McLerran}},\ }%
  \bibfield{journal}{%
  \Doi{10.1103/PhysRevD.34.783}{\bibinfo {journal} {Phys. Rev.}}\ }%
  \textbf{\bibinfo {volume} {D34}},\ \bibinfo {pages} {783} (\bibinfo {year}
  {1986})%
  \bibAnnoteFile{NoStop}{Matsui:1985eu}%
%%CITATION = PHRVA,D34,783;%%
\bibitem{Biro:1993qt}%
  \BibitemOpen
  \bibfield{author}{%
  \bibinfo {author} {\bibfnamefont{T.~S.}\ \bibnamefont{Biro}}, \bibinfo
  {author} {\bibfnamefont{E.}~\bibnamefont{van Doorn}}, \bibinfo {author}
  {\bibfnamefont{B.}~\bibnamefont{M\"uller}}, \bibinfo {author}
  {\bibfnamefont{M.~H.}\ \bibnamefont{Thoma}},\ and\ \bibinfo {author}
  {\bibfnamefont{X.~N.}\ \bibnamefont{Wang}},\ }%
  \bibfield{journal}{%
  \Doi{10.1103/PhysRevC.48.1275}{\bibinfo {journal} {Phys. Rev.}}\ }%
  \textbf{\bibinfo {volume} {C48}},\ \bibinfo {pages} {1275} (\bibinfo {year}
  {1993}),\
  \Eprint{http://arxiv.org/abs/nucl-th/9303004}{arXiv:nucl-th/9303004}%
  \bibAnnoteFile{NoStop}{Biro:1993qt}%
%%CITATION = NUCL-TH/9303004;%%
\bibitem{Zhang:2008zzc}%
  \BibitemOpen
  \bibfield{author}{%
  \bibinfo {author} {\bibfnamefont{B.-W.}\ \bibnamefont{Zhang}}, \bibinfo
  {author} {\bibfnamefont{C.-M.}\ \bibnamefont{Ko}},\ and\ \bibinfo {author}
  {\bibfnamefont{W.}~\bibnamefont{Liu}},\ }%
  \bibfield{journal}{%
  \Doi{10.1103/PhysRevC.77.024901}{\bibinfo {journal} {Phys. Rev.}}\ }%
  \textbf{\bibinfo {volume} {C77}},\ \bibinfo {pages} {024901} (\bibinfo {year}
  {2008}),\ \Eprint{http://arxiv.org/abs/0709.1684}{arXiv:0709.1684 [nucl-th]}%
  \bibAnnoteFile{NoStop}{Zhang:2008zzc}%
%%CITATION = 0709.1684;%%
\bibitem{Thews:2000rj}%
  \BibitemOpen
  \bibfield{author}{%
  \bibinfo {author} {\bibfnamefont{R.~L.}\ \bibnamefont{Thews}}, \bibinfo
  {author} {\bibfnamefont{M.}~\bibnamefont{Schroedter}},\ and\ \bibinfo
  {author} {\bibfnamefont{J.}~\bibnamefont{Rafelski}},\ }%
  \bibfield{journal}{%
  \Doi{10.1103/PhysRevC.63.054905}{\bibinfo {journal} {Phys. Rev.}}\ }%
  \textbf{\bibinfo {volume} {C63}},\ \bibinfo {pages} {054905} (\bibinfo {year}
  {2001}),\ \Eprint{http://arxiv.org/abs/hep-ph/0007323}{arXiv:hep-ph/0007323}%
  \bibAnnoteFile{NoStop}{Thews:2000rj}%
%%CITATION = HEP-PH/0007323;%%
\bibitem{Soff:2000eh}%
  \BibitemOpen
  \bibfield{author}{%
  \bibinfo {author} {\bibfnamefont{S.}~\bibnamefont{Soff}}, \bibinfo {author}
  {\bibfnamefont{S.~A.}\ \bibnamefont{Bass}},\ and\ \bibinfo {author}
  {\bibfnamefont{A.}~\bibnamefont{Dumitru}},\ }%
  \bibfield{journal}{%
  \Doi{10.1103/PhysRevLett.86.3981}{\bibinfo {journal} {Phys. Rev. Lett.}}\ }%
  \textbf{\bibinfo {volume} {86}},\ \bibinfo {pages} {3981} (\bibinfo {year}
  {2001}),\
  \Eprint{http://arxiv.org/abs/nucl-th/0012085}{arXiv:nucl-th/0012085}%
  \bibAnnoteFile{NoStop}{Soff:2000eh}%
%%CITATION = NUCL-TH/0012085;%%
\bibitem{Fries:2002kt}%
  \BibitemOpen
  \bibfield{author}{%
  \bibinfo {author} {\bibfnamefont{R.~J.}\ \bibnamefont{Fries}}, \bibinfo
  {author} {\bibfnamefont{B.}~\bibnamefont{M\"uller}},\ and\ \bibinfo {author}
  {\bibfnamefont{D.~K.}\ \bibnamefont{Srivastava}},\ }%
  \bibfield{journal}{%
  \Doi{10.1103/PhysRevLett.90.132301}{\bibinfo {journal} {Phys. Rev. Lett.}}\
  }%
  \textbf{\bibinfo {volume} {90}},\ \bibinfo {pages} {132301} (\bibinfo {year}
  {2003}),\
  \Eprint{http://arxiv.org/abs/nucl-th/0208001}{arXiv:nucl-th/0208001}%
  \bibAnnoteFile{NoStop}{Fries:2002kt}%
%%CITATION = NUCL-TH/0208001;%%
\bibitem{Turbide:2005fk}%
  \BibitemOpen
  \bibfield{author}{%
  \bibinfo {author} {\bibfnamefont{S.}~\bibnamefont{Turbide}}, \bibinfo
  {author} {\bibfnamefont{C.}~\bibnamefont{Gale}}, \bibinfo {author}
  {\bibfnamefont{S.}~\bibnamefont{Jeon}},\ and\ \bibinfo {author}
  {\bibfnamefont{G.~D.}\ \bibnamefont{Moore}},\ }%
  \bibfield{journal}{%
  \Doi{10.1103/PhysRevC.72.014906}{\bibinfo {journal} {Phys. Rev.}}\ }%
  \textbf{\bibinfo {volume} {C72}},\ \bibinfo {pages} {014906} (\bibinfo {year}
  {2005}),\ \Eprint{http://arxiv.org/abs/hep-ph/0502248}{arXiv:hep-ph/0502248}%
  \bibAnnoteFile{NoStop}{Turbide:2005fk}%
%%CITATION = HEP-PH/0502248;%%
\bibitem{Rapp:2000pe}%
  \BibitemOpen
  \bibfield{author}{%
  \bibinfo {author} {\bibfnamefont{R.}~\bibnamefont{Rapp}},\ }%
  \bibfield{journal}{%
  \Doi{10.1103/PhysRevC.63.054907}{\bibinfo {journal} {Phys. Rev.}}\ }%
  \textbf{\bibinfo {volume} {C63}},\ \bibinfo {pages} {054907} (\bibinfo {year}
  {2001}),\ \Eprint{http://arxiv.org/abs/hep-ph/0010101}{arXiv:hep-ph/0010101}%
  \bibAnnoteFile{NoStop}{Rapp:2000pe}%
%%CITATION = HEP-PH/0010101;%%
\bibitem{Cooper:2002td}%
  \BibitemOpen
  \bibfield{author}{%
  \bibinfo {author} {\bibfnamefont{F.}~\bibnamefont{Cooper}}, \bibinfo {author}
  {\bibfnamefont{E.}~\bibnamefont{Mottola}},\ and\ \bibinfo {author}
  {\bibfnamefont{G.~C.}\ \bibnamefont{Nayak}},\ }%
  \bibfield{journal}{%
  \Doi{10.1016/S0370-2693(03)00080-7}{\bibinfo {journal} {Phys. Lett.}}\ }%
  \textbf{\bibinfo {volume} {B555}},\ \bibinfo {pages} {181} (\bibinfo {year}
  {2003}),\ \Eprint{http://arxiv.org/abs/hep-ph/0210391}{arXiv:hep-ph/0210391}%
  \bibAnnoteFile{NoStop}{Cooper:2002td}%
%%CITATION = HEP-PH/0210391;%%
\bibitem{Wang:1996yf}%
  \BibitemOpen
  \bibfield{author}{%
  \bibinfo {author} {\bibfnamefont{X.-N.}\ \bibnamefont{Wang}},\ }%
  \bibfield{journal}{%
  \Doi{10.1016/S0370-1573(96)00022-1}{\bibinfo {journal} {Phys. Rept.}}\ }%
  \textbf{\bibinfo {volume} {280}},\ \bibinfo {pages} {287} (\bibinfo {year}
  {1997}),\ \Eprint{http://arxiv.org/abs/hep-ph/9605214}{arXiv:hep-ph/9605214}%
  \bibAnnoteFile{NoStop}{Wang:1996yf}%
%%CITATION = HEP-PH/9605214;%%
\bibitem{wong}%
  \BibitemOpen
  \bibfield{author}{%
  \bibinfo {author} {\bibfnamefont{C.-Y.}\ \bibnamefont{Wong}},\ }%
  \emph{\bibinfo {title} {Introduction to High-Energy Heavy-Ion Collisions}}\
  (\bibinfo {publisher} {{World Scientific Publishing Co. Pte. Ltd.}},\
  \bibinfo {year} {1994})%
  \bibAnnoteFile{NoStop}{wong}%
\bibitem{Sarcevic:1994ma}%
  \BibitemOpen
  \bibfield{author}{%
  \bibinfo {author} {\bibfnamefont{I.}~\bibnamefont{Sarcevic}}\ and\ \bibinfo
  {author} {\bibfnamefont{P.}~\bibnamefont{Valerio}},\ }%
  \bibfield{journal}{%
  \Doi{10.1103/PhysRevC.51.1433}{\bibinfo {journal} {Phys. Rev.}}\ }%
  \textbf{\bibinfo {volume} {C51}},\ \bibinfo {pages} {1433} (\bibinfo {year}
  {1995}),\ \Eprint{http://arxiv.org/abs/hep-ph/9411317}{arXiv:hep-ph/9411317}%
  \bibAnnoteFile{NoStop}{Sarcevic:1994ma}%
%%CITATION = HEP-PH/9411317;%%
\bibitem{misko_overlap}%
  \BibitemOpen
  \bibfield{author}{%
  \bibinfo {author} {\bibfnamefont{D.}~\bibnamefont{Miskowiec}}\ and\ \bibinfo
  {author} {\bibfnamefont{J.}~\bibnamefont{Elgeti}},\ }%
  \enquote{\bibinfo {title} {Nuclear overlap calculation},}\ \bibinfo
  {howpublished} {\url{http://www-linux.gsi.de/~misko/overlap/}} (\bibinfo
  {year} {2001})%
  \bibAnnoteFile{NoStop}{misko_overlap}%
\bibitem{Adams:2004cb}%
  \BibitemOpen
  \bibfield{author}{%
  \bibinfo {author} {\bibfnamefont{J.}~\bibnamefont{Adams}} \emph{et~al.}
  (\bibinfo {collaboration} {STAR}),\ }%
  \bibfield{journal}{%
  \Doi{10.1103/PhysRevC.70.054907}{\bibinfo {journal} {Phys. Rev.}}\ }%
  \textbf{\bibinfo {volume} {C70}},\ \bibinfo {pages} {054907} (\bibinfo {year}
  {2004}),\
  \Eprint{http://arxiv.org/abs/nucl-ex/0407003}{arXiv:nucl-ex/0407003}%
  \bibAnnoteFile{NoStop}{Adams:2004cb}%
%%CITATION = NUCL-EX/0407003;%%
\bibitem{Emel'yanov:1999bn}%
  \BibitemOpen
  \bibfield{author}{%
  \bibinfo {author} {\bibfnamefont{V.}~\bibnamefont{Emel'yanov}}, \bibinfo
  {author} {\bibfnamefont{A.}~\bibnamefont{Khodinov}}, \bibinfo {author}
  {\bibfnamefont{S.~R.}\ \bibnamefont{Klein}},\ and\ \bibinfo {author}
  {\bibfnamefont{R.}~\bibnamefont{Vogt}},\ }%
  \bibfield{journal}{%
  \Doi{10.1103/PhysRevC.61.044904}{\bibinfo {journal} {Phys. Rev.}}\ }%
  \textbf{\bibinfo {volume} {C61}},\ \bibinfo {pages} {044904} (\bibinfo {year}
  {2000}),\ \Eprint{http://arxiv.org/abs/hep-ph/9909427}{arXiv:hep-ph/9909427}%
  \bibAnnoteFile{NoStop}{Emel'yanov:1999bn}%
%%CITATION = HEP-PH/9909427;%%
\bibitem{Accardi:2004be}%
  \BibitemOpen
  \bibfield{author}{%
  \bibinfo {author} {\bibfnamefont{A.}~\bibnamefont{Accardi}} \emph{et~al.}}%
   (\bibinfo {year} {2004}),\
  \Eprint{http://arxiv.org/abs/hep-ph/0308248}{arXiv:hep-ph/0308248}%
  \bibAnnoteFile{NoStop}{Accardi:2004be}%
%%CITATION = HEP-PH/0308248;%%
\bibitem{El:2007vg}%
  \BibitemOpen
  \bibfield{author}{%
  \bibinfo {author} {\bibfnamefont{A.}~\bibnamefont{El}}, \bibinfo {author}
  {\bibfnamefont{Z.}~\bibnamefont{Xu}},\ and\ \bibinfo {author}
  {\bibfnamefont{C.}~\bibnamefont{Greiner}},\ }%
  \bibfield{journal}{%
  \Doi{10.1016/j.nuclphysa.2008.03.005}{\bibinfo {journal} {Nucl. Phys.}}\ }%
  \textbf{\bibinfo {volume} {A806}},\ \bibinfo {pages} {287} (\bibinfo {year}
  {2008}),\ \Eprint{http://arxiv.org/abs/0712.3734}{arXiv:0712.3734 [hep-ph]}%
  \bibAnnoteFile{NoStop}{El:2007vg}%
%%CITATION = 0712.3734;%%
\bibitem{Xu:2008zi}%
  \BibitemOpen
  \bibfield{author}{%
  \bibinfo {author} {\bibfnamefont{Z.}~\bibnamefont{Xu}}, \bibinfo {author}
  {\bibfnamefont{L.}~\bibnamefont{Cheng}}, \bibinfo {author}
  {\bibfnamefont{A.}~\bibnamefont{El}}, \bibinfo {author}
  {\bibfnamefont{K.}~\bibnamefont{Gallmeister}},\ and\ \bibinfo {author}
  {\bibfnamefont{C.}~\bibnamefont{Greiner}},\ }%
  \bibfield{journal}{%
  \Doi{10.1088/0954-3899/36/6/064035}{\bibinfo {journal} {J. Phys.}}\ }%
  \textbf{\bibinfo {volume} {G36}},\ \bibinfo {pages} {064035} (\bibinfo {year}
  {2009}),\ \Eprint{http://arxiv.org/abs/0812.3839}{arXiv:0812.3839 [hep-ph]}%
  \bibAnnoteFile{NoStop}{Xu:2008zi}%
%%CITATION = 0812.3839;%%
\bibitem{Wang:1991hta}%
  \BibitemOpen
  \bibfield{author}{%
  \bibinfo {author} {\bibfnamefont{X.-N.}\ \bibnamefont{Wang}}\ and\ \bibinfo
  {author} {\bibfnamefont{M.}~\bibnamefont{Gyulassy}},\ }%
  \bibfield{journal}{%
  \Doi{10.1103/PhysRevD.44.3501}{\bibinfo {journal} {Phys. Rev.}}\ }%
  \textbf{\bibinfo {volume} {D44}},\ \bibinfo {pages} {3501} (\bibinfo {year}
  {1991})%
  \bibAnnoteFile{NoStop}{Wang:1991hta}%
%%CITATION = PHRVA,D44,3501;%%
\bibitem{Xu:2005wv}%
  \BibitemOpen
  \bibfield{author}{%
  \bibinfo {author} {\bibfnamefont{Z.}~\bibnamefont{Xu}}\ and\ \bibinfo
  {author} {\bibfnamefont{C.}~\bibnamefont{Greiner}},\ }%
  \bibfield{journal}{%
  \Doi{10.1140/epja/i2005-10294-8}{\bibinfo {journal} {Eur. Phys. J.}}\ }%
  \textbf{\bibinfo {volume} {A29}},\ \bibinfo {pages} {33} (\bibinfo {year}
  {2006}),\ \Eprint{http://arxiv.org/abs/hep-ph/0511145}{arXiv:hep-ph/0511145}%
  \bibAnnoteFile{NoStop}{Xu:2005wv}%
%%CITATION = HEP-PH/0511145;%%
\bibitem{Drescher:2006pi}%
  \BibitemOpen
  \bibfield{author}{%
  \bibinfo {author} {\bibfnamefont{A.}~\bibnamefont{Adil}}, \bibinfo {author}
  {\bibfnamefont{H.-J.}\ \bibnamefont{Drescher}}, \bibinfo {author}
  {\bibfnamefont{A.}~\bibnamefont{Dumitru}}, \bibinfo {author}
  {\bibfnamefont{A.}~\bibnamefont{Hayashigaki}},\ and\ \bibinfo {author}
  {\bibfnamefont{Y.}~\bibnamefont{Nara}},\ }%
  \bibfield{journal}{%
  \Doi{10.1103/PhysRevC.74.044905}{\bibinfo {journal} {Phys. Rev.}}\ }%
  \textbf{\bibinfo {volume} {C74}},\ \bibinfo {pages} {044905} (\bibinfo {year}
  {2006}),\
  \Eprint{http://arxiv.org/abs/nucl-th/0605012}{arXiv:nucl-th/0605012}%
  \bibAnnoteFile{NoStop}{Drescher:2006pi}%
%%CITATION = NUCL-TH/0605012;%%
\bibitem{Gribov:1984tu}%
  \BibitemOpen
  \bibfield{author}{%
  \bibinfo {author} {\bibfnamefont{L.~V.}\ \bibnamefont{Gribov}}, \bibinfo
  {author} {\bibfnamefont{E.~M.}\ \bibnamefont{Levin}},\ and\ \bibinfo {author}
  {\bibfnamefont{M.~G.}\ \bibnamefont{Ryskin}},\ }%
  \bibfield{journal}{%
  \Doi{10.1016/0370-1573(83)90022-4}{\bibinfo {journal} {Phys. Rept.}}\ }%
  \textbf{\bibinfo {volume} {100}},\ \bibinfo {pages} {1} (\bibinfo {year}
  {1983})%
  \bibAnnoteFile{NoStop}{Gribov:1984tu}%
%%CITATION = PRPLC,100,1;%%
\bibitem{Kharzeev:2000ph}%
  \BibitemOpen
  \bibfield{author}{%
  \bibinfo {author} {\bibfnamefont{D.}~\bibnamefont{Kharzeev}}\ and\ \bibinfo
  {author} {\bibfnamefont{M.}~\bibnamefont{Nardi}},\ }%
  \bibfield{journal}{%
  \Doi{10.1016/S0370-2693(01)00457-9}{\bibinfo {journal} {Phys. Lett.}}\ }%
  \textbf{\bibinfo {volume} {B507}},\ \bibinfo {pages} {121} (\bibinfo {year}
  {2001}),\
  \Eprint{http://arxiv.org/abs/nucl-th/0012025}{arXiv:nucl-th/0012025}%
  \bibAnnoteFile{NoStop}{Kharzeev:2000ph}%
%%CITATION = NUCL-TH/0012025;%%
\bibitem{Kharzeev:2002ei}%
  \BibitemOpen
  \bibfield{author}{%
  \bibinfo {author} {\bibfnamefont{D.}~\bibnamefont{Kharzeev}}, \bibinfo
  {author} {\bibfnamefont{E.}~\bibnamefont{Levin}},\ and\ \bibinfo {author}
  {\bibfnamefont{M.}~\bibnamefont{Nardi}},\ }%
  \bibfield{journal}{%
  \Doi{10.1016/j.nuclphysa.2004.06.022}{\bibinfo {journal} {Nucl. Phys.}}\ }%
  \textbf{\bibinfo {volume} {A730}},\ \bibinfo {pages} {448} (\bibinfo {year}
  {2004}),\ \Eprint{http://arxiv.org/abs/hep-ph/0212316}{arXiv:hep-ph/0212316}%
  \bibAnnoteFile{NoStop}{Kharzeev:2002ei}%
%%CITATION = HEP-PH/0212316;%%
\bibitem{Armesto:2008fj}%
  \BibitemOpen
  \bibfield{author}{%
  \bibinfo {author} {\bibfnamefont{N.}~\bibnamefont{Armesto}},\ }%
  \bibfield{journal}{%
  \Doi{10.1088/0954-3899/35/10/104042}{\bibinfo {journal} {J. Phys.}}\ }%
  \textbf{\bibinfo {volume} {G35}},\ \bibinfo {pages} {104042} (\bibinfo {year}
  {2008}),\ \Eprint{http://arxiv.org/abs/0804.4158}{arXiv:0804.4158 [hep-ph]}%
  \bibAnnoteFile{NoStop}{Armesto:2008fj}%
%%CITATION = 0804.4158;%%
\bibitem{Smith:1996sb}%
  \BibitemOpen
  \bibfield{author}{%
  \bibinfo {author} {\bibfnamefont{J.}~\bibnamefont{Smith}}\ and\ \bibinfo
  {author} {\bibfnamefont{R.}~\bibnamefont{Vogt}},\ }%
  \bibfield{journal}{%
  \Doi{10.1007/s002880050470}{\bibinfo {journal} {Z. Phys.}}\ }%
  \textbf{\bibinfo {volume} {C75}},\ \bibinfo {pages} {271} (\bibinfo {year}
  {1997}),\ \Eprint{http://arxiv.org/abs/hep-ph/9609388}{arXiv:hep-ph/9609388}%
  \bibAnnoteFile{NoStop}{Smith:1996sb}%
%%CITATION = HEP-PH/9609388;%%
\bibitem{Eskola:2003fk}%
  \BibitemOpen
  \bibfield{author}{%
  \bibinfo {author} {\bibfnamefont{K.~J.}\ \bibnamefont{Eskola}}, \bibinfo
  {author} {\bibfnamefont{V.~J.}\ \bibnamefont{Kolhinen}},\ and\ \bibinfo
  {author} {\bibfnamefont{R.}~\bibnamefont{Vogt}},\ }%
  \bibfield{journal}{%
  \Doi{10.1016/j.physletb.2003.11.077}{\bibinfo {journal} {Phys. Lett.}}\ }%
  \textbf{\bibinfo {volume} {B582}},\ \bibinfo {pages} {157} (\bibinfo {year}
  {2004}),\ \Eprint{http://arxiv.org/abs/hep-ph/0310111}{arXiv:hep-ph/0310111}%
  \bibAnnoteFile{NoStop}{Eskola:2003fk}%
%%CITATION = HEP-PH/0310111;%%
\bibitem{Vogt:2007aw}%
  \BibitemOpen
  \bibfield{author}{%
  \bibinfo {author} {\bibfnamefont{R.}~\bibnamefont{Vogt}},\ }%
  \bibfield{journal}{%
  \Doi{10.1140/epjst/e2008-00603-5}{\bibinfo {journal} {Eur. Phys. J. ST}}\ }%
  \textbf{\bibinfo {volume} {155}},\ \bibinfo {pages} {213} (\bibinfo {year}
  {2008}),\ \Eprint{http://arxiv.org/abs/0709.2531}{arXiv:0709.2531 [hep-ph]}%
  \bibAnnoteFile{NoStop}{Vogt:2007aw}%
%%CITATION = 0709.2531;%%
\bibitem{Whalley:2005nh}%
  \BibitemOpen
  \bibfield{author}{%
  \bibinfo {author} {\bibfnamefont{M.~R.}\ \bibnamefont{Whalley}}, \bibinfo
  {author} {\bibfnamefont{D.}~\bibnamefont{Bourilkov}},\ and\ \bibinfo {author}
  {\bibfnamefont{R.~C.}\ \bibnamefont{Group}}}%
   (\bibinfo {year} {2005}),\
  \Eprint{http://arxiv.org/abs/hep-ph/0508110}{arXiv:hep-ph/0508110}%
  \bibAnnoteFile{NoStop}{Whalley:2005nh}%
%%CITATION = HEP-PH/0508110;%%
\bibitem{:2008asa_PHENIX}%
  \BibitemOpen
  \bibfield{author}{%
  \bibinfo {author} {\bibfnamefont{A.}~\bibnamefont{Adare}} \emph{et~al.}
  (\bibinfo {collaboration} {PHENIX}),\ }%
  \bibfield{journal}{%
  \Doi{10.1016/j.physletb.2008.10.064}{\bibinfo {journal} {Phys. Lett.}}\ }%
  \textbf{\bibinfo {volume} {B670}},\ \bibinfo {pages} {313} (\bibinfo {year}
  {2009}),\ \Eprint{http://arxiv.org/abs/0802.0050}{arXiv:0802.0050 [hep-ex]}%
  \bibAnnoteFile{NoStop}{:2008asa_PHENIX}%
%%CITATION = 0802.0050;%%
\bibitem{Adare:2006hc_PHENIX_dsigmadY}%
  \BibitemOpen
  \bibfield{author}{%
  \bibinfo {author} {\bibfnamefont{A.}~\bibnamefont{Adare}} \emph{et~al.}
  (\bibinfo {collaboration} {PHENIX}),\ }%
  \bibfield{journal}{%
  \Doi{10.1103/PhysRevLett.97.252002}{\bibinfo {journal} {Phys. Rev. Lett.}}\
  }%
  \textbf{\bibinfo {volume} {97}},\ \bibinfo {pages} {252002} (\bibinfo {year}
  {2006}),\ \Eprint{http://arxiv.org/abs/hep-ex/0609010}{arXiv:hep-ex/0609010}%
  \bibAnnoteFile{NoStop}{Adare:2006hc_PHENIX_dsigmadY}%
%%CITATION = HEP-EX/0609010;%%
\bibitem{Adams:2004fc_STAR_dcsdY_cstot}%
  \BibitemOpen
  \bibfield{author}{%
  \bibinfo {author} {\bibfnamefont{J.}~\bibnamefont{Adams}} \emph{et~al.}
  (\bibinfo {collaboration} {STAR}),\ }%
  \bibfield{journal}{%
  \Doi{10.1103/PhysRevLett.94.062301}{\bibinfo {journal} {Phys. Rev. Lett.}}\
  }%
  \textbf{\bibinfo {volume} {94}},\ \bibinfo {pages} {062301} (\bibinfo {year}
  {2005}),\
  \Eprint{http://arxiv.org/abs/nucl-ex/0407006}{arXiv:nucl-ex/0407006}%
  \bibAnnoteFile{NoStop}{Adams:2004fc_STAR_dcsdY_cstot}%
%%CITATION = NUCL-EX/0407006;%%
\bibitem{Pop:2009sd}%
  \BibitemOpen
  \bibfield{author}{%
  \bibinfo {author} {\bibfnamefont{V.~T.}\ \bibnamefont{Pop}}, \bibinfo
  {author} {\bibfnamefont{J.}~\bibnamefont{Barrette}},\ and\ \bibinfo {author}
  {\bibfnamefont{M.}~\bibnamefont{Gyulassy}},\ }%
  \bibfield{journal}{%
  \Doi{10.1103/PhysRevLett.102.232302}{\bibinfo {journal} {Phys. Rev. Lett.}}\
  }%
  \textbf{\bibinfo {volume} {102}},\ \bibinfo {pages} {232302} (\bibinfo {year}
  {2009}),\ \Eprint{http://arxiv.org/abs/0902.4028}{arXiv:0902.4028 [hep-ph]}%
  \bibAnnoteFile{NoStop}{Pop:2009sd}%
%%CITATION = 0902.4028;%%
\bibitem{Lai:1999wy}%
  \BibitemOpen
  \bibfield{author}{%
  \bibinfo {author} {\bibfnamefont{H.~L.}\ \bibnamefont{Lai}} \emph{et~al.}
  (\bibinfo {collaboration} {CTEQ}),\ }%
  \bibfield{journal}{%
  \Doi{10.1007/s100529900196}{\bibinfo {journal} {Eur. Phys. J.}}\ }%
  \textbf{\bibinfo {volume} {C12}},\ \bibinfo {pages} {375} (\bibinfo {year}
  {2000}),\ \Eprint{http://arxiv.org/abs/hep-ph/9903282}{arXiv:hep-ph/9903282}%
  \bibAnnoteFile{NoStop}{Lai:1999wy}%
%%CITATION = HEP-PH/9903282;%%
\bibitem{Pumplin:2002vw_CTEQ6}%
  \BibitemOpen
  \bibfield{author}{%
  \bibinfo {author} {\bibfnamefont{J.}~\bibnamefont{Pumplin}} \emph{et~al.},\
  }%
  \bibfield{journal}{%
  \bibinfo {journal} {JHEP}\ }%
  \textbf{\bibinfo {volume} {07}},\ \bibinfo {pages} {012} (\bibinfo {year}
  {2002}),\ \Eprint{http://arxiv.org/abs/hep-ph/0201195}{arXiv:hep-ph/0201195}%
  \bibAnnoteFile{NoStop}{Pumplin:2002vw_CTEQ6}%
%%CITATION = HEP-PH/0201195;%%
\bibitem{Martin:2002dr}%
  \BibitemOpen
  \bibfield{author}{%
  \bibinfo {author} {\bibfnamefont{A.~D.}\ \bibnamefont{Martin}}, \bibinfo
  {author} {\bibfnamefont{R.~G.}\ \bibnamefont{Roberts}}, \bibinfo {author}
  {\bibfnamefont{W.~J.}\ \bibnamefont{Stirling}},\ and\ \bibinfo {author}
  {\bibfnamefont{R.~S.}\ \bibnamefont{Thorne}},\ }%
  \bibfield{journal}{%
  \Doi{10.1016/S0370-2693(02)01483-1}{\bibinfo {journal} {Phys. Lett.}}\ }%
  \textbf{\bibinfo {volume} {B531}},\ \bibinfo {pages} {216} (\bibinfo {year}
  {2002}),\ \Eprint{http://arxiv.org/abs/hep-ph/0201127}{arXiv:hep-ph/0201127}%
  \bibAnnoteFile{NoStop}{Martin:2002dr}%
%%CITATION = HEP-PH/0201127;%%
\bibitem{Sherstnev:2007nd}%
  \BibitemOpen
  \bibfield{author}{%
  \bibinfo {author} {\bibfnamefont{A.}~\bibnamefont{Sherstnev}}\ and\ \bibinfo
  {author} {\bibfnamefont{R.~S.}\ \bibnamefont{Thorne}},\ }%
  \bibfield{journal}{%
  \Doi{10.1140/epjc/s10052-008-0610-x}{\bibinfo {journal} {Eur. Phys. J.}}\ }%
  \textbf{\bibinfo {volume} {C55}},\ \bibinfo {pages} {553} (\bibinfo {year}
  {2008}),\ \Eprint{http://arxiv.org/abs/0711.2473}{arXiv:0711.2473 [hep-ph]}%
  \bibAnnoteFile{NoStop}{Sherstnev:2007nd}%
%%CITATION = 0711.2473;%%
\bibitem{Nagano:2008ip}%
  \BibitemOpen
  \bibfield{author}{%
  \bibinfo {author} {\bibfnamefont{K.}~\bibnamefont{Nagano}} (\bibinfo
  {collaboration} {H1})}%
   (\bibinfo {year} {2008}),\
  \Eprint{http://arxiv.org/abs/0808.3797}{arXiv:0808.3797 [hep-ex]}%
  \bibAnnoteFile{NoStop}{Nagano:2008ip}%
%%CITATION = 0808.3797;%%
\bibitem{Gluck:2007ck}%
  \BibitemOpen
  \bibfield{author}{%
  \bibinfo {author} {\bibfnamefont{M.}~\bibnamefont{Gl\"uck}}, \bibinfo
  {author} {\bibfnamefont{P.}~\bibnamefont{Jimenez-Delgado}},\ and\ \bibinfo
  {author} {\bibfnamefont{E.}~\bibnamefont{Reya}},\ }%
  \bibfield{journal}{%
  \Doi{10.1140/epjc/s10052-007-0462-9}{\bibinfo {journal} {Eur. Phys. J.}}\ }%
  \textbf{\bibinfo {volume} {C53}},\ \bibinfo {pages} {355} (\bibinfo {year}
  {2008}),\ \Eprint{http://arxiv.org/abs/0709.0614}{arXiv:0709.0614 [hep-ph]}%
  \bibAnnoteFile{NoStop}{Gluck:2007ck}%
%%CITATION = 0709.0614;%%
\bibitem{Gluck:2008gs}%
  \BibitemOpen
  \bibfield{author}{%
  \bibinfo {author} {\bibfnamefont{M.}~\bibnamefont{Gl\"uck}}, \bibinfo
  {author} {\bibfnamefont{P.}~\bibnamefont{Jimenez-Delgado}}, \bibinfo {author}
  {\bibfnamefont{E.}~\bibnamefont{Reya}},\ and\ \bibinfo {author}
  {\bibfnamefont{C.}~\bibnamefont{Schuck}},\ }%
  \bibfield{journal}{%
  \Doi{10.1016/j.physletb.2008.04.063}{\bibinfo {journal} {Phys. Lett.}}\ }%
  \textbf{\bibinfo {volume} {B664}},\ \bibinfo {pages} {133} (\bibinfo {year}
  {2008}),\ \Eprint{http://arxiv.org/abs/0801.3618}{arXiv:0801.3618 [hep-ph]}%
  \bibAnnoteFile{NoStop}{Gluck:2008gs}%
%%CITATION = 0801.3618;%%
\bibitem{Gluck:1998xa}%
  \BibitemOpen
  \bibfield{author}{%
  \bibinfo {author} {\bibfnamefont{M.}~\bibnamefont{Gl\"uck}}, \bibinfo
  {author} {\bibfnamefont{E.}~\bibnamefont{Reya}},\ and\ \bibinfo {author}
  {\bibfnamefont{A.}~\bibnamefont{Vogt}},\ }%
  \bibfield{journal}{%
  \Doi{10.1007/s100520050289}{\bibinfo {journal} {Eur. Phys. J.}}\ }%
  \textbf{\bibinfo {volume} {C5}},\ \bibinfo {pages} {461} (\bibinfo {year}
  {1998}),\ \Eprint{http://arxiv.org/abs/hep-ph/9806404}{arXiv:hep-ph/9806404}%
  \bibAnnoteFile{NoStop}{Gluck:1998xa}%
%%CITATION = HEP-PH/9806404;%%
\bibitem{Duraes:2004zt}%
  \BibitemOpen
  \bibfield{author}{%
  \bibinfo {author} {\bibfnamefont{F.~O.}\ \bibnamefont{Duraes}}, \bibinfo
  {author} {\bibfnamefont{F.~S.}\ \bibnamefont{Navarra}},\ and\ \bibinfo
  {author} {\bibfnamefont{M.}~\bibnamefont{Nielsen}},\ }%
  \bibfield{journal}{%
  \bibinfo {journal} {Braz. J. Phys.}\ }%
  \textbf{\bibinfo {volume} {34}},\ \bibinfo {pages} {290} (\bibinfo {year}
  {2004})%
  \bibAnnoteFile{NoStop}{Duraes:2004zt}%
%%CITATION = BJPHE,34,290;%%
\bibitem{Gavin:1996bx}%
  \BibitemOpen
  \bibfield{author}{%
  \bibinfo {author} {\bibfnamefont{S.}~\bibnamefont{Gavin}}, \bibinfo {author}
  {\bibfnamefont{P.~L.}\ \bibnamefont{McGaughey}}, \bibinfo {author}
  {\bibfnamefont{P.~V.}\ \bibnamefont{Ruuskanen}},\ and\ \bibinfo {author}
  {\bibfnamefont{R.}~\bibnamefont{Vogt}},\ }%
  \bibfield{journal}{%
  \Doi{10.1103/PhysRevC.54.2606}{\bibinfo {journal} {Phys. Rev.}}\ }%
  \textbf{\bibinfo {volume} {C54}},\ \bibinfo {pages} {2606} (\bibinfo {year}
  {1996}),\ \Eprint{http://arxiv.org/abs/hep-ph/9604369}{arXiv:hep-ph/9604369}%
  \bibAnnoteFile{NoStop}{Gavin:1996bx}%
%%CITATION = HEP-PH/9604369;%%
\bibitem{Vogt:2002vx}%
  \BibitemOpen
  \bibfield{author}{%
  \bibinfo {author} {\bibfnamefont{R.}~\bibnamefont{Vogt}}}%
   (\bibinfo {year} {2002}),\
  \Eprint{http://arxiv.org/abs/hep-ph/0203151}{arXiv:hep-ph/0203151}%
  \bibAnnoteFile{NoStop}{Vogt:2002vx}%
%%CITATION = HEP-PH/0203151;%%
\bibitem{Xu:2008av}%
  \BibitemOpen
  \bibfield{author}{%
  \bibinfo {author} {\bibfnamefont{Z.}~\bibnamefont{Xu}}\ and\ \bibinfo
  {author} {\bibfnamefont{C.}~\bibnamefont{Greiner}},\ }%
  \bibfield{journal}{%
  \Doi{10.1103/PhysRevC.79.014904}{\bibinfo {journal} {Phys. Rev.}}\ }%
  \textbf{\bibinfo {volume} {C79}},\ \bibinfo {pages} {014904} (\bibinfo {year}
  {2009}),\ \Eprint{http://arxiv.org/abs/0811.2940}{arXiv:0811.2940 [hep-ph]}%
  \bibAnnoteFile{NoStop}{Xu:2008av}%
%%CITATION = 0811.2940;%%
\bibitem{Adler:2004ta}%
  \BibitemOpen
  \bibfield{author}{%
  \bibinfo {author} {\bibfnamefont{S.~S.}\ \bibnamefont{Adler}} \emph{et~al.}
  (\bibinfo {collaboration} {PHENIX}),\ }%
  \bibfield{journal}{%
  \Doi{10.1103/PhysRevLett.94.082301}{\bibinfo {journal} {Phys. Rev. Lett.}}\
  }%
  \textbf{\bibinfo {volume} {94}},\ \bibinfo {pages} {082301} (\bibinfo {year}
  {2005}),\
  \Eprint{http://arxiv.org/abs/nucl-ex/0409028}{arXiv:nucl-ex/0409028}%
  \bibAnnoteFile{NoStop}{Adler:2004ta}%
%%CITATION = NUCL-EX/0409028;%%
\bibitem{:2008hja}%
  \BibitemOpen
  \bibfield{author}{%
  \bibinfo {author} {\bibfnamefont{B.~I.}\ \bibnamefont{Abelev}} \emph{et~al.}
  (\bibinfo {collaboration} {STAR})}%
   (\bibinfo {year} {2008}),\
  \Eprint{http://arxiv.org/abs/0805.0364}{arXiv:0805.0364 [nucl-ex]}%
  \bibAnnoteFile{NoStop}{:2008hja}%
%%CITATION = 0805.0364;%%
\bibitem{Cheng:2006qk}%
  \BibitemOpen
  \bibfield{author}{%
  \bibinfo {author} {\bibfnamefont{M.}~\bibnamefont{Cheng}} \emph{et~al.},\ }%
  \bibfield{journal}{%
  \Doi{10.1103/PhysRevD.74.054507}{\bibinfo {journal} {Phys. Rev.}}\ }%
  \textbf{\bibinfo {volume} {D74}},\ \bibinfo {pages} {054507} (\bibinfo {year}
  {2006}),\
  \Eprint{http://arxiv.org/abs/hep-lat/0608013}{arXiv:hep-lat/0608013}%
  \bibAnnoteFile{NoStop}{Cheng:2006qk}%
%%CITATION = HEP-LAT/0608013;%%
\bibitem{Aoki:2006br}%
  \BibitemOpen
  \bibfield{author}{%
  \bibinfo {author} {\bibfnamefont{Y.}~\bibnamefont{Aoki}}, \bibinfo {author}
  {\bibfnamefont{Z.}~\bibnamefont{Fodor}}, \bibinfo {author}
  {\bibfnamefont{S.~D.}\ \bibnamefont{Katz}},\ and\ \bibinfo {author}
  {\bibfnamefont{K.~K.}\ \bibnamefont{Szabo}},\ }%
  \bibfield{journal}{%
  \Doi{10.1016/j.physletb.2006.10.021}{\bibinfo {journal} {Phys. Lett.}}\ }%
  \textbf{\bibinfo {volume} {B643}},\ \bibinfo {pages} {46} (\bibinfo {year}
  {2006}),\
  \Eprint{http://arxiv.org/abs/hep-lat/0609068}{arXiv:hep-lat/0609068}%
  \bibAnnoteFile{NoStop}{Aoki:2006br}%
%%CITATION = HEP-LAT/0609068;%%
\bibitem{Aoki:2009sc}%
  \BibitemOpen
  \bibfield{author}{%
  \bibinfo {author} {\bibfnamefont{Y.}~\bibnamefont{Aoki}} \emph{et~al.},\ }%
  \bibfield{journal}{%
  \Doi{10.1088/1126-6708/2009/06/088}{\bibinfo {journal} {JHEP}}\ }%
  \textbf{\bibinfo {volume} {06}},\ \bibinfo {pages} {088} (\bibinfo {year}
  {2009}),\ \Eprint{http://arxiv.org/abs/0903.4155}{arXiv:0903.4155 [hep-lat]}%
  \bibAnnoteFile{NoStop}{Aoki:2009sc}%
%%CITATION = 0903.4155;%%
\bibitem{Andronic:2007ff}%
  \BibitemOpen
  \bibfield{author}{%
  \bibinfo {author} {\bibfnamefont{A.}~\bibnamefont{Andronic}}, \bibinfo
  {author} {\bibfnamefont{P.}~\bibnamefont{Braun-Munzinger}}, \bibinfo {author}
  {\bibfnamefont{K.}~\bibnamefont{Redlich}},\ and\ \bibinfo {author}
  {\bibfnamefont{J.}~\bibnamefont{Stachel}},\ }%
  \bibfield{journal}{%
  \bibinfo {journal} {PoS}\ }%
  \textbf{\bibinfo {volume} {CPOD07}},\ \bibinfo {pages} {044} (\bibinfo {year}
  {2007}),\ \Eprint{http://arxiv.org/abs/0710.1851}{arXiv:0710.1851 [nucl-th]}%
  \bibAnnoteFile{NoStop}{Andronic:2007ff}%
%%CITATION = 0710.1851;%%
\bibitem{BraunMunzinger:2000dv}%
  \BibitemOpen
  \bibfield{author}{%
  \bibinfo {author} {\bibfnamefont{P.}~\bibnamefont{Braun-Munzinger}}\ and\
  \bibinfo {author} {\bibfnamefont{K.}~\bibnamefont{Redlich}},\ }%
  \bibfield{journal}{%
  \Doi{10.1007/s100520000356}{\bibinfo {journal} {Eur. Phys. J.}}\ }%
  \textbf{\bibinfo {volume} {C16}},\ \bibinfo {pages} {519} (\bibinfo {year}
  {2000}),\ \Eprint{http://arxiv.org/abs/hep-ph/0001008}{arXiv:hep-ph/0001008}%
  \bibAnnoteFile{NoStop}{BraunMunzinger:2000dv}%
%%CITATION = HEP-PH/0001008;%%
\bibitem{Grandchamp:2005yw}%
  \BibitemOpen
  \bibfield{author}{%
  \bibinfo {author} {\bibfnamefont{L.}~\bibnamefont{Grandchamp}}, \bibinfo
  {author} {\bibfnamefont{S.}~\bibnamefont{Lumpkins}}, \bibinfo {author}
  {\bibfnamefont{D.}~\bibnamefont{Sun}}, \bibinfo {author}
  {\bibfnamefont{H.}~\bibnamefont{van Hees}},\ and\ \bibinfo {author}
  {\bibfnamefont{R.}~\bibnamefont{Rapp}},\ }%
  \bibfield{journal}{%
  \Doi{10.1103/PhysRevC.73.064906}{\bibinfo {journal} {Phys. Rev.}}\ }%
  \textbf{\bibinfo {volume} {C73}},\ \bibinfo {pages} {064906} (\bibinfo {year}
  {2006}),\ \Eprint{http://arxiv.org/abs/hep-ph/0507314}{arXiv:hep-ph/0507314}%
  \bibAnnoteFile{NoStop}{Grandchamp:2005yw}%
%%CITATION = HEP-PH/0507314;%%
\bibitem{Djordjevic:2003zk}%
  \BibitemOpen
  \bibfield{author}{%
  \bibinfo {author} {\bibfnamefont{M.}~\bibnamefont{Djordjevic}}\ and\ \bibinfo
  {author} {\bibfnamefont{M.}~\bibnamefont{Gyulassy}},\ }%
  \bibfield{journal}{%
  \Doi{10.1016/j.nuclphysa.2003.12.020}{\bibinfo {journal} {Nucl. Phys.}}\ }%
  \textbf{\bibinfo {volume} {A733}},\ \bibinfo {pages} {265} (\bibinfo {year}
  {2004}),\
  \Eprint{http://arxiv.org/abs/nucl-th/0310076}{arXiv:nucl-th/0310076}%
  \bibAnnoteFile{NoStop}{Djordjevic:2003zk}%
%%CITATION = NUCL-TH/0310076;%%
\bibitem{Djordjevic:2005db}%
  \BibitemOpen
  \bibfield{author}{%
  \bibinfo {author} {\bibfnamefont{M.}~\bibnamefont{Djordjevic}}, \bibinfo
  {author} {\bibfnamefont{M.}~\bibnamefont{Gyulassy}}, \bibinfo {author}
  {\bibfnamefont{R.}~\bibnamefont{Vogt}},\ and\ \bibinfo {author}
  {\bibfnamefont{S.}~\bibnamefont{Wicks}},\ }%
  \bibfield{journal}{%
  \Doi{10.1016/j.physletb.2005.09.087}{\bibinfo {journal} {Phys. Lett.}}\ }%
  \textbf{\bibinfo {volume} {B632}},\ \bibinfo {pages} {81} (\bibinfo {year}
  {2006}),\
  \Eprint{http://arxiv.org/abs/nucl-th/0507019}{arXiv:nucl-th/0507019}%
  \bibAnnoteFile{NoStop}{Djordjevic:2005db}%
%%CITATION = NUCL-TH/0507019;%%
\bibitem{Zhang:2005ni}%
  \BibitemOpen
  \bibfield{author}{%
  \bibinfo {author} {\bibfnamefont{B.}~\bibnamefont{Zhang}}, \bibinfo {author}
  {\bibfnamefont{L.-W.}\ \bibnamefont{Chen}},\ and\ \bibinfo {author}
  {\bibfnamefont{C.-M.}\ \bibnamefont{Ko}},\ }%
  \bibfield{journal}{%
  \Doi{10.1103/PhysRevC.72.024906}{\bibinfo {journal} {Phys. Rev.}}\ }%
  \textbf{\bibinfo {volume} {C72}},\ \bibinfo {pages} {024906} (\bibinfo {year}
  {2005}),\
  \Eprint{http://arxiv.org/abs/nucl-th/0502056}{arXiv:nucl-th/0502056}%
  \bibAnnoteFile{NoStop}{Zhang:2005ni}%
%%CITATION = NUCL-TH/0502056;%%
\bibitem{Peshier:2008bg}%
  \BibitemOpen
  \bibfield{author}{%
  \bibinfo {author} {\bibfnamefont{A.}~\bibnamefont{Peshier}}}%
   (\bibinfo {year} {2008}),\
  \Eprint{http://arxiv.org/abs/0801.0595}{arXiv:0801.0595 [hep-ph]}%
  \bibAnnoteFile{NoStop}{Peshier:2008bg}%
%%CITATION = 0801.0595;%%
\bibitem{Gossiaux:2009mk}%
  \BibitemOpen
  \bibfield{author}{%
  \bibinfo {author} {\bibfnamefont{P.~B.}\ \bibnamefont{Gossiaux}}, \bibinfo
  {author} {\bibfnamefont{R.}~\bibnamefont{Bierkandt}},\ and\ \bibinfo {author}
  {\bibfnamefont{J.}~\bibnamefont{Aichelin}},\ }%
  \bibfield{journal}{%
  \Doi{10.1103/PhysRevC.79.044906}{\bibinfo {journal} {Phys. Rev.}}\ }%
  \textbf{\bibinfo {volume} {C79}},\ \bibinfo {pages} {044906} (\bibinfo {year}
  {2009}),\ \Eprint{http://arxiv.org/abs/0901.0946}{arXiv:0901.0946 [hep-ph]}%
  \bibAnnoteFile{NoStop}{Gossiaux:2009mk}%
%%CITATION = 0901.0946;%%
\bibitem{Fochler:2008ts}%
  \BibitemOpen
  \bibfield{author}{%
  \bibinfo {author} {\bibfnamefont{O.}~\bibnamefont{Fochler}}, \bibinfo
  {author} {\bibfnamefont{Z.}~\bibnamefont{Xu}},\ and\ \bibinfo {author}
  {\bibfnamefont{C.}~\bibnamefont{Greiner}},\ }%
  \bibfield{journal}{%
  \Doi{10.1103/PhysRevLett.102.202301}{\bibinfo {journal} {Phys. Rev. Lett.}}\
  }%
  \textbf{\bibinfo {volume} {102}},\ \bibinfo {pages} {202301} (\bibinfo {year}
  {2009}),\ \Eprint{http://arxiv.org/abs/0806.1169}{arXiv:0806.1169 [hep-ph]}%
  \bibAnnoteFile{NoStop}{Fochler:2008ts}%
%%CITATION = 0806.1169;%%
\bibitem{Xu:2010cq}%
  \BibitemOpen
  \bibfield{author}{%
  \bibinfo {author} {\bibfnamefont{Z.}~\bibnamefont{Xu}}\ and\ \bibinfo
  {author} {\bibfnamefont{C.}~\bibnamefont{Greiner}}}%
   (\bibinfo {year} {2010}),\
  \Eprint{http://arxiv.org/abs/1001.2912}{arXiv:1001.2912 [hep-ph]}%
  \bibAnnoteFile{NoStop}{Xu:2010cq}%
%%CITATION = 1001.2912;%%
\bibitem{Greco:2003vf}%
  \BibitemOpen
  \bibfield{author}{%
  \bibinfo {author} {\bibfnamefont{V.}~\bibnamefont{Greco}}, \bibinfo {author}
  {\bibfnamefont{C.~M.}\ \bibnamefont{Ko}},\ and\ \bibinfo {author}
  {\bibfnamefont{R.}~\bibnamefont{Rapp}},\ }%
  \bibfield{journal}{%
  \Doi{10.1016/j.physletb.2004.06.064}{\bibinfo {journal} {Phys. Lett.}}\ }%
  \textbf{\bibinfo {volume} {B595}},\ \bibinfo {pages} {202} (\bibinfo {year}
  {2004}),\
  \Eprint{http://arxiv.org/abs/nucl-th/0312100}{arXiv:nucl-th/0312100}%
  \bibAnnoteFile{NoStop}{Greco:2003vf}%
%%CITATION = NUCL-TH/0312100;%%
\bibitem{Lin:2003jy}%
  \BibitemOpen
  \bibfield{author}{%
  \bibinfo {author} {\bibfnamefont{Z.-w.}\ \bibnamefont{Lin}}\ and\ \bibinfo
  {author} {\bibfnamefont{D.}~\bibnamefont{Molnar}},\ }%
  \bibfield{journal}{%
  \Doi{10.1103/PhysRevC.68.044901}{\bibinfo {journal} {Phys. Rev.}}\ }%
  \textbf{\bibinfo {volume} {C68}},\ \bibinfo {pages} {044901} (\bibinfo {year}
  {2003}),\
  \Eprint{http://arxiv.org/abs/nucl-th/0304045}{arXiv:nucl-th/0304045}%
  \bibAnnoteFile{NoStop}{Lin:2003jy}%
%%CITATION = NUCL-TH/0304045;%%
\bibitem{CasalderreySolana:2006rq}%
  \BibitemOpen
  \bibfield{author}{%
  \bibinfo {author} {\bibfnamefont{J.}~\bibnamefont{Casalderrey-Solana}}\ and\
  \bibinfo {author} {\bibfnamefont{D.}~\bibnamefont{Teaney}},\ }%
  \bibfield{journal}{%
  \Doi{10.1103/PhysRevD.74.085012}{\bibinfo {journal} {Phys. Rev.}}\ }%
  \textbf{\bibinfo {volume} {D74}},\ \bibinfo {pages} {085012} (\bibinfo {year}
  {2006}),\ \Eprint{http://arxiv.org/abs/hep-ph/0605199}{arXiv:hep-ph/0605199}%
  \bibAnnoteFile{NoStop}{CasalderreySolana:2006rq}%
%%CITATION = HEP-PH/0605199;%%
\bibitem{Herzog:2006gh}%
  \BibitemOpen
  \bibfield{author}{%
  \bibinfo {author} {\bibfnamefont{C.~P.}\ \bibnamefont{Herzog}}, \bibinfo
  {author} {\bibfnamefont{A.}~\bibnamefont{Karch}}, \bibinfo {author}
  {\bibfnamefont{P.}~\bibnamefont{Kovtun}}, \bibinfo {author}
  {\bibfnamefont{C.}~\bibnamefont{Kozcaz}},\ and\ \bibinfo {author}
  {\bibfnamefont{L.~G.}\ \bibnamefont{Yaffe}},\ }%
  \bibfield{journal}{%
  \bibinfo {journal} {JHEP}\ }%
  \textbf{\bibinfo {volume} {07}},\ \bibinfo {pages} {013} (\bibinfo {year}
  {2006}),\ \Eprint{http://arxiv.org/abs/hep-th/0605158}{arXiv:hep-th/0605158}%
  \bibAnnoteFile{NoStop}{Herzog:2006gh}%
%%CITATION = HEP-TH/0605158;%%
\end{thebibliography}%

\end{document}